\documentclass{emulateapj}
\usepackage{amsmath}
\usepackage{amssymb}
\usepackage{ulem}
\usepackage{longtable}
\usepackage{graphicx}
\usepackage{epstopdf}
\usepackage{natbib}
\usepackage[colorlinks, citecolor=blue]{hyperref}
\usepackage{hyperref}
\usepackage[all]{hypcap}

\shorttitle{Time-resolved photosphere spectra from structured jet}
\input{tcilatex}
\begin{document}

\title{The time-resolved spectra of photospheric emission from a structured
jet for gamma-ray bursts }
\author{Yan-Zhi Meng\altaffilmark{1,2}, Liang-Duan Liu\altaffilmark{3,4},
Jun-Jie Wei\altaffilmark{1}, Xue-Feng Wu\altaffilmark{1}, Bin-Bin Zhang%
\altaffilmark{3,4}}

\begin{abstract}
The quasi-thermal components found in many \textit{Fermi} gamma-ray bursts
(GRBs) imply that the photosphere emission indeed contributes to the prompt
emission of many GRBs. But whether the observed spectra empirically fitted
by the Band function or cutoff power law, especially the spectral and peak
energy ($E_{p}$) evolutions can be explained by the photosphere emission
model alone needs further discussion. In this work, we investigate in detail
the time-resolved spectra and $E_{p}$ evolutions of photospheric emission
from a structured jet, with an inner-constant and outer-decreasing angular
Lorentz factor profile. Also, a continuous wind with a time-dependent wind
luminosity has been considered. We show that the photosphere spectrum near
the peak luminosity is similar to the cutoff power-law spectrum. The
spectrum can have the observed average low-energy spectral index $\alpha $ $%
\sim -1$, and the distribution of the low-energy spectral index in our
photosphere model is similar to that observed ($-2\lesssim $ $\alpha
\lesssim 0$). Furthermore, the two kinds of spectral evolutions during the
decay phase, separated by the width of the core ($\theta _{c}$), are
consistent with the time-resolved spectral analysis results of several 
\textit{Fermi} multi-pulse GRBs and single-pulse GRBs, respectively. Also,
for this photosphere model we can reproduce the two kinds of observed $E_{p}$
evolution patterns rather well. Thus, by considering the photospheric
emission from a structured jet, we reproduce the observations well for the
GRBs best fitted by the cutoff power-law model for the peak-flux spectrum or
the time-integrated spectrum.
\end{abstract}


\affil{\altaffilmark{1}Purple Mountain Observatory, Chinese Academy
of Sciences, Nanjing 210008, China; xfwu@pmo.ac.cn} 
\affil{\altaffilmark{2}University of Chinese Academy of Sciences,
Beijing 100049, China} 
\affil{\altaffilmark{3}School of Astronomy and Space Science,
Nanjing University, Nanjing 210093, China} 
\affil{\altaffilmark{4}Key Laboratory of Modern Astronomy and
Astrophysics (Nanjing University), Ministry of Education, China}

\section{INTRODUCTION}

\label{sec:intro}

After decades of investigations, the radiation mechanism of gamma-ray burst
(GRB) prompt emission remains unclear. Optically thin synchrotron emission
caused by internal shocks \citep{Ree1994} has been the most widely discussed
model for many years, since it can naturally explain the non-thermal nature
of the observed typical spectrum, which is a smoothly joint broken power law
called the \textquotedblleft Band\textquotedblright\ function %
\citep{Band1993}. Observationally, the typical low-energy photon index $%
\alpha $ of the Band function is around $-1$ %
\citep{Pree2000,Nava2011,ZhaBB2011}. However, this model is found to face
several difficulties in recent years. First, many of the observed bursts
have a harder low-energy slope than the death line $\alpha =-2/3$, which
cannot be obtained by basic synchrotron theory %
\citep{Cri1997,Pree1998,Kan2006,Gold2012}. Second, the spectral width for a
large fraction of GRBs is found so narrow that it cannot be explained by
synchrotron radiation \citep{AxBo2015,Yu2015}. Third,\ the narrow
distribution at a few hundred keV of the observed peak energies cannot be
well explained. Finally, since only the relative kinetic energy between
different shells in the internal shock model can be released, the radiation
efficiency is rather low \citep{Koba1997,Lazz1999,Gue2001,Kin2004}, which
contradicts with the observed high efficiency of a few tens of percent %
\citep{Fan2006,ZhaB2007,Beni2015}.

Due to these difficulties for the internal shock model\footnote{%
Some scenarios within the synchrotron radiation model have been proposed to
alleviate these difficulties\textbf{\ }\citep[e.g.,][]{Zhang2011,Geng18}.},
the photospheric emission model seems to be a promising scenario %
\citep[e.g.,][]{Thom1994,Me2000,Ree2005,Pe2011,Toma2011,Fan2012,Lazz2013,Lund2013,Ru2013,Deng2014,Be2015,Gao2015,Pe2015,Ry2017,Acun2018,Hou2018,Meng2018,Li2019}%
. The photospheric emission is the reasonable result of the original
fireball model \citep{Good1986,Pac1986}, since at the base of the outflow
the optical depth is much greater than unity \citep[e.g.][]{Pi1999}. As the
fireball expands and becomes transparent, the internally trapped photons are
eventually released at the photosphere. The photospheric emission model
naturally explains the clustering of the peak energies and the high
radiation efficiency observed.

Indeed, a quasi-thermal component has been found in tens of BATSE GRBs %
\citep{Ry2004,Ry2005,Ry2009} and some \textit{Fermi} GRBs (GRB 090902B, %
\citealt{Abdo2009}, \citealt{Ry2010}, \citealt{ZhaB2011}; GRB100724B, %
\citealt{Gui2011}; GRB 110721A, \citealt{Axel2012}; GRB 100507, %
\citealt{Ghir2013}; GRB 101219B, \citealt{Lar2015}; and the short GRB
120323A, \citealt{Gui2013}). Especially in the case of GRB 090902B, the
photospheric emission dominates the observed emission. But whether the whole
observed Band function or cutoff power law is of a photosphere origin
remains unknown. If they are, the quasi-thermal spectrum needs to be
broadened. Two different ways of broadening have been considered currently:
subphotospheric dissipation \citep{Ree2005,Gian2006,Belo2016,Vur2016} and
geometric broadening \citep{Pe2008,Ito2013,Lund2013,Deng2014}.

Generally, photosphere is defined as a surface where the Thompson scattering
optical depth $\tau $ for a photon is $\tau =1$. Consistent with %
\citet{Abra1991}, \citet{Pe2008} found that the photospheric radius $R_{%
\text{ph}}$ is angle-dependent for a relativistic, spherically symmetric
wind, $R_{\text{ph}}(\theta )\propto $ $(\theta ^{2}/3+1/\Gamma ^{2})$,
where $\theta $ is the angle measured from the line of sight (LOS) and $%
\Gamma $ is the outflow bulk Lorentz factor. But in principle, the photons
can be last scattered at any position $(r,\Omega )$ inside the outflow,
where $r$ is the distance from the explosion center and $\Omega (\theta
,\phi )$ is the angular coordinates. Thus, a probability function $%
P(r,\Omega )$ is brought in to describe the possibility of last scattering
at any location \citep{Pe2008,Belo2011,Pe2011}. Also, the observed spectrum
is a superposition of a collection of blackbodies with different
temperature, therefore it is broadened (namely geometric broadening).

Based on the geometric broadening, \citet{Deng2014} performed a detailed
study of the photosphere emission spectrum for the spherically symmetric
wind. They showed that the spectrum below $E_{p}$ can be modified to $F_{\nu
}\sim \nu ^{1.5}$ ($\alpha \sim +0.5$), which is not consistent well with
the observation ($\alpha \sim -1.0$). Also, for the $E_{p}$ evolution as a
function of photosphere luminosity, the $L_{\text{ph}}-E_{p}$
anti-correlation is clearly shown. The observed hard-to-soft evolution and $%
E_{p}$-intensity tracking \citep{Ford1995,Lia1996,Ghir2010,Lu2010,Lu2012}
cannot be reproduced well. They thus claimed that a more complicated
photosphere model may be needed. Here in this work, by considering the
photosphere emission for a jet with lateral structure, we show that the
observed typical low-energy photon index ($\alpha \sim -1.0$) and $E_{p}$
evolutions (hard-to-soft evolution or $E_{p}$-intensity tracking) can be
obtained. Indeed within the collapsar model \citep{Mac1999}, as the jet gets
through the collapsing progenitor star, the pressure of the surrounding gas
collimates it \citep[e.g.,][]{ZhaWoo2003,Mor2007,Mizu2011}, thus the jet may
have angular profiles of energy flux and Lorentz factor, namely a structured
jet \citep[e.g.,][]{Dai2001,Ro2002,ZhaB2002,ZhaB2004,Beni2019,BeNa2019}.
Noteworthily, this structured jet is well believed to exist in GRB 170817A
(the first joint detection of short GRB and gravitational wave, %
\citealt{Lazzati2018,Lyman2018,Meng2018,Mooley2018,ZhangBB18b,Ghir2019}),
whose unusual performance of the prompt emission and the afterglow has
invoked hot debate %
\citep[e.g.,][]{Ai2018,Geng2018,LiB2018,Lin2018,Geng2019,Lan2019,LiL2019,Wang2019}%
. Previously, \citet{Lund2013} showed that, with the collimated and
steady-state jet the photospheric spectrum can reproduce the observed
average low-energy photon index $\alpha \approx -1$. But for the
photospheric emission from a structured jet, the time-resolved spectra, the
spectral evolutions, more reasonable energy injection of continuous wind,
the effect of variable wind luminosity and the evolution of $E_{p}$ need to
be further explored, which are the main contents in our work.

We calculate the time-resolved photosphere spectra from a structured jet for
progressively more reasonable energy injections: impulsive injection,
continuous wind with a constant wind luminosity and continuous wind with a
variable wind luminosity. We also perform the time-resolved spectral
analysis of several GRBs observed by \textit{Fermi} GBM and possessing a
pulse that has a rather good profile, and then compare the spectral
evolutions of this analysis and the model. In addition, we discuss the
luminosity profiles and $E_{p}$ evolution based on the model calculation. We
show that the photosphere spectrum around the peak luminosity is close to
the spectrum of the cutoff power-law model, which is the best-fit model for
large amounts of the time-resolved spectra in GRBs %
\citep[e.g.,][]{Kan2006,Yu2016}. Also, the spectrum can get a flattened
shape ($\alpha $ $\sim -1$) below the peak, and the distribution of the
low-energy spectral index is similar to that observed ($-2\lesssim $ $\alpha
\lesssim 0$). Based on the model calculation, the two types of spectral
evolutions (decided by the width of the core) during the decay phase are
consistent with the time-resolved spectral analysis results of several 
\textit{Fermi} multi-pulse GRBs and single-pulse GRBs, respectively.
Finally, for this photosphere model we can reproduce the two types of
observed $E_{p}$ evolution patterns rather well.

The paper is organized as follows. In Section \ref{sec:assum}, we describe
the basic assumptions in our photosphere model. We then present the
calculations of the time-resolved photosphere spectra for progressively more
reasonable energy injections, the time-resolved spectral analysis of several 
\textit{Fermi }GRBs and the discussion on $E_{p}$ evolution in Section \ref%
{sec:res}. The conclusions are drawn in Section \ref{sec:con}.

\section{Basic Assumptions}

\label{sec:assum}

\begin{figure}[tbph]
\label{Fig_1} \centering\includegraphics[angle=0,height=2.1in]{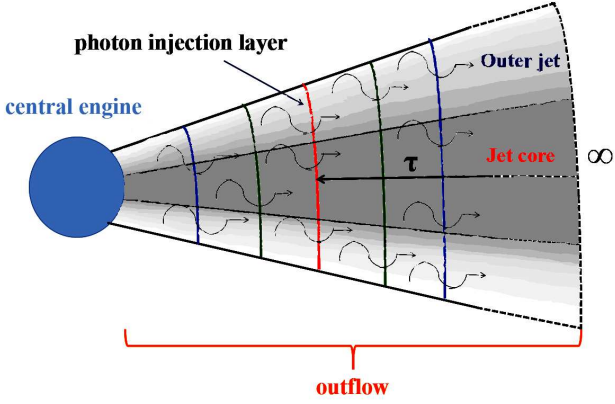}
\caption{{}Basic picture of our model. Long-lasting central engine with\ a
variable wind luminosity and an inner constant and outer decreased angular
Lorentz factor profile are assumed. Different colors of the layers represent
different wind luminosity (the maximum for the red, the minimum for the
blue). Also, the grayscale in the lateral directions shows the Lorentz
factor profile (darker for larger Lorentz factor). }
\end{figure}

The basic physical picture of our paper is shown in Figure 1. The photons
are continuously emitted into a series of layers released by a long-lasting
central engine, and the wind luminosity of the central engine is
time-dependent, similar to those in \citet{Deng2014} (see Figure 1 therein).
But in our model, the jet is structured, with an inner-constant and
outer-decreasing angular Lorentz factor profile (as seen in Figure 1 of %
\citealt{Lund2013}) and an angle-independent luminosity\footnote{%
As shown in \citet{Lund2013}, for a prompt GRB spectrum the part expected to
be observed is formed by the photons making their final scattering at
approximately $\lesssim 5/$ $\Gamma _{0}$, where $dL/d\Omega \approx $ const
(see the top panels of Figures 8 and 9 in \citealt{ZhaWoo2003}).}. The
angular Lorentz factor profile takes the form 
\begin{equation}
(\Gamma -\Gamma _{\min })^{2}=\frac{(\Gamma _{0}-\Gamma _{\min })^{2}}{%
(\theta /\theta _{c})^{2p}+1},
\end{equation}%
where $\Gamma _{0}$ is the constant Lorentz factor in the jet core, $\theta
_{c}$ is the half-opening angle for the jet core, $p$ is the power-law index
of the profile, and $\Gamma _{\min }=1.2$ is the minimum value of the
Lorentz factor.

Also, the effect of the non-zero viewing angle $(\theta _{\text{v}})$ is the
same as that shown in Figure 2 of \citet{Lund2013}. There are two sets of
spherical coordinates in the jet: the spherical coordinates $(r,\theta ,\phi
)$ with the polar axis parallel to the jet axis of symmetry and the
spherical coordinates $(r,\theta _{\text{LOS}},\phi _{\text{LOS}})$ with the
polar axis parallel to the LOS. The radial coordinate $(r)$ is along the
axis at an angle $\theta _{\text{LOS}}$ to the LOS and $\theta $ to the jet
axis.

\section{THE TIME-RESOLVED SPECTRA OF MODEL AND OBSERVATION}

\label{sec:res}

In this section, we firstly present calculations of the time-resolved
photosphere spectra for progressively more reasonable energy injections,
impulsive injection in Section 3.1, continuous wind with a constant wind
luminosity in Section 3.2, and continuous wind with a variable wind
luminosity in Section 3.3. Then in Section 3.4, we compare them with the
time-resolved spectral analysis results of several GRBs observed by \textit{%
Fermi} GBM and possessing a pulse that has a rather good profile. Finally,
discussions on the luminosity profiles and $E_{p}$ evolution patterns are
presented in Section 3.5.

\subsection{Impulsive Injection}

\begin{figure}[bh]
\label{Fig_2} \centering\includegraphics[angle=0,scale=0.18]{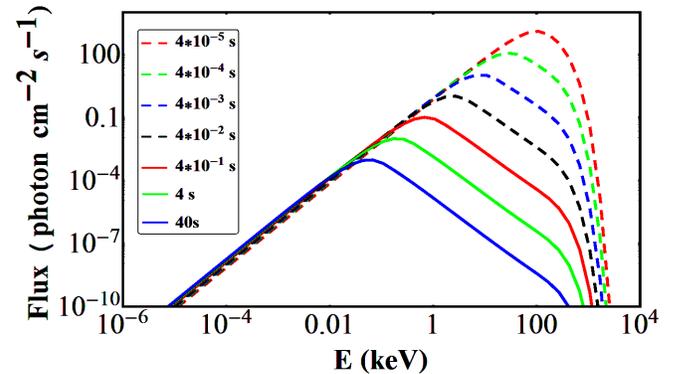}
\caption{Instantaneous photosphere spectra from a structural jet with
impulsive injection of energy. The Lorentz factor profile $\Gamma _{0}=400$, 
$\protect\theta _{c}\Gamma _{0}=1$ and $p=1$ is used along with $\protect%
\theta _{\text{v}}$ $=0$. A total outflow luminosity of $L=10^{52}$ erg s$%
^{-1}$ is assumed,\ base outflow radius $r_{0}=10^{8}$ cm, and luminosity
distance $d_{\text{L}}=$ $4.85\times 10^{28}$ cm ($z=2$). Different line
styles represent different observational times.\ \ \ }
\end{figure}

\begin{figure*}[th]
\label{Fig_3} \centering\includegraphics[angle=0,height=2.0in]{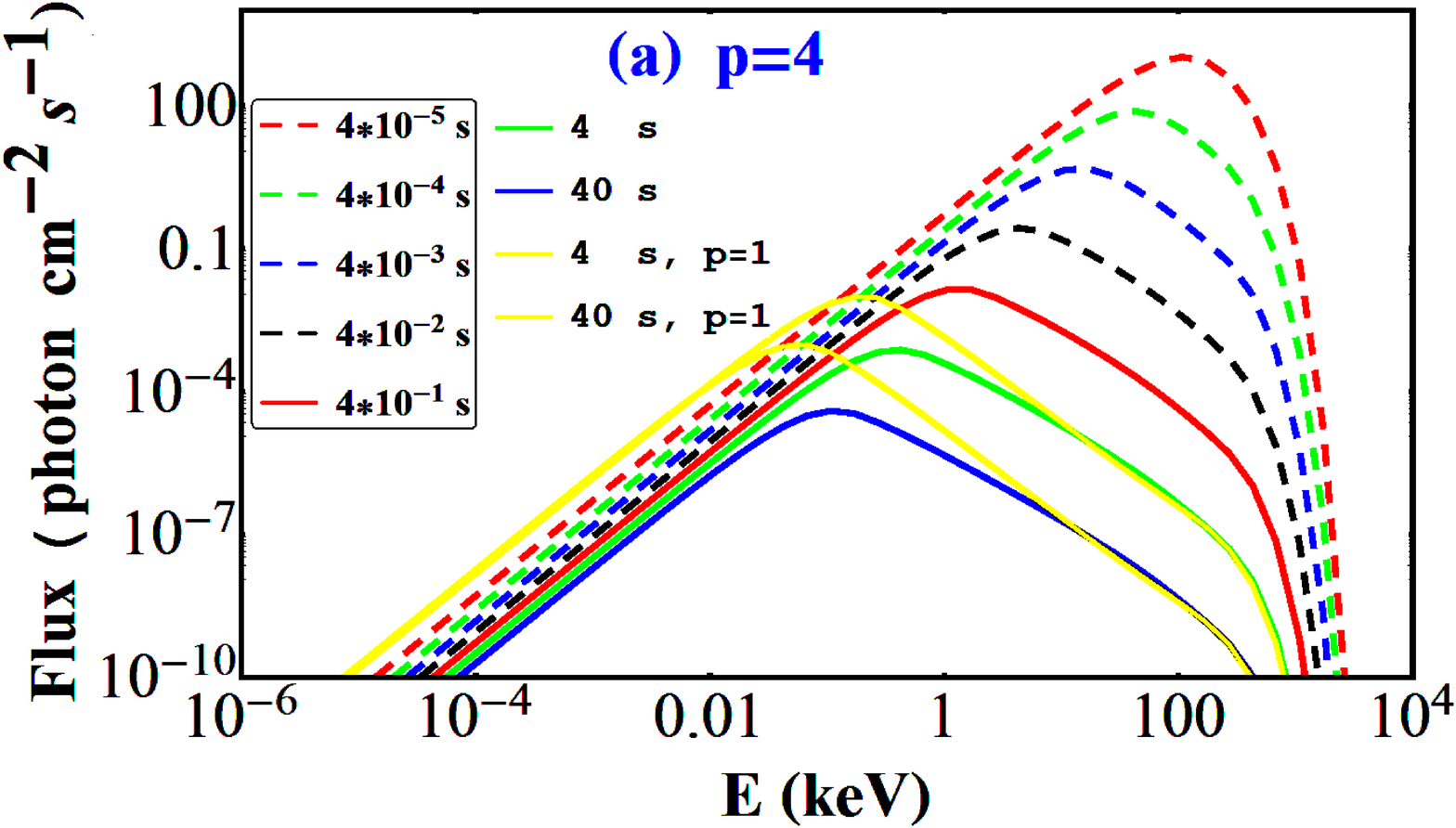} \ \ %
\centering\includegraphics[angle=0,height=2.0in]{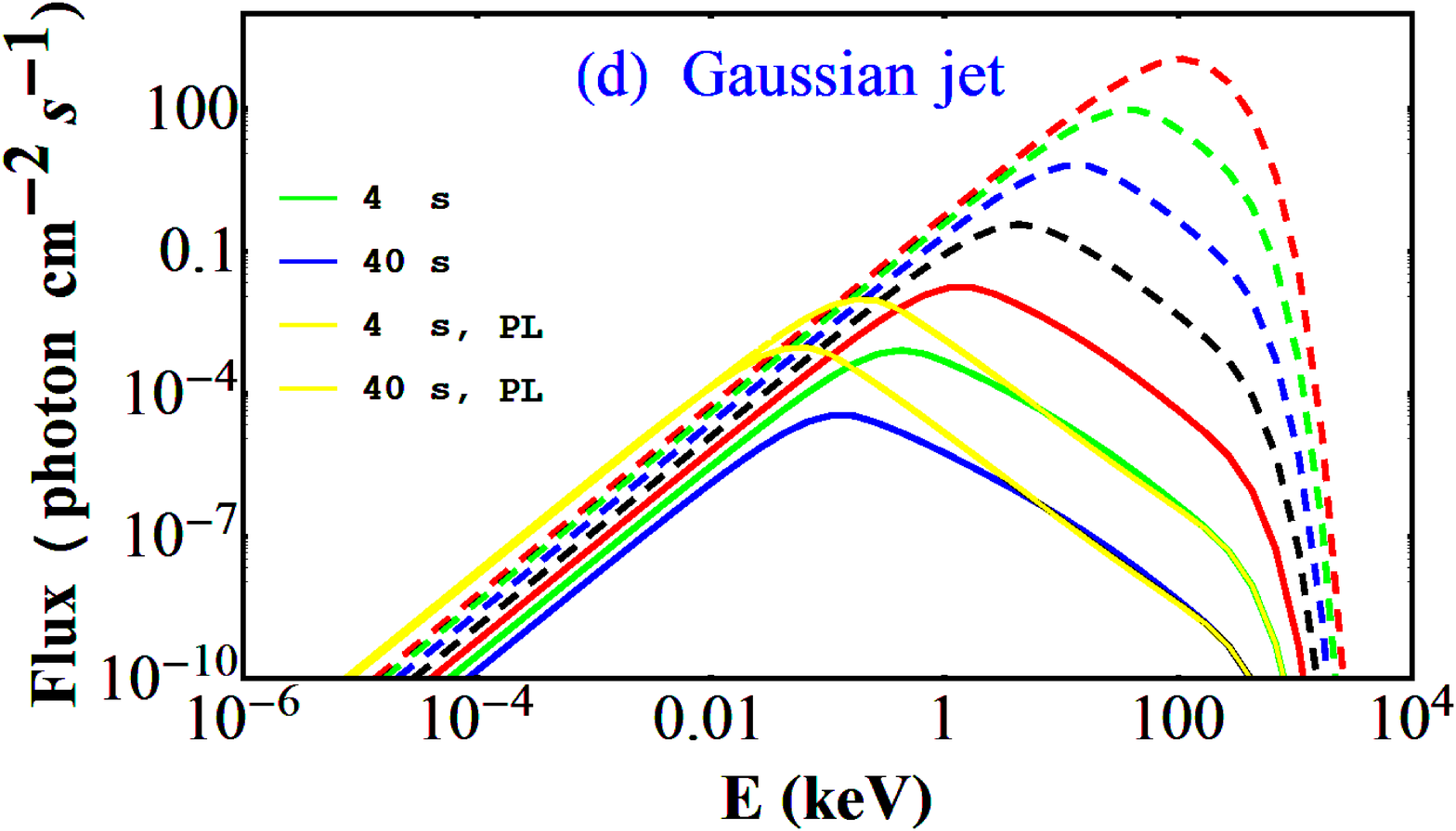} \ \ \centering%
\includegraphics[angle=0,height=2.0in]{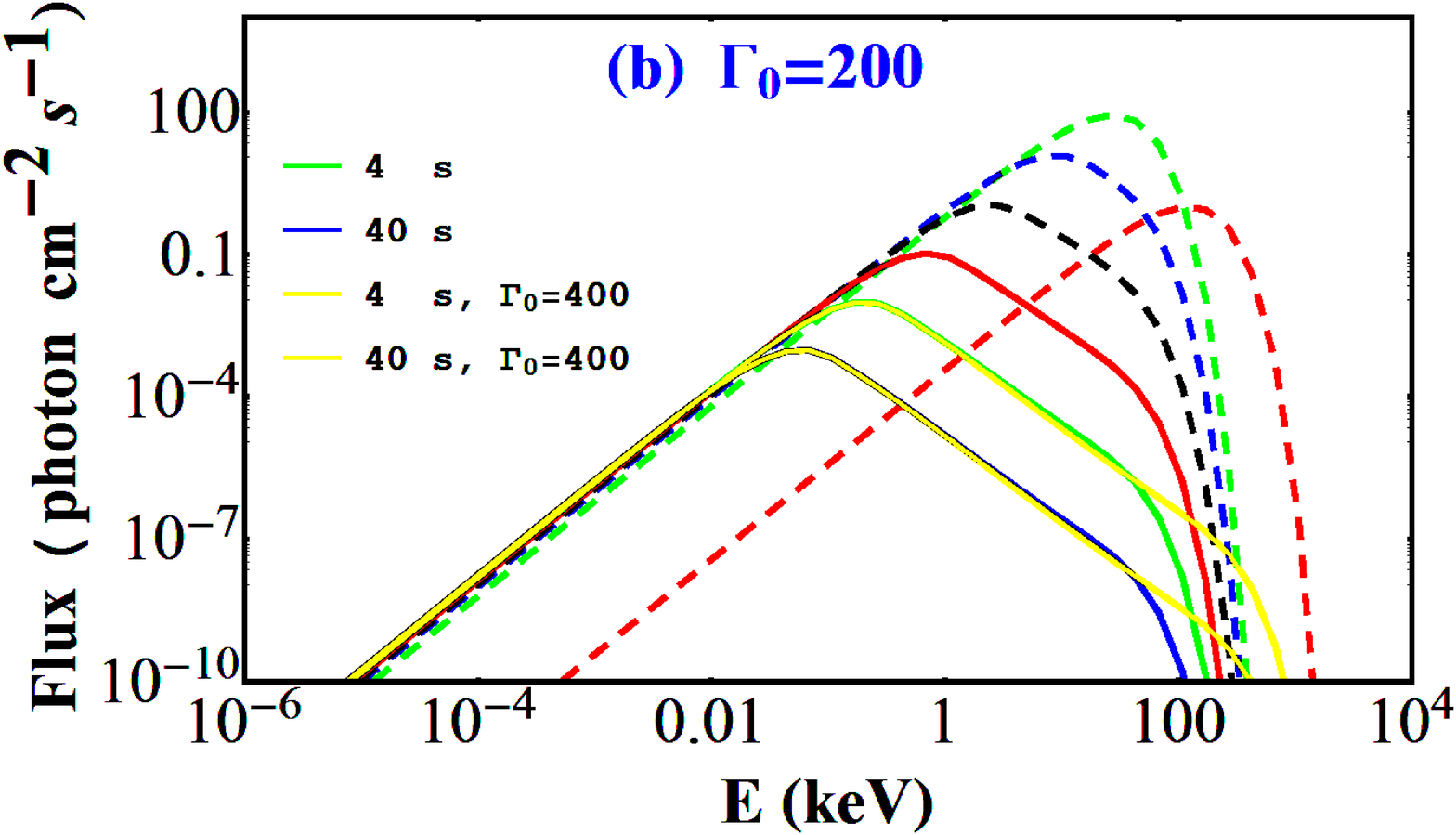} \ \ \centering%
\includegraphics[angle=0,height=2.0in]{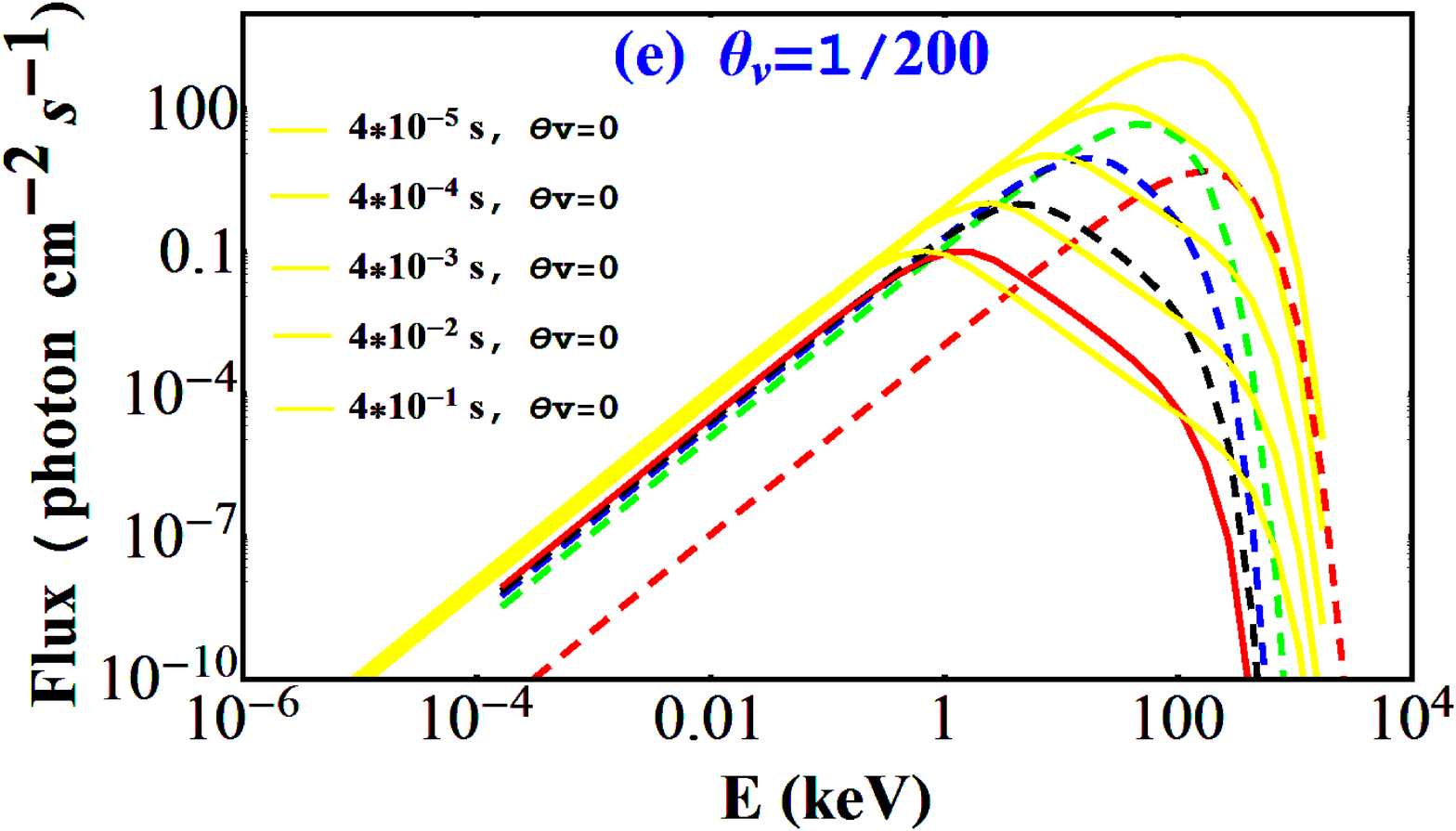} \ \ \ \centering%
\includegraphics[angle=0,height=2.0in]{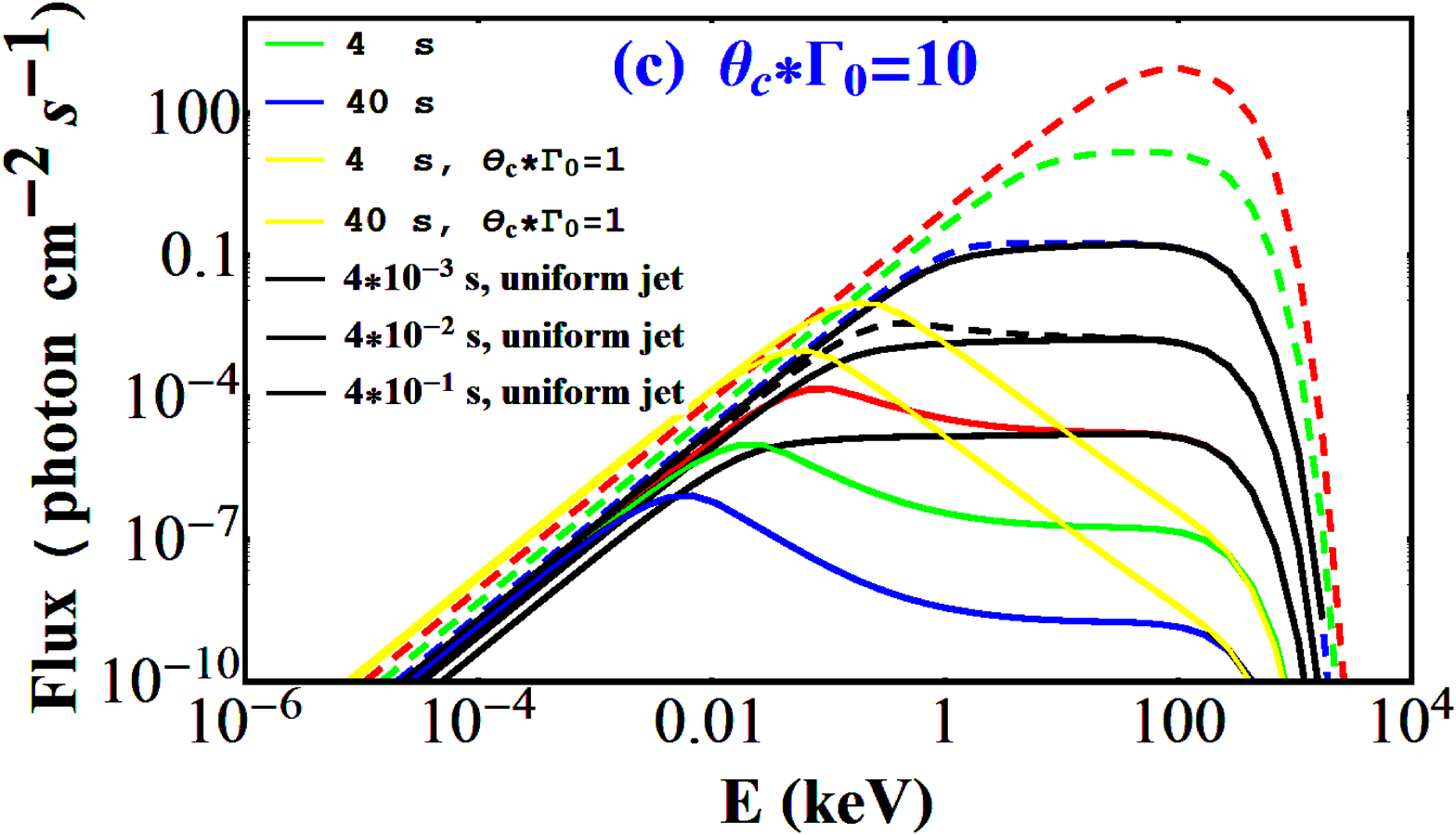} .
\caption{Time-resolved spectra for different Lorentz factor profiles or
viewing angle. (a) A larger Lorentz factor gradient $p=4$,\ along with $%
\Gamma _{0}=400$, $\protect\theta _{c}\Gamma _{0}=1$ and $\protect\theta _{%
\text{v}}=0$. (b) $\ \Gamma _{0}=200$, with $\protect\theta _{c}=1/$ $\Gamma
_{0}$, $p=1$ and $\protect\theta _{\text{v}}=0$. \ (c) \ A wider jet core $%
\protect\theta _{c}=10/\Gamma _{0}=1/40$, with $\Gamma _{0}=400$, $p=1$ and $%
\protect\theta _{\text{v}}=0$. \ (d) \ Gaussian jet\ with $\Gamma _{0}=400$, 
$\protect\theta _{c}=1/400$, and $\protect\theta _{\text{v}}=0$. \ (e) \
Non-zero viewing angle $\protect\theta _{\text{v}}=1/200$, with $\Gamma
_{0}=400$, $\protect\theta _{c}=1/400$ and $p=1$. The yellow spectra in each
figure represent the spectra of the corresponding time in Figure 2. While
the black solid spectra in Figure 3c are for the uniform jet. Note that the
line styles and the corresponding observational times within the box in
Figure 3a are applied to all the figures (from 3a to 3e).}
\end{figure*}

In this section, to calculate the time-resolved spectra, we modify Equation
(10) in \citet{Lund2013} which calculates the time-integrated photospheric
spectrum:$\ $%
\begin{eqnarray}
F_{E}^{\text{ob}}(\theta _{\text{v}}) &=&\frac{1}{4\pi d_{\text{L}}^{2}}%
\iint (1+\beta )D^{2}\frac{d\dot{N}_{\gamma }}{d\Omega }\times \frac{R_{%
\text{dcp}}}{r^{2}}\exp \left( -\frac{R_{\text{ph}}}{r}\right)  \notag \\
&&\times \left\{ E\frac{dP}{dE}\right\} d\Omega dr,  \label{a}
\end{eqnarray}%
where both the velocity $\beta $ and the Doppler factor $D=[\Gamma (1-\beta
\cos \theta _{\text{LOS}})]^{-1}$ depend on the angle $\theta $ to the jet
axis of symmetry [$\beta =\beta (\theta )$, $D=D(\theta ,\theta _{\text{LOS}%
})$]. $\theta _{\text{LOS}}$ is\ the angle to the LOS. While the viewing
angle $\theta _{\text{v}}$ is the angle of the jet axis of symmetry to the
LOS. If $\theta _{\text{v}}=0$, then $\theta =\theta _{\text{LOS}}$. Else if
\ $\theta _{\text{v}}>0$, we have 
\begin{align}
\theta & =\theta (\theta _{\text{LOS}},\phi _{\text{LOS}})  \notag \\
& =\arccos [\cos (\theta _{\text{LOS}})\cos (\theta _{\text{v}})+\sin
(\theta _{\text{LOS}})\sin (\theta _{\text{v}})\cos \phi _{\text{LOS}}].%
\text{ \ \ \ \ \ \ \ \ }
\end{align}%
Thus, $\beta (\theta _{\text{LOS}},\phi _{\text{LOS}})=\beta \lbrack \theta
(\theta _{\text{LOS}},\phi _{\text{LOS}})]$ and $D(\theta _{\text{LOS}},\phi
_{\text{LOS}})=D[\theta (\theta _{\text{LOS}},\phi _{\text{LOS}}),$ $\theta
_{\text{LOS}}]$.

Since the outflow luminosity is assumed to be angle-independent, the photon
emission rate at the base of the outflow, $r=r_{0}$, is also independent of
angle, $d\dot{N}_{\gamma }/d\Omega =\dot{N}_{\gamma }/4\pi $ and $\dot{N}%
_{\gamma }=L/2.7k_{\text{B}}T_{0}$, where $L$ is the total outflow
luminosity and $T_{0}=(L/4\pi r_{0}^{2}ac)^{1/4}$ is the base outflow
temperature.

The decoupling radius,$\ R_{\text{dcp}}$, is defined as the radius where the
optical depth to scattering a photon moving in the radial direction becomes
unity. While the photospheric radius, $R_{\text{ph}}$, is the radius where
the optical depth to scattering a photon moving towards the observer becomes
unity. Their difference is the direction where the photon propagates.

The decoupling radius,$\ R_{\text{dcp}}$, can be calculated by 
\begin{equation}
R_{\text{dcp}}=\frac{1}{(1+\beta )\beta \Gamma ^{2}}\frac{\sigma _{\text{T}}%
}{m_{\text{p}}c}\frac{d\dot{M}}{d\Omega },  \label{a1}
\end{equation}%
where\ $d\dot{M}(\theta )/d\Omega =L/4\pi c^{2}\Gamma (\theta )$ is the
angle-dependent mass outflow rate per solid angle.\ Thus, the$\ $decoupling
radius is also angle-dependent.

While the photospheric radius, $R_{\text{ph}}$, can be written as 
\begin{equation}
R_{\text{ph}}=\frac{\sigma _{\text{T}}}{m_{\text{p}}c\sin \theta _{\text{LOS}%
}}\int\nolimits_{0}^{\theta _{\text{LOS}}}\frac{(1-\beta \cos \tilde{\theta}%
_{_{\text{LOS}}})}{\beta }\frac{d\dot{M}}{d\Omega }\text{ }d\tilde{\theta}%
_{_{\text{LOS}}}.\text{ \ \ \ \ \ \ \ \ \ \ \ \ \ }
\end{equation}

In the former term of Equation $(\ref{a})$, $[(1+\beta )D^{2}/4\pi ]\cdot
(R_{\text{dcp}}/r^{2})\exp \left( -R_{\text{ph}}/r\right) $ represents the
probability density function for the final scattering to happen at the
radius $r$ and the angular coordinate $\Omega (\theta ,\phi )$, namely, $%
P(r,\Omega )$. Also, this term is very similar to the probability density
function used in \citet{Pe2011} and that introduced in \citet{Belo2011}.

For the latter term $E\cdot (dP/dE)$, $dP/dE$ describes the probability for
a photon to have an observer frame energy between $E$ and $E+$ $dE$ within
volume element $dV$. It is derived as 
\begin{equation}
\frac{dP}{dE}=\frac{1}{2.40(k_{\text{B}}T^{\text{ob}})^{3}}\frac{E^{2}}{\exp
(E/k_{\text{B}}T^{\text{ob}})-1},
\end{equation}%
where $T^{\text{ob}}(r,\Omega )=D(\Omega )\cdot $ $T^{^{\prime }}(r,\Omega )$
is the observer frame temperature, $T^{^{\prime }}(r,\Omega )$ is the
comoving temperature. Notice that the comoving temperature also depends on
the angle, since the Lorentz factor $\Gamma $, the saturation radius $R_{s}$
and the photospheric radius $R_{\text{ph}}$ are all angle-dependent, i.e., 
\begin{equation}
T^{\prime }(r,\Omega )=\left\{ 
\begin{array}{c}
\frac{T_{0}}{\Gamma (\Omega )},\text{ \ \ \ \ \ \ \ \ \ \ \ \ \ \ \ \ \ \ \ }%
r<R_{s}(\Omega )<R_{\text{ph}}(\Omega ), \\ 
\frac{T_{0}[r/R_{s}(\Omega )]^{-2/3}}{\Gamma (\Omega )},\text{ \ \ \ \ \ \ \ 
}R_{s}(\Omega )<r<R_{\text{ph}}(\Omega ), \\ 
\frac{T_{0}[R_{\text{ph}}(\Omega )/R_{s}(\Omega )]^{-2/3}}{\Gamma (\Omega )},%
\text{ }R_{s}(\Omega )<R_{\text{ph}}(\Omega )<r.%
\end{array}%
\right.
\end{equation}

Adding a $\delta $-function \ $\delta (t-ru/\beta c)$ to Equation $(\ref{a})$%
, $u=(1-\beta \cos \theta _{\text{LOS}})$ here, we get the formula that
calculates the instantaneous spectrum at the observer time $t$: 
\begin{eqnarray}
F_{E}^{\text{ob}}(\theta _{\text{v}},t) &=&\frac{1}{4\pi d_{\text{L}}^{2}}%
\iint (1+\beta )D^{2}\frac{d\dot{N}_{\gamma }}{d\Omega }\times \frac{R_{%
\text{dcp}}}{r^{2}}\exp \left( -\frac{R_{\text{ph}}}{r}\right)  \notag \\
&&\left\{ E\frac{dP}{dE}\right\} \times \delta (t-\frac{ru}{\beta c})\text{ }%
d\Omega dr.
\end{eqnarray}

Since $\beta c/u$ is independent of $t$, we have $\delta (t-ru/\beta
c)=(\beta c/u)\times \delta (r=\beta ct/u)$. Meanwhile, taking into account
the effect of redshift, we have 
\begin{gather}
F_{E^{\text{ob}}}^{\text{ob}}(\theta _{\text{v}},t)=\frac{1}{4\pi d_{\text{L}%
}^{2}}\int (1+\beta )D^{2}\frac{d\dot{N}_{\gamma }}{d\Omega }\times \frac{R_{%
\text{dcp}}}{r^{2}}\exp \left( -\frac{R_{\text{ph}}}{r}\right)  \notag \\
\text{ \ \ \ \ \ \ \ \ \ }\left\{ E\frac{dP}{dE}\right\} \times \frac{\beta c%
}{u}\text{ }d\Omega \text{, \ \ \ \ }r=\frac{\beta ct}{u}\text{, }E=(1+z)E^{%
\text{ob}}.  \label{a0}
\end{gather}

So, through\ the numerical integration of Equation $(\ref{a0})$, we obtain
the time-resolved spectra. The results are presented in Figs 2 and 3. We
have considered a big set of the parameter space region: $\Gamma _{0}=200$, $%
400$; $p=1$, $4$; $\theta _{c}\Gamma _{0}=1$, $10$ and $\theta _{\text{v}%
}/\theta _{c}=0$, $2$. The Gaussian jet is considered, too. Also, a total
outflow luminosity of $L=10^{52}$ erg s$^{-1}$ and the base outflow radius $%
r_{0}=10^{8}$ cm are assumed. A luminosity distance of $d_{\text{L}}=$ $%
4.85\times 10^{28}$ cm ($z=2$, which is the peak of the GRB formation rate
according to \citealt{Pes2016}) is used for spectrum normalization and
redshift effect.

In Figure 2, we consider a narrow jet core ($\theta _{c}=1/\Gamma _{0}$)
with the Lorentz factor gradient $p=1$ observed at $\theta _{\text{v}}$ $=0$%
. Obviously, in this case the time-resolved photosphere spectrum evolves
from the pure blackbody (early on) to a power law with negative index ($%
F_{\nu }\sim \nu ^{-1.75}$). The late-time spectrum is quite different from
the flattened shape for the uniform jet \citep{Pe2011,Deng2014}. In
addition, the power law has an exponential tail of blackbody emission at the
high-energy end, which is the same as the case of the uniform jet.

From Figure 3, we can compare the spectral evolutions for different Lorentz
factor profiles or viewing angle. As shown in Figure 3a, with a larger
Lorentz factor gradient $p=4$, the late-time power law is more flat. This is
because that the outer jet region becomes more narrow and thus contributes
less to the spectra. Also, Figure 3d is quite similar to Figure 3a, because
the Lorentz factor falls down very quickly for the Gaussian profile, too.
For $p=4$, when $\Gamma \sim 1$, we have $\theta \sim (\Gamma
_{0}/1)^{1/4}\cdot \theta _{c}\sim 4.5\theta _{c}$; while for Gaussian jet, $%
\theta \sim \sqrt{2\ln (\Gamma _{0}/1)}\cdot \theta _{c}\sim 3.5\theta _{c}$%
. Figure 3b shows that, when $\Gamma _{0}$ is smaller the slope of the power
law has no change, but the cut-off energy on the high-energy end decreases.
Surely the peak energy of the early-time blackbody decreases too, and the
blackbody arrives later. Furthermore, Figure 3e is close to Figure 3b, which
means if the viewing angle is non-zero the spectral evolution is similar.
Finally, Figure 3c presents the time-resolved spectra for a wider jet core $%
\theta _{c}=10/\Gamma _{0}$. Compared with the yellow spectra (for the
narrow jet core), we find that the late-time power law flattens
significantly, but not completely (compared with the black solid spectra for
the uniform jet).

\subsection{Continuous Wind with a Constant Wind Luminosity}

GRBs are observed to have a duration, we thus consider the more reasonable
case that the central engine produces a continuous wind. In this section,
the wind luminosity and the baryon loading rate at different time are
assumed to be constant, thus the Lorentz factor is also constant: 
\begin{eqnarray}
L_{w}(\hat{t}) &=&L_{0},  \notag \\
\dot{M}(\hat{t}) &=&\dot{M}_{0},  \notag \\
\eta (\hat{t}) &=&\Gamma (\hat{t})=\Gamma (\hat{t}=0),
\end{eqnarray}%
where $\hat{t}$ indicates the central-engine time since the very first layer
of the wind was injected.

We may consider that the wind consists of many thin layers, with each layer
indicated by its injection time $\hat{t}$ . For a layer ejected from $\hat{t}
$ to $\hat{t}$ $+$ $d\hat{t}$, the spectrum at the observer time $t$ (for $t>%
\hat{t}$) is 
\begin{gather}
F_{E^{\text{ob}}}^{\text{ob}}(\theta _{\text{v}},t,\hat{t})=\frac{1}{4\pi d_{%
\text{L}}^{2}}\int (1+\beta )D^{2}\frac{d\dot{N}_{\gamma }}{d\Omega }\times 
\frac{R_{\text{dcp}}}{r^{2}}  \notag \\
\exp \left( -\frac{R_{\text{ph}}}{r}\right) \text{\ }\left\{ E\frac{dP}{dE}%
\right\} \times \frac{\beta c}{u}\text{ }d\Omega \text{, \ \ }  \notag \\
r=\frac{\beta c(t-\hat{t})}{u}\text{, }E=(1+z)E^{\text{ob}}\text{.}
\end{gather}%
Compared with Equation $(\ref{a0})$, the only difference is \ $r=\beta c(t-%
\hat{t})/u$. Then, by integrating over all the layers, we get the spectrum
at $t$, i.e.,%
\begin{equation}
F_{E^{\text{ob}}}^{\text{ob}}(\theta _{\text{v}},t)=\int%
\nolimits_{0}^{t}F_{E^{\text{ob}}}^{\text{ob}}(\theta _{\text{v}},t,\hat{t})d%
\hat{t}.  \label{c2}
\end{equation}

\begin{figure}[tbp]
\label{Fig_4} \centering\includegraphics[angle=0,height=2.1in]{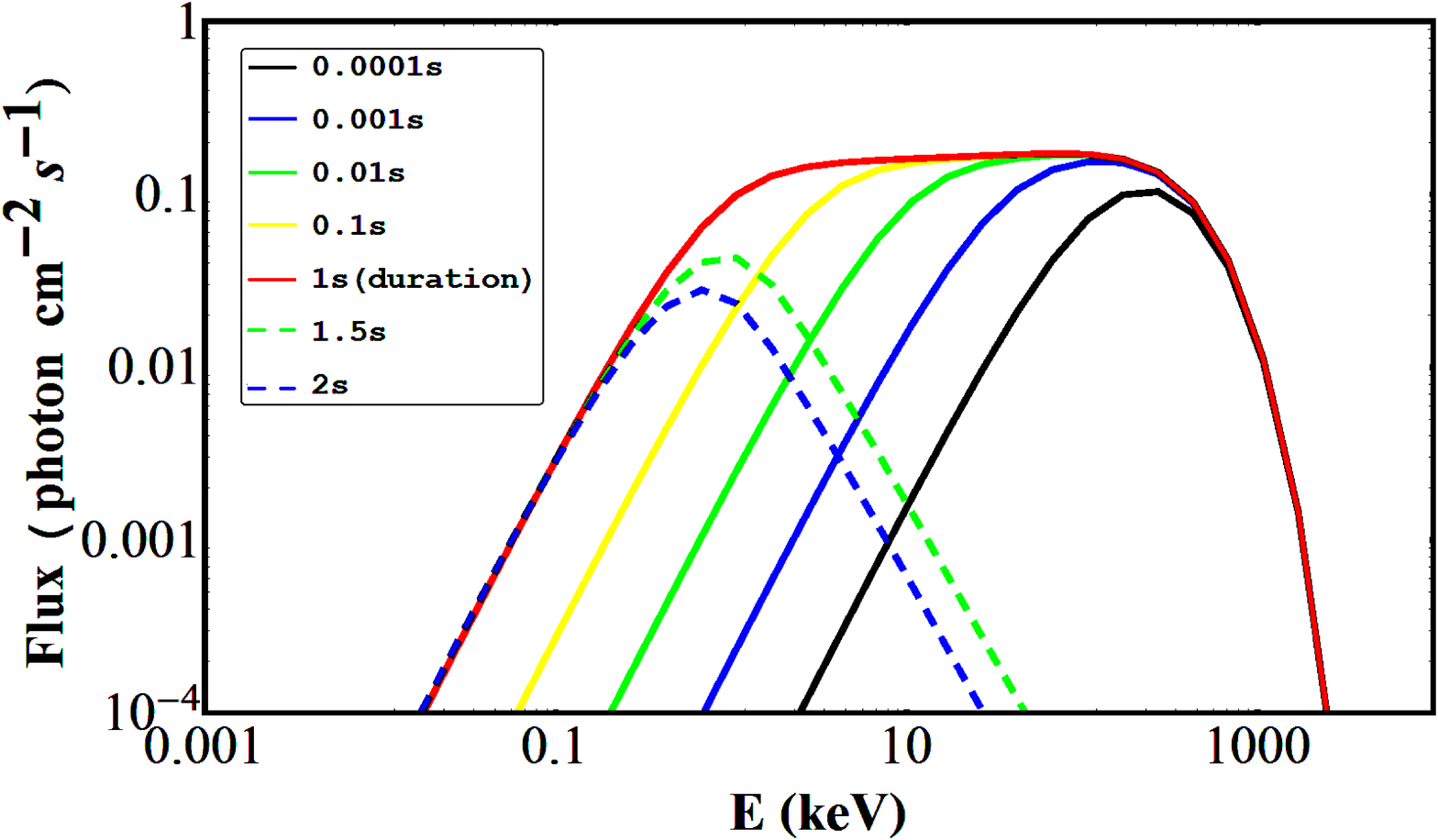}
\caption{{}Time-resolved photosphere spectra for the continuous wind with
abrupt shut-down at $1$ s (red). The parameters are the same as Figure 2.}
\end{figure}

\begin{figure}[tbp]
\label{Fig_5} \centering\includegraphics[angle=0,height=2.1in]{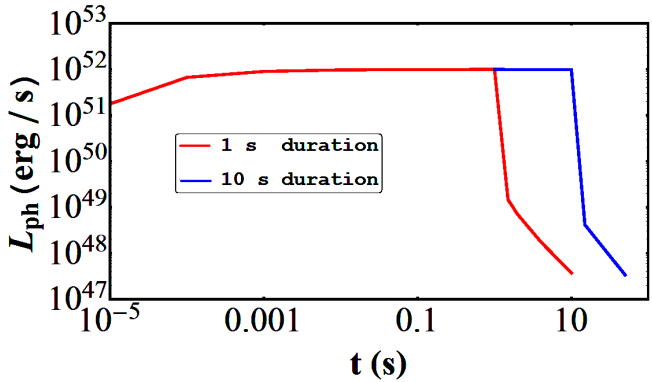}
\caption{{}Photosphere luminosity light curves of continuous winds, which
shut down at $1$ s (red) and $10$ s (blue). After the shut-down, the light
curves fall quickly before entering the $t^{-2}$ phase.}
\end{figure}

Figure 4 presents the time-resolved photosphere spectra of the continuous
wind with time-independent luminosity and Lorentz factor (before $1$ s), and
the spectral evolution after an abrupt shut-down (at $1$ s). Here, we take
the same parameters as Figure 2. Before $1$ s, the spectrum evolves from a
pure blackbody ($t=10^{-4}$ s) to the spectrum with a flattened shape ($%
F_{\nu }\sim \nu ^{0}$) below the peak. This is caused by the superposition
of emission from all layers, since the spectrum from the old layer is a
power law with negative index ($F_{\nu }\sim \nu ^{-1.75}$) as shown in
Figure 2. The flattened shape ($F_{\nu }\sim \nu ^{0}$) below the peak is
consistent with the average low-energy spectral index for the time-resolved
spectra observed in GRBs \citep[e.g.,][]{Kan2006,Yu2016}. After an abrupt
shut-down (at $1$ s) the power law with negative index shows up quickly, but
the flux is predominantly low, meaning that there is a rapid falling phase
(see the photosphere luminosity light curves in Figure 5). This is the same
as the case of the uniform jet.

\subsection{Continuous Wind with a Variable Wind Luminosity}

\begin{figure*}[th]
\label{Fig_6} \centering\includegraphics[angle=0,height=2.0in]{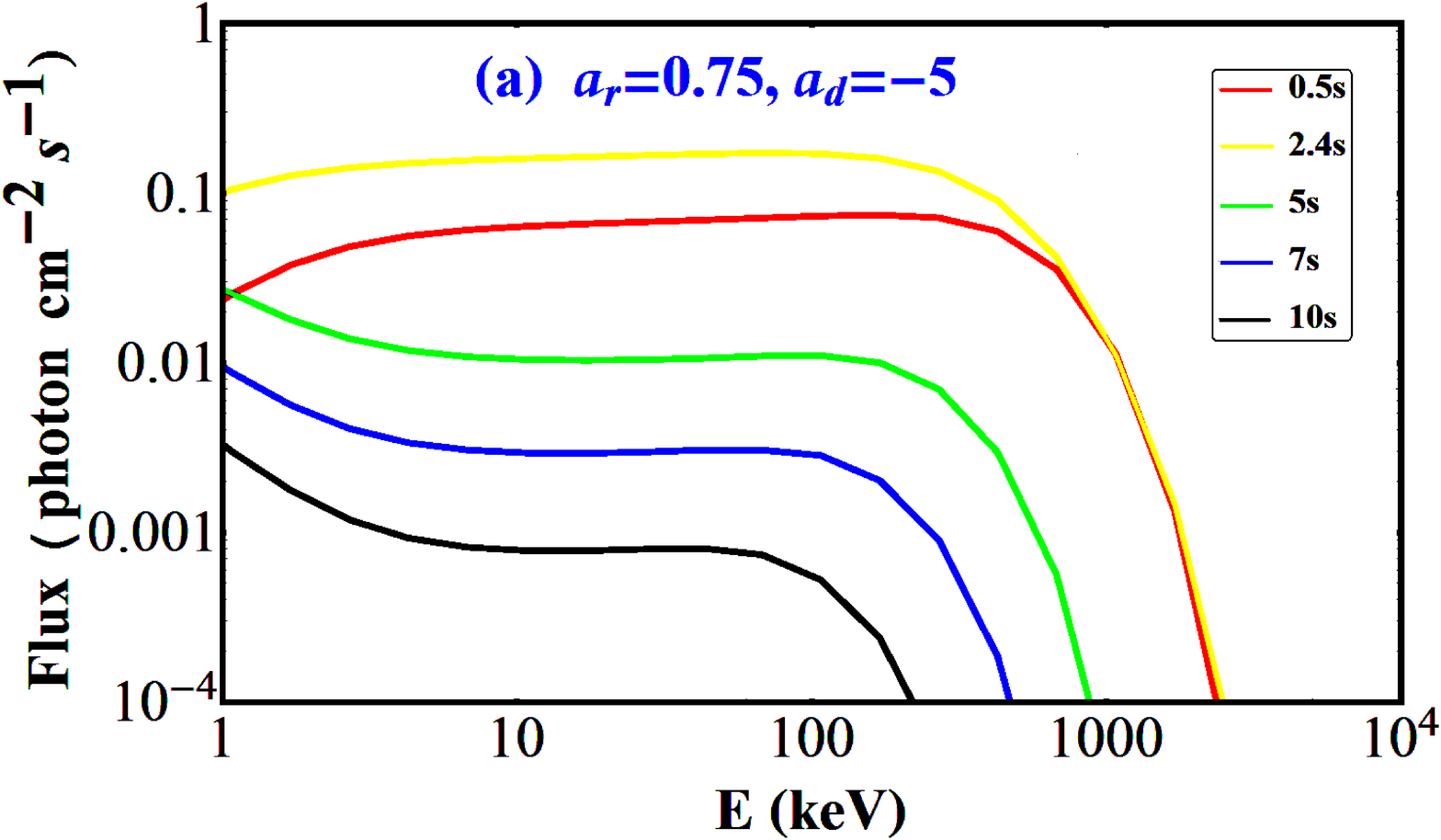} \ \ %
\centering\includegraphics[angle=0,height=2.0in]{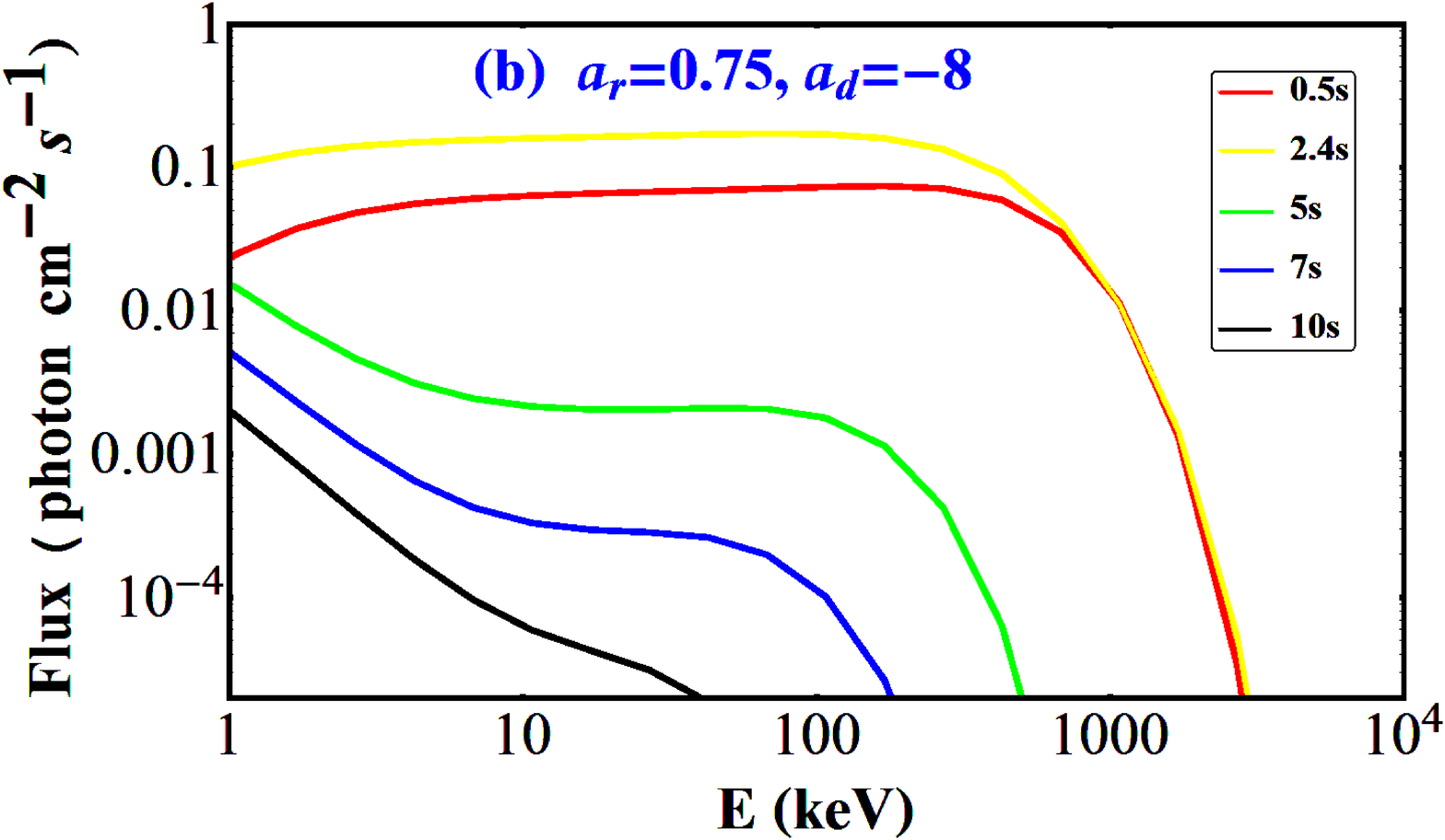} \ \ \centering%
\includegraphics[angle=0,height=2.0in]{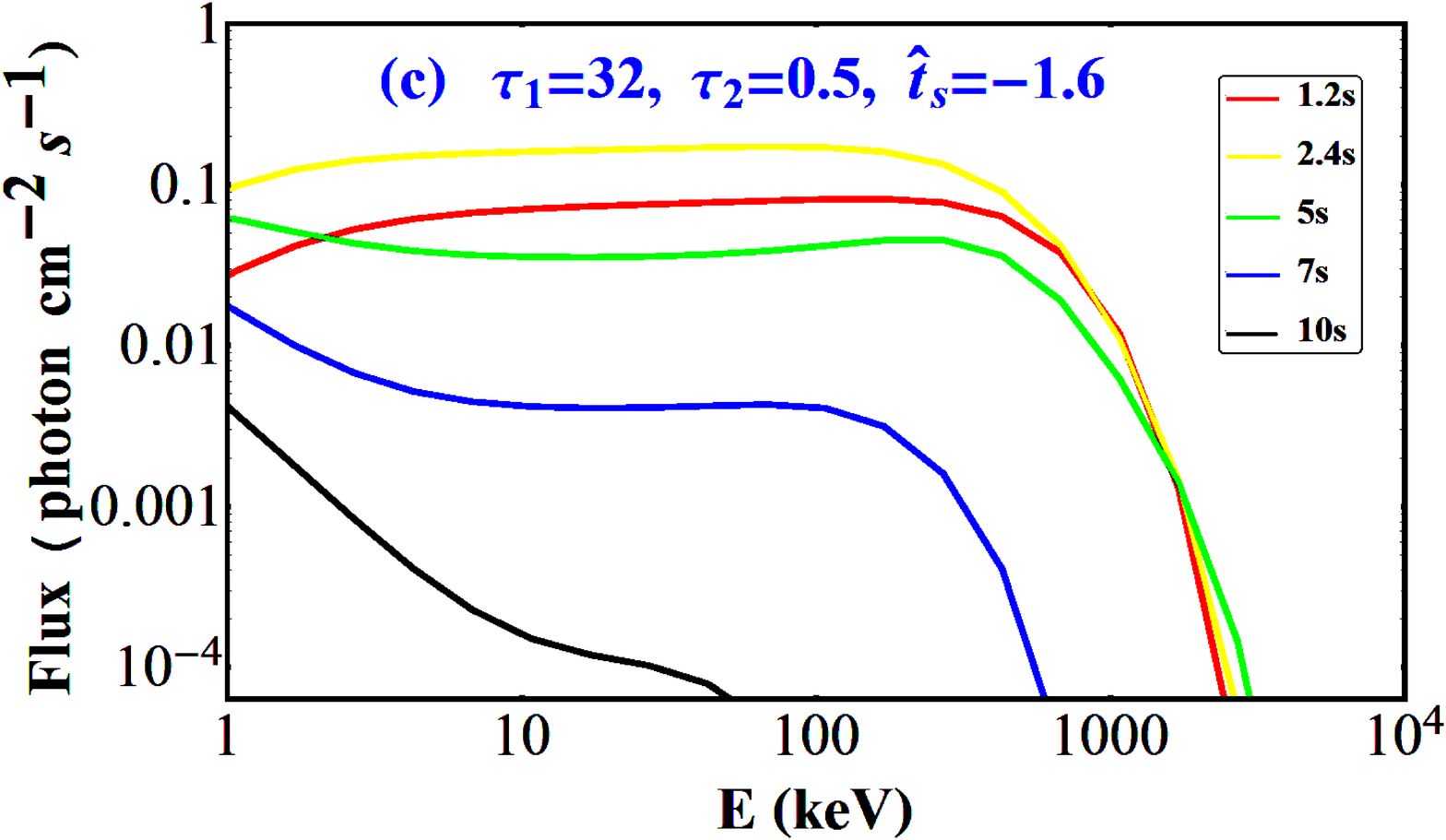} \ \ 
\caption{Time-resolved spectra of winds with variable luminosity for
different luminosity histories. A peak time $\hat{t}_{p}=2.4$ s and a peak
luminosity $L_{w,\text{ }p}$ $=10^{52}$ erg s$^{-1}$ are adopted for all
cases. \ (a) The broken power law model with $a_{r}=0.75$ and $a_{d}=$ $-5$.
\ (b) The broken power law model with $a_{r}=0.75$ and $a_{d}=$ $-8$. \ (c)
\ The exponential model with $\protect\tau _{1}=32$, $\ \protect\tau _{2}$ $%
=0.5\ \ $and$\ \ \hat{t}_{s}=$ $-1.6$. The other parameters are the same as
Figure 2. The Lorentz factor profile $\Gamma _{0}=400$, $\protect\theta %
_{c}\Gamma _{0}=1$ and $p=1$ is used along with $\protect\theta _{\text{v}%
}=0 $. The base outflow radius $r_{0}=10^{8}$ cm is assumed, and luminosity
distance $d_{\text{L}}=$ $4.85\times 10^{28}$ cm ($z=2$). Different colors
represent different observational times.}
\end{figure*}

\begin{figure*}[th]
\label{Fig_7} \centering\includegraphics[angle=0,height=1.9in]{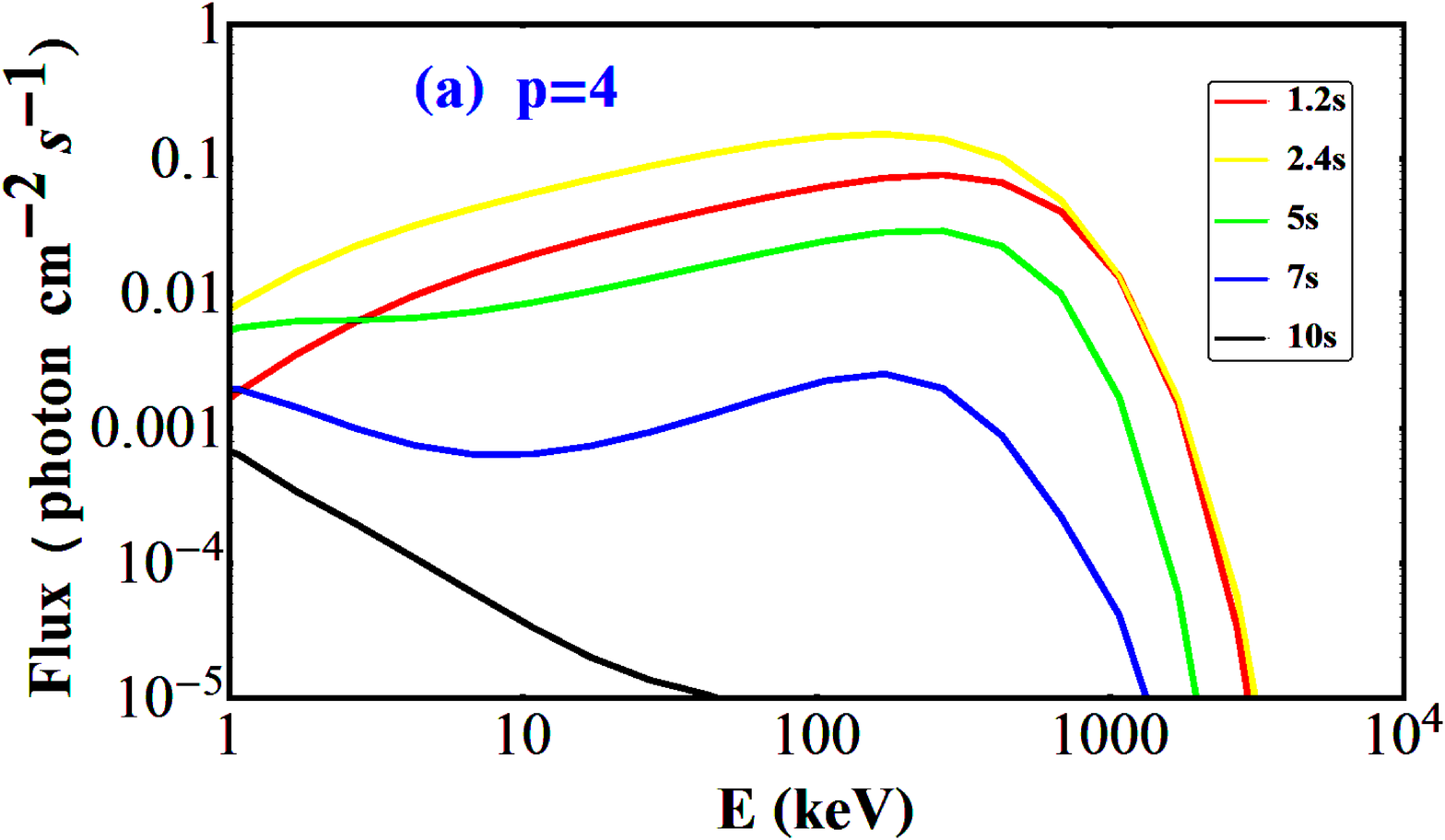} \ \ %
\centering\includegraphics[angle=0,height=1.9in]{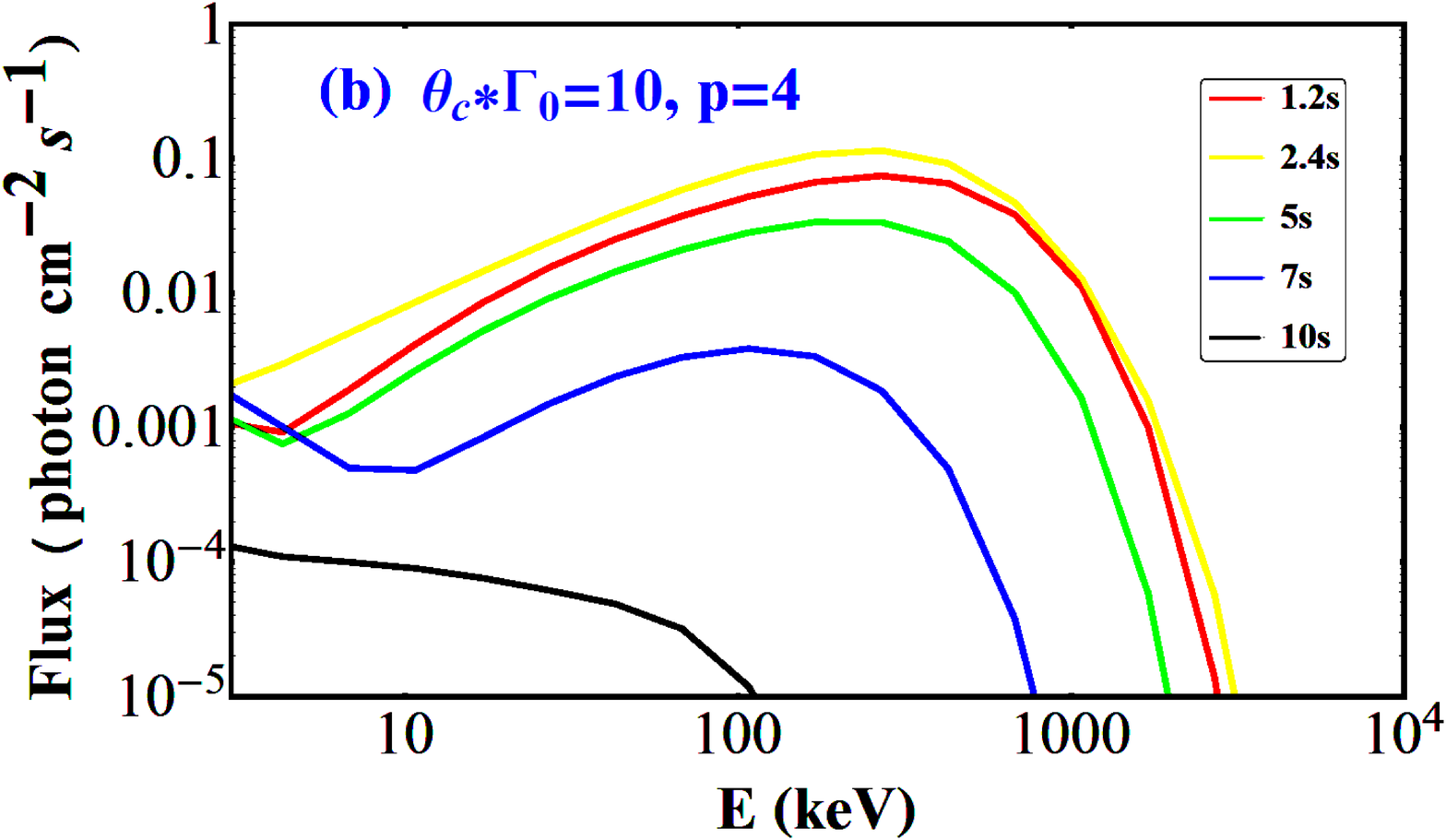} \ \ \centering%
\includegraphics[angle=0,height=1.9in]{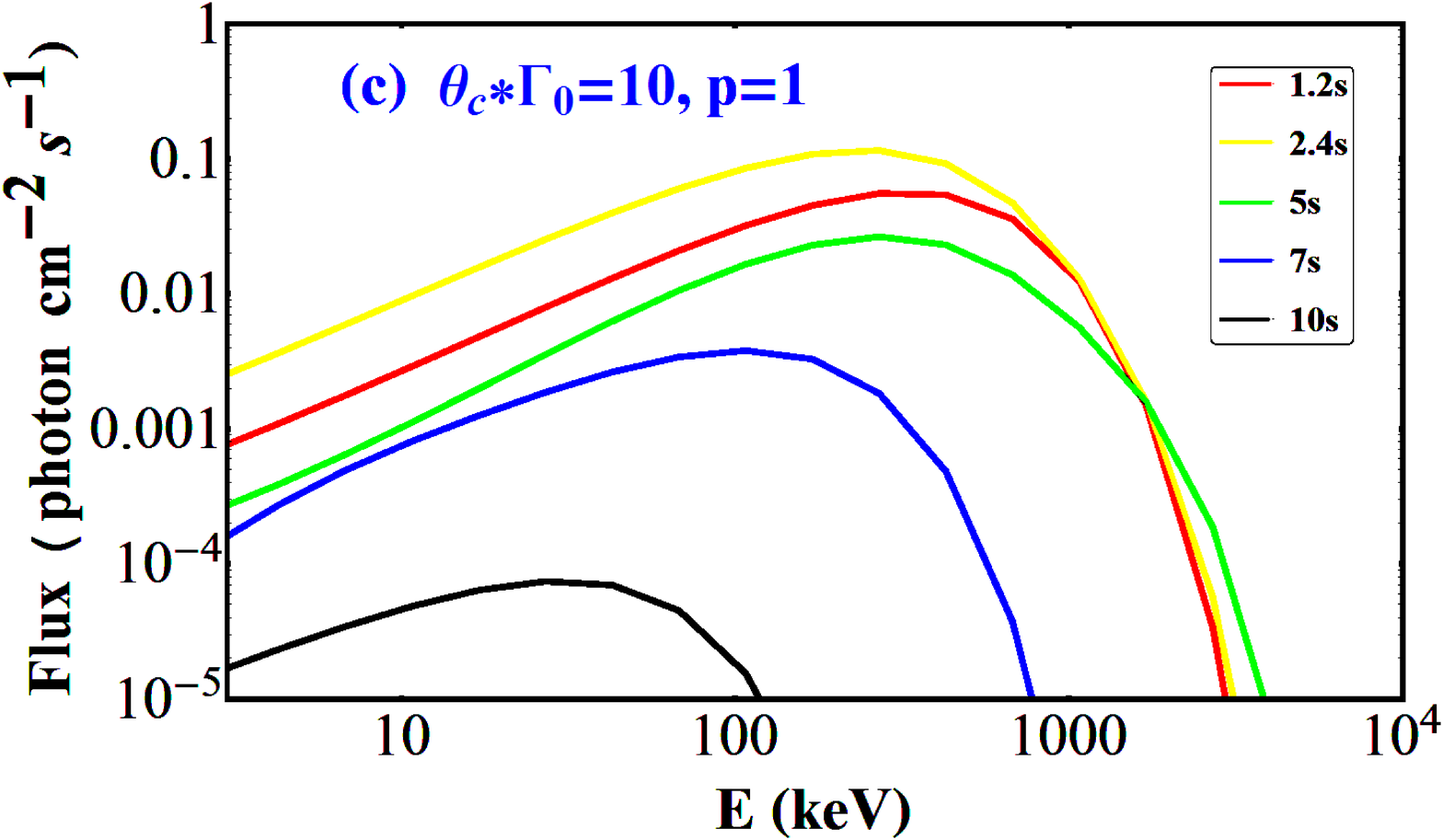} \ \ \centering%
\includegraphics[angle=0,height=1.9in]{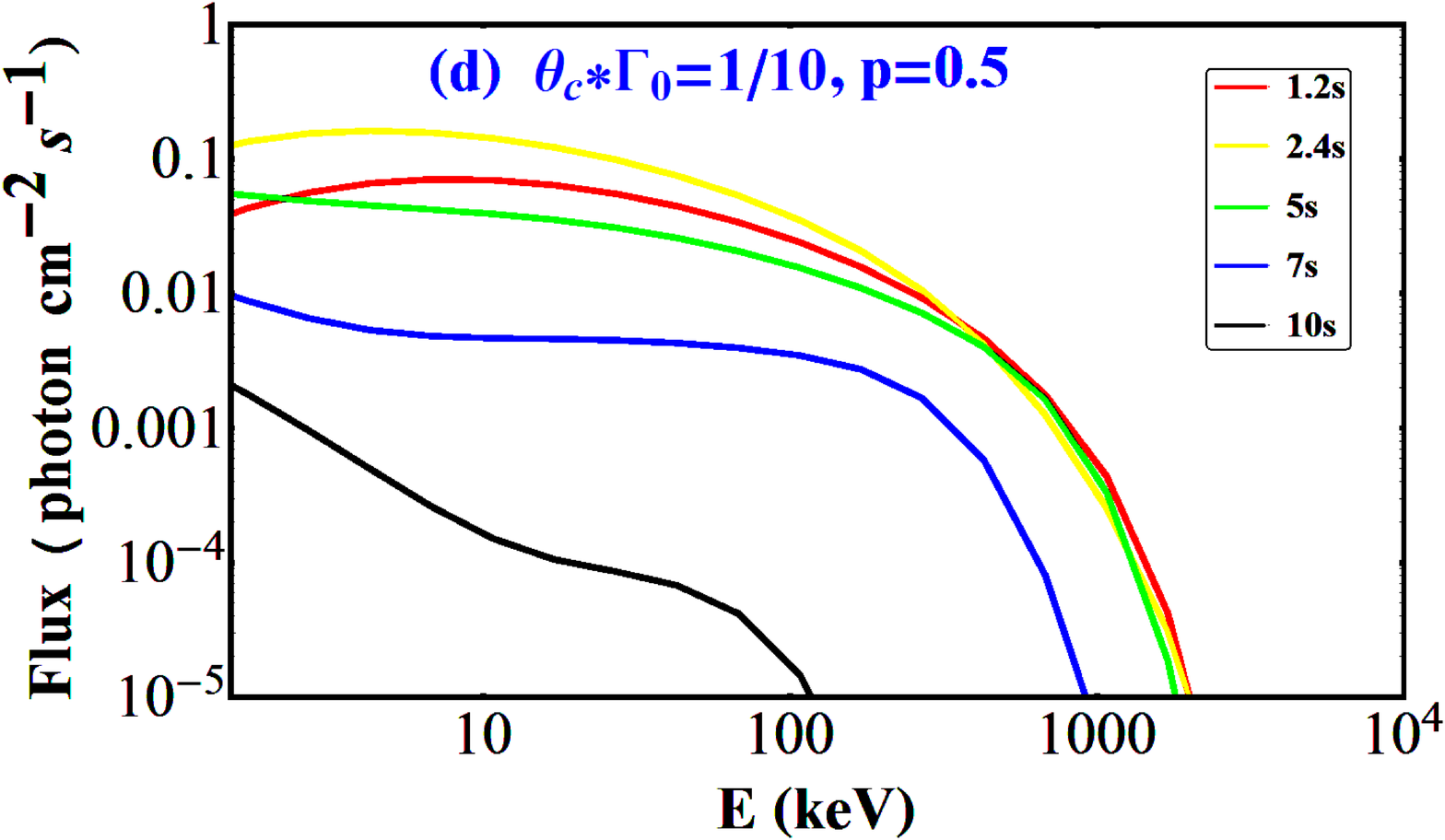} \ \ \ 
\caption{Time-resolved spectra of winds with variable luminosity for
different Lorentz factor profiles. \ (a) $p=4$. \ (b) $\ \protect\theta %
_{c}\Gamma _{0}=10$, $p=4$. \ (c) $\ \protect\theta _{c}\Gamma _{0}=10$, $%
p=1 $. \ (d) $\ \protect\theta _{c}\Gamma _{0}=1/10$, $p=0.5$. The other
parameters are the same as Figure 6c. Different colors represent different
observational times.}
\end{figure*}

\begin{figure*}[ph]
\label{Fig_8} \centering\includegraphics[angle=0,height=1.9in]{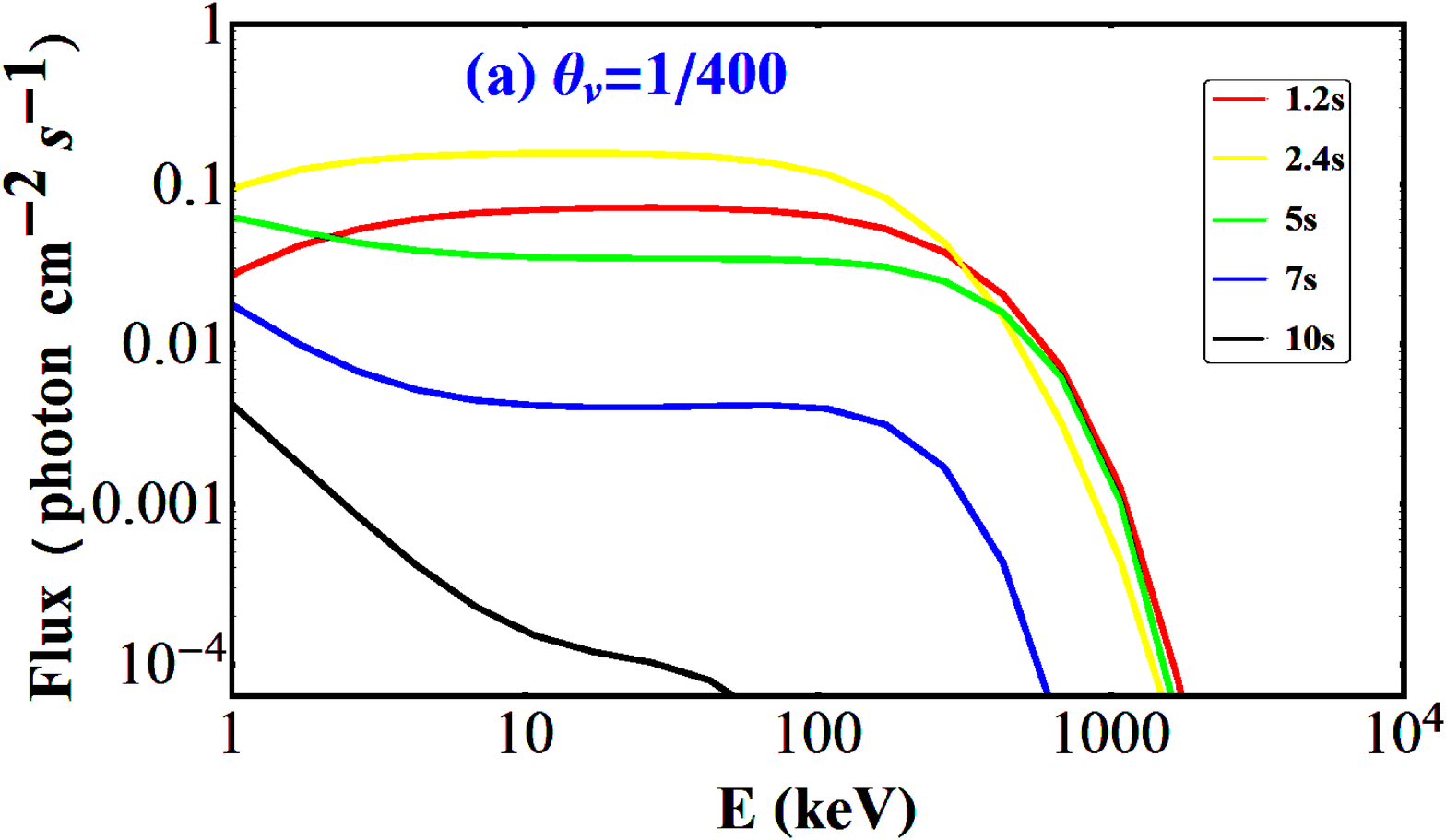} \ \ %
\centering\includegraphics[angle=0,height=1.9in]{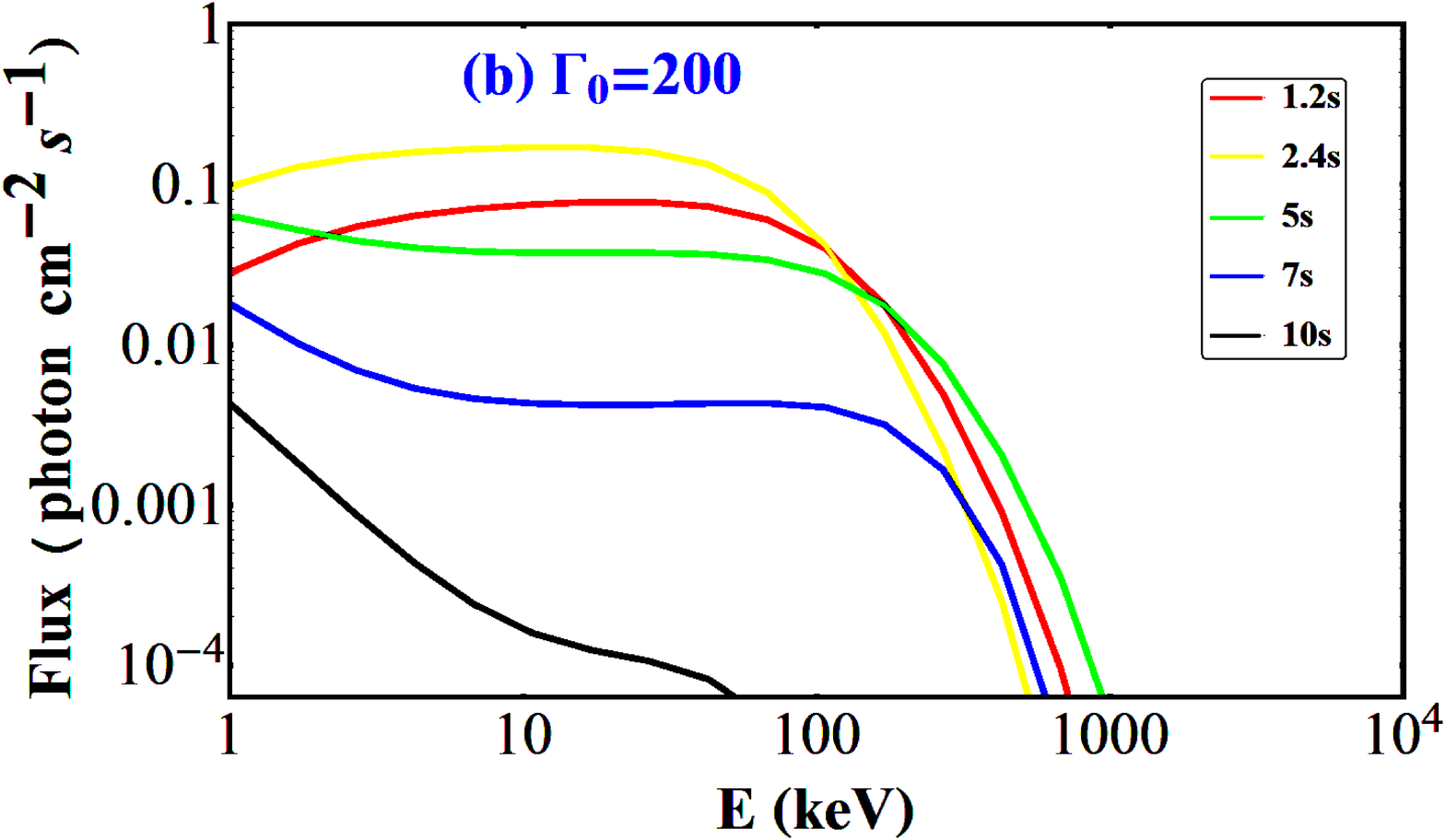} \ \ \centering%
\includegraphics[angle=0,height=1.9in]{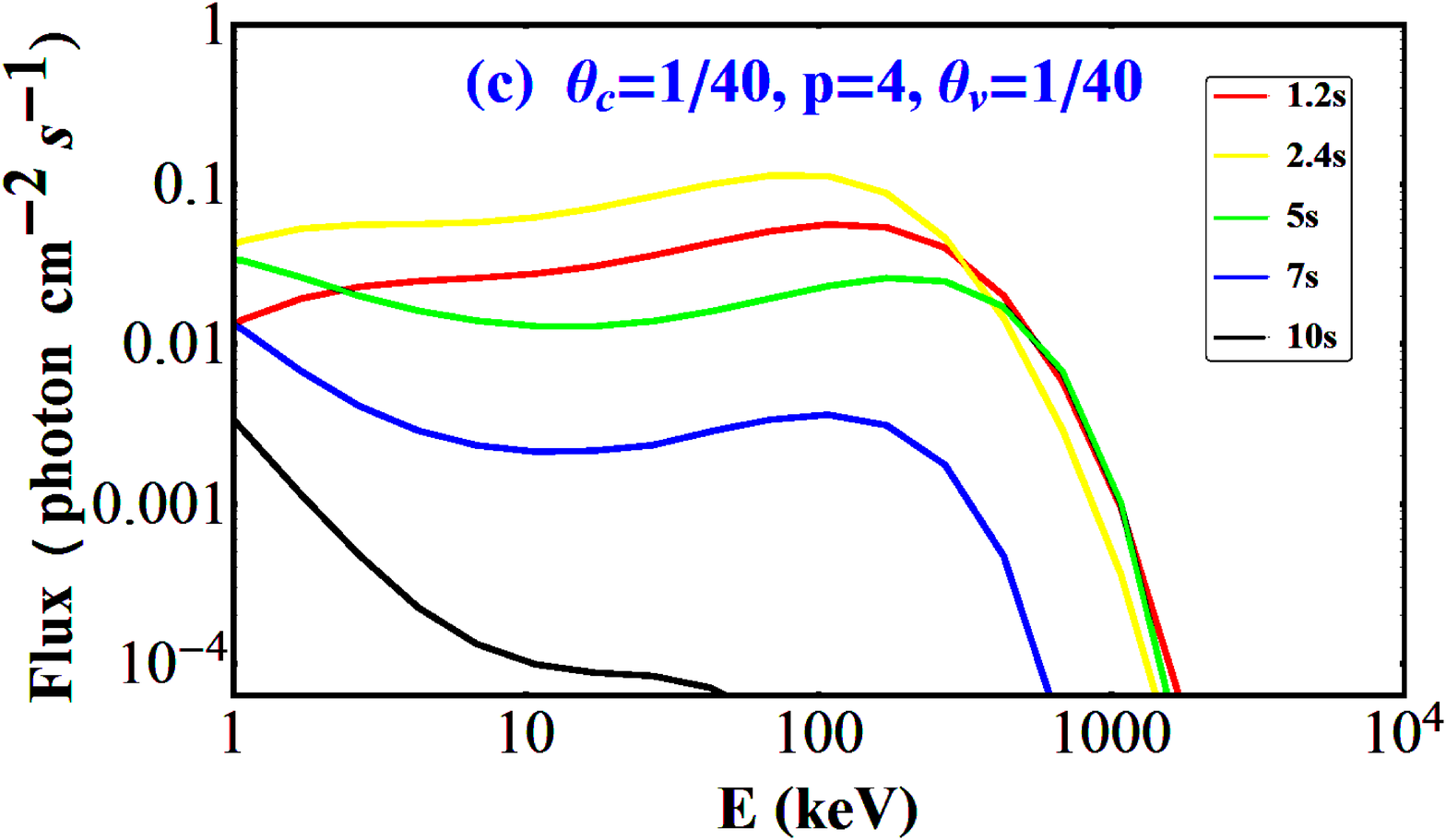} \ \ 
\caption{Time-resolved spectra of winds with variable luminosity for
non-zero viewing angle or smaller $\Gamma _{0}$. \ (a) $\ \protect\theta _{%
\text{v}}$ $=1/400$. \ (b) $\Gamma _{0}=200$, $\protect\theta _{c}=1/200$. \
(c) $\ \protect\theta _{c}=1/40$, $p=4$ and $\protect\theta _{\text{v}}$ $%
=1/40$. The other parameters are the same as Figure 6c. Different colors
represent different observational times.}
\end{figure*}

\begin{figure*}[ph]
\label{Fig_9} \centering\includegraphics[angle=0,height=1.9in]{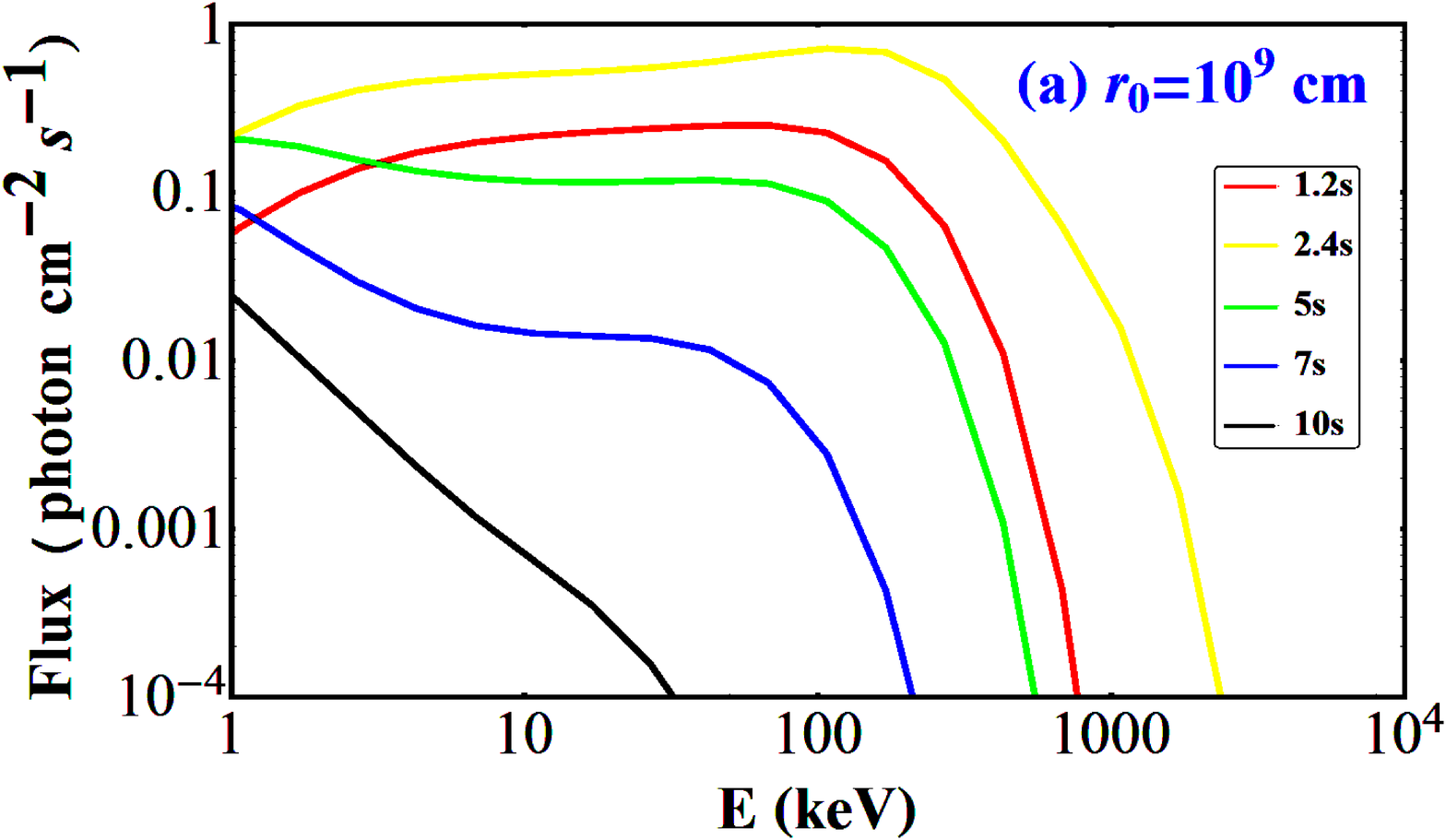} \ \ %
\centering\includegraphics[angle=0,height=1.9in]{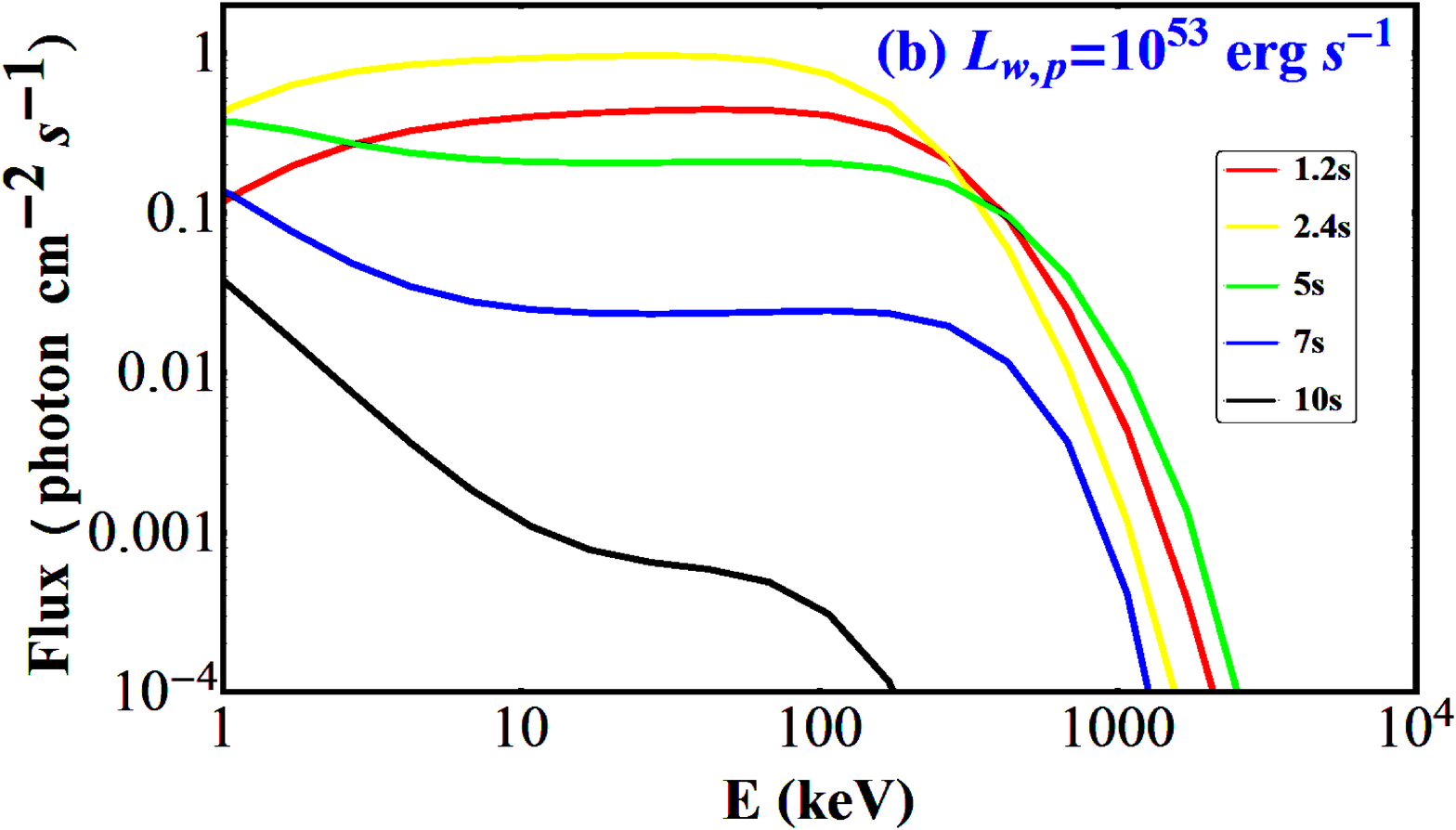} \ \ 
\caption{Time-resolved spectra of winds with variable luminosity for
different $r_{0}$ or $L_{w,\text{ }p}$. \ (a) $\ r_{0}=10^{9}$ cm. \ (b) $\
L_{w,\text{ }p}$ $=10^{53}$ erg s$^{-1}$. The other parameters are the same
as Figure 6c. Different colors represent different observational times.}
\end{figure*}

Since the light curves of the GRBs show relatively slow change in
luminosity, unlike the steep rise and fall (almost within $10^{-4}$ s) for
the case of constant wind luminosity, the wind luminosity may vary with
time, rising and then falling gradually.

\subsubsection{Wind Luminosity History}

Generally, GRB pulses can be fitted well with the exponential model %
\citep{Norr2005} or the smoothly joint broken power law model %
\citep{Koce2003}. So, we approximate the wind luminosity history with the
broken power law model and the exponential model, respectively.

For the broken power law model, the rising and decaying indices are $a_{r}$
and $a_{d}$ respectively, along with a peak luminosity $L_{w,p}$ at $\hat{t}%
_{p}$. Then in the rising phase ($\hat{t}$ $<$ $\hat{t}_{p}$), the
luminosity history can be written as 
\begin{equation}
\log L_{w}(\hat{t})=a_{r}\log \hat{t}+b_{r}\text{,}
\end{equation}%
while in the decaying phase ($\hat{t}$ $>$ $\hat{t}_{p}$), as 
\begin{equation}
\log L_{w}(\hat{t})=a_{d}\log \hat{t}+b_{d}\text{,}
\end{equation}%
where $b_{r}=\log L_{w,p}-a_{r}\log \hat{t}_{p}$ and $b_{d}=\log
L_{w,p}-a_{d}\log \hat{t}_{p}$ are normalization parameters.

For the exponential model, the luminosity history can be 
\begin{eqnarray}
L_{w}(\hat{t} &>&\hat{t}_{s})=L_{w,p}\times \exp \left[ 2\left( \tau
_{1}/\tau _{2}\right) ^{1/2}\right]  \notag \\
&&\text{ \ \ \ \ \ \ \ \ }\times \exp \left( -\frac{\tau _{1}}{\hat{t}-\hat{t%
}_{s}}-\frac{\hat{t}-\hat{t}_{s}}{\tau _{2}}\right) \text{,}
\end{eqnarray}%
where $\hat{t}_{s}$ is the start time, $\tau _{1}$ and $\tau _{2}$ are
respectively the characteristic time scales indicating the rise and decay
periods, $L_{w,p}$ is still the peak of luminosity at $\hat{t}_{p}$, and $%
\hat{t}_{p}=\hat{t}_{s}+\left( \tau _{1}\cdot \tau _{2}\right) ^{1/2}$.

\subsubsection{Spectra Calculation}

From Equation $(\ref{a1})$, we have $R_{\text{dcp}}\propto $ $L_{w}$%
\footnote{%
As shown in Figure 4 in \citet{Lund2013}, $R_{\text{ph}}\thicksim R_{\text{%
dcp}}$. Thus, we only use $R_{\text{dcp}}$ to judge how large $L_{w}$ is.}.
In addition, $R_{s}=\Gamma r_{0}$ is independent of $L_{w}$. So we normally
have $R_{\text{dcp}}\geq R_{s}$ for a relatively large $L_{w}$ ($%
L_{w}=10^{52}$ erg s$^{-1}$); but as the wind luminosity may rise and then
fall, we may have $R_{\text{dcp}}<R_{s}$ in the relatively early and late
periods of the pulse. In the following, we deal with these two cases,
respectively.

\paragraph{ $\ \ \ R_{\text{dcp}}\geq R_{s}$ Case}

Since the outflow luminosity varies with time, $d\dot{N}_{\gamma }/d\Omega $
is surely time-dependent, 
\begin{gather}
\frac{d\dot{N}_{\gamma }(\hat{t})}{d\Omega }=\frac{\dot{N}_{\gamma }(\hat{t})%
}{4\pi },  \notag \\
\dot{N}_{\gamma }(\hat{t})=\frac{L_{w}(\hat{t})}{2.7k_{\text{B}}T_{0}(\hat{t}%
)}\text{, \ }T_{0}(\hat{t})=\left( \frac{L_{w}(\hat{t})}{4\pi r_{0}^{2}ac}%
\right) ^{1/4}.  \label{d1}
\end{gather}%
Note that $T_{0}$ is time-dependent, too.

When calculating\ the decoupling radius$\ R_{\text{dcp}}$, \ the optical
depth can be written as 
\begin{equation}
\tau =\frac{1}{(1+\beta )\beta \Gamma ^{2}}\frac{\sigma _{\text{T}}}{m_{%
\text{p}}c}\int\nolimits_{R_{\text{dcp}}}^{\infty }\frac{d\dot{M}}{d\Omega }%
\frac{1}{r^{2}}dr\text{.}  \label{b1}
\end{equation}%
Here, $d\dot{M}(\theta )/d\Omega $ should depend both on $\hat{t}$ and $r$ $%
(r>R_{\text{dcp}})$. Also, $d\dot{M}/d\Omega $\ and $\Gamma $\ are both
angle-dependent. We omit writing angular dependences here and below for
clarity. $d\dot{M}(\hat{t},R_{\text{dcp}})/d\Omega =d\dot{M}(\hat{t}%
)/d\Omega $ $=L_{w}(\hat{t})/4\pi c^{2}\Gamma $ is the mass outflow rate at
the central engine time $\hat{t}$, while 
\begin{equation}
\frac{d\dot{M}(\hat{t},r)}{d\Omega }=\frac{L_{w}(\hat{t}-\Delta \hat{t})}{%
4\pi c^{2}\Gamma }
\end{equation}%
denotes the mass outflow rate at $\hat{t}-\Delta \hat{t}$ (much earlier than 
$\hat{t}$), $\Delta \hat{t}=(r-R_{\text{dcp}})/2\Gamma ^{2}\beta c$. In
addition, since $\tau \thicksim 1/r$, we can set the upper limit in Equation 
$(\ref{b1})$ as 11$R_{\text{dcp}}$. Then, for $L_{w}=10^{52}$ erg s$^{-1}$
and $\Gamma =200$, $\Delta \hat{t}$ $\mathbf{=}$ $10(R_{\text{dcp}}/2\Gamma
^{2}\beta c)\thicksim $ $3\times 10^{-3}$ s $\ll 1$ s. Thus, we may consider 
$d\dot{M}/d\Omega $ to be independent of $r$, which means 
\begin{equation}
R_{\text{dcp}}(\hat{t})=\frac{1}{(1+\beta )\beta \Gamma ^{2}}\frac{\sigma _{%
\text{T}}}{m_{\text{p}}c}\frac{d\dot{M}(\hat{t})}{d\Omega }\text{.}
\end{equation}%
And the other way round, when the photon emitted from the layer ($\hat{t}$)
catches up with the only slightly earlier layer ($\hat{t}-\Delta \hat{t}$, $%
\Delta \hat{t}$ $\thicksim $ $3\times 10^{-3}$ s), it has reached a quite
large radius 11$R_{\text{dcp}}$. This means that the assumption of infinity
outer boundary is reasonable.

For the same reason, we take $d\dot{M}(\hat{t},r)/d\Omega \simeq d\dot{M}(%
\hat{t},R_{\text{ph}})/d\Omega =d\dot{M}(\hat{t})/d\Omega $ $(r>R_{\text{ph}%
})$, regardless of the angle-dependent $\Delta \hat{t}$. Thus, the
time-dependent photospheric radius $R_{\text{ph}}(\hat{t})$ can be written
as 
\begin{equation}
R_{\text{ph}}(\hat{t})=\frac{\sigma _{\text{T}}}{m_{\text{p}}c\sin \theta _{%
\text{LOS}}}\int\nolimits_{0}^{\theta _{\text{LOS}}}\frac{1-\beta \cos 
\tilde{\theta}_{_{\text{LOS}}}}{\beta }\frac{d\dot{M}(\hat{t})}{d\Omega }d%
\tilde{\theta}_{_{\text{LOS}}}.  \label{e}
\end{equation}%
Then, the comoving temperature $T^{^{\prime }}(r,\Omega ,\hat{t})$ can be
obtained by (omitting angular dependences): 
\begin{equation}
T^{\prime }(r,\hat{t})=\left\{ 
\begin{array}{c}
T_{0}(\hat{t})/\Gamma ,\text{ \ \ \ \ \ \ \ \ \ \ \ \ \ \ \ \ \ \ \ \ \ \ \ }%
r<R_{s}<R_{\text{ph}}(\hat{t}),\text{ } \\ 
T_{0}(\hat{t})[r/R_{s}]^{-2/3}/\Gamma ,\text{ \ \ \ \ \ \ \ \ \ }R_{s}<r<R_{%
\text{ph}}(\hat{t}),\text{ } \\ 
\text{ }T_{0}(\hat{t})[R_{\text{ph}}(\hat{t})/R_{s}]^{-2/3}/\Gamma ,\text{ \ 
}R_{s}<R_{\text{ph}}(\hat{t})\text{ }<r.\text{ }%
\end{array}%
\right.
\end{equation}%
Similar to the case of constant wind luminosity, we have 
\begin{gather}
F_{E^{\text{ob}}}^{\text{ob}}(\theta _{\text{v}},t,\hat{t})=\frac{1}{4\pi d_{%
\text{L}}^{2}}\int (1+\beta )D^{2}\frac{d\dot{N}_{\gamma }(\hat{t})}{d\Omega 
}\times \frac{R_{\text{dcp}}(\hat{t})}{r^{2}}  \notag \\
\exp \left( -\frac{R_{\text{ph}}(\hat{t})}{r}\right) \left\{ \frac{1}{2.40}%
\frac{[E/k_{\text{B}}DT^{\prime }(r,\hat{t})]^{3}}{\exp [E/k_{\text{B}%
}DT^{\prime }(r,\hat{t})]-1}\right\} \times \frac{\beta c}{u}d\Omega ,\text{%
\ }  \notag \\
\text{\ }r=\frac{\beta c(t-\hat{t})}{u}\text{, }E=(1+z)E^{\text{ob}}\text{.}
\label{f}
\end{gather}

Then, using Equation $(\ref{c2})$ to integrate over all the layers, we get
the spectrum at $t$. Note that we must judge whether we have $R_{\text{dcp}}(%
\hat{t})\geq R_{s}$ for the layer ejected at $\hat{t}$; if not, we should
calculate as following.

\paragraph{ $\ \ \ R_{\text{dcp}}<R_{s}$ Case}

In this condition, $d\dot{N}_{\gamma }(\hat{t})/d\Omega $ is still
calculated by Equation $(\ref{d1})$. As for the decoupling radius$\ R_{\text{%
dcp}}$, we have 
\begin{equation}
R_{\text{dcp}}(\hat{t})=\left[ \frac{\sigma _{\text{T}}}{6m_{\text{p}}c}%
\frac{d\dot{M}(\hat{t})}{d\Omega }r_{0}^{2}\right] ^{1/3}\text{.}
\end{equation}%
Meanwhile, firstly $\Gamma $\ depends on both $r$\ and $\theta $, thus it is
hard to calculate $R_{\text{ph}}(\hat{t})$. Secondly, $R_{\text{ph}%
}\thicksim R_{\text{dcp}}$(see the Figure 4 in \citealt{Lund2013}). Thirdly,
since $R_{\text{dcp}}<R_{s}$\ the observed temperature is close to $T_{0}$\
if $\theta $\ is not too large. While as shown in \citet{Lund2013}, for a
prompt GRB spectrum the part expected to be observed is formed by the
photons making their final scattering at approximately $\lesssim 5/$\ $%
\Gamma _{0}$. For simplicity\textbf{, }we take 
\begin{equation}
R_{\text{ph}}(\hat{t})=R_{\text{dcp}}(\hat{t})\text{. }
\end{equation}%
Besides, the comoving temperature $T^{^{\prime }}(r,\hat{t})$ is given by 
\begin{equation}
T^{\prime }(r,\hat{t})=T_{0}(\hat{t})/\Gamma (\hat{t})\text{,}
\end{equation}%
where $\Gamma (\hat{t})=R_{\text{dcp}}(\hat{t})/r_{0}$. Then, $F_{E^{\text{ob%
}}}^{\text{ob}}(\theta _{\text{v}},t,\hat{t})$ is still calculated by
Equation $(\ref{f})$, except that 
\begin{eqnarray}
\beta &=&\beta (\hat{t})=\{1-[\Gamma (\hat{t})]^{-2}\}^{1/2}\text{,}  \notag
\\
D &=&D(\hat{t})=\frac{1}{\Gamma (\hat{t})}\frac{1}{1-\beta (\hat{t})\cos
\theta _{\text{LOS}}}\text{.}
\end{eqnarray}

\subsubsection{Results}

Figures 6a-6c show the calculated, instantaneous spectra of winds with
variable luminosity for different luminosity histories. We fix $\hat{t}%
_{p}=2.4$ s, $L_{w,p}$ $=10^{52}$ erg s$^{-1}$, $r_{0}=10^{8}$ cm and $d_{%
\text{L}}=$ $4.85\times 10^{28}$ cm ($z=2$), and use the Lorentz factor
profile $\Gamma _{0}=400$, $\theta _{c}\Gamma _{0}=1$, $p=1$ along with $%
\theta _{\text{v}}$ $=0$. We investigate three different luminosity
profiles, the broken power law model with $(a_{r},a_{d})=(0.75,-5)$, $%
(0.75,-8)$ and the exponential model with $\tau _{1}=32$, $\ \tau _{2}$ $%
=0.5\ \ $and$\ \ \hat{t}_{s}=$ $-1.6$. For each plot, different colors show
different observational times. Obviously, during the rising phase ($t=0.5$ s
or $1.2$ s, $2.4$ s) the resulting spectra are quite similar to those for
the case of the constant luminosity (Figure 4), i.e., the spectra have a
flattened shape ($F_{\nu }\sim \nu ^{0}$) below the peak, consistent with
the average low-energy spectral index for the time-resolved spectra observed
in GRBs \citep[e.g.,][]{Kan2006,Yu2016}, and caused by the superposition of
emission from the layers injected at different times. During the decay phase
($t=5$ s, $7$ s, $10$ s), the power law with negative index shows up
gradually. The reason is that the high-latitude emission becomes more
dominant (see Figure 2), since it comes from the much earlier layers that
have higher luminosities. The steeper the decay phase gets, the more
significant the power law with negative index is. For the broken power law
model with $(a_{r},a_{d})=(0.75,-8)$ and the exponential model, the spectrum
at $t=10$ s\ is fully a power law with negative index.

In Figures 7a-7d, we compare the resulting time-resolved spectra for
different Lorentz factor profiles. The other parameters are the same as
Figure 6c. As shown in Figure 7a, with a larger Lorentz factor gradient $p=4$%
, the resulting spectra during the rising phase ($t=$ $1.2$ s, $2.4$ s) are
a little harder ($F_{\nu }\sim \nu ^{0.5}$) below the peak, during the decay
phase ($t=5$ s, $7$ s, $10$ s), a power law with negative index still shows
up gradually. At $t=7$ s, the spectrum is the mix of a power law with
negative index on the low-energy end and a modified blackbody with a
shallower low-energy spectral index \ ($\alpha \sim -0.5$) on the
high-energy end. While the spectrum at $t=10$ s\ is fully a power law with
negative index. Figure 7b presents the time-resolved spectra for a wider jet
core with $\theta _{c}\Gamma _{0}=10$ and $p=4$, the resulting spectra
during the rising phase are similar to Figure 7a, with $F_{\nu }\sim \nu
^{1.0}$ below the peak. At $t=7$ s, the spectrum is still the mix of a power
law with negative index and a modified blackbody with a shallower low-energy
spectral index \ ($\alpha \sim 0.0$). However, the spectrum at $t=10$ s\ is
not a power law with negative index, but a modified blackbody with a
flattened shape ($F_{\nu }\sim \nu ^{0}$) below the peak. Figure 7c shows
that, with $\theta _{c}\Gamma _{0}=10$ and $p=1$, the resulting spectra
during the rising phase and the decay phase are almost the same, i.e., a
modified blackbody with a shallower low-energy spectral index ($\alpha \sim
0.0$). Finally, Figure 7d presents the time-resolved spectra for a narrower
jet core with $\theta _{c}\Gamma _{0}=1/10$ and $p=0.5$, the resulting
spectra during the rising phase are much softer ($F_{\nu }\sim \nu ^{-0.8}$)
below the peak, which is consistent with the lowest low-energy spectral
index for the time-resolved spectra observed in GRBs %
\citep[e.g.,][]{Kan2006,Yu2016}. During the decay phase, a power law with
negative index shows up gradually also.

Figures 8a-8c show the resulting time-resolved spectra for non-zero viewing
angle or smaller $\Gamma _{0}$. Figure 8a shows that, for a non-zero viewing
angle $\theta _{\text{v}}$ $=1/400$, the resulting spectra during the rising
phase are similar to those for $\theta _{\text{v}}$ $=0$ (see Figure 6c),
i.e., the spectra have a flattened shape ($F_{\nu }\sim \nu ^{0}$) below the
peak, but the $E_{p}$ is much smaller. During the decay phase, the resulting
spectra are similar to those for $\theta _{\text{v}}$ $=0$, too. For $\Gamma
_{0}=200$ (smaller), $\theta _{c}=1/200$ in Figure 8b and $\theta _{c}=1/40$%
, $p=4$, $\theta _{\text{v}}$ $=1/40$ in Figure 8c, the resulting spectra
during the rising and decay phases are similar to Figure 8a.

In Figures 9a and 9b, we consider the influence of different $r_{0}$ or $%
L_{w,p}$ on the time-resolved spectra, respectively. For$\ r_{0}=10^{9}$ cm
and$\ L_{w,p}$ $=10^{53}$ erg s$^{-1}$, the resulting spectra during the
rising phase change little (compared with Figure 6c). But during the decay
phase, the power law with negative index shows up more quickly for $\
r_{0}=10^{9}$ cm, while same with Figure 6c for $L_{w,p}$ $=10^{53}$ erg s$%
^{-1}$.

\subsection{Time-resolved spectral analysis of GRBs observed by \textit{Fermi%
} GBM}

\begin{figure*}[th]
\label{Fig_10} \ \ \ \ \ \ \ \ \ \ \ \ \ \centering%
\includegraphics[angle=0,height=1.8in]{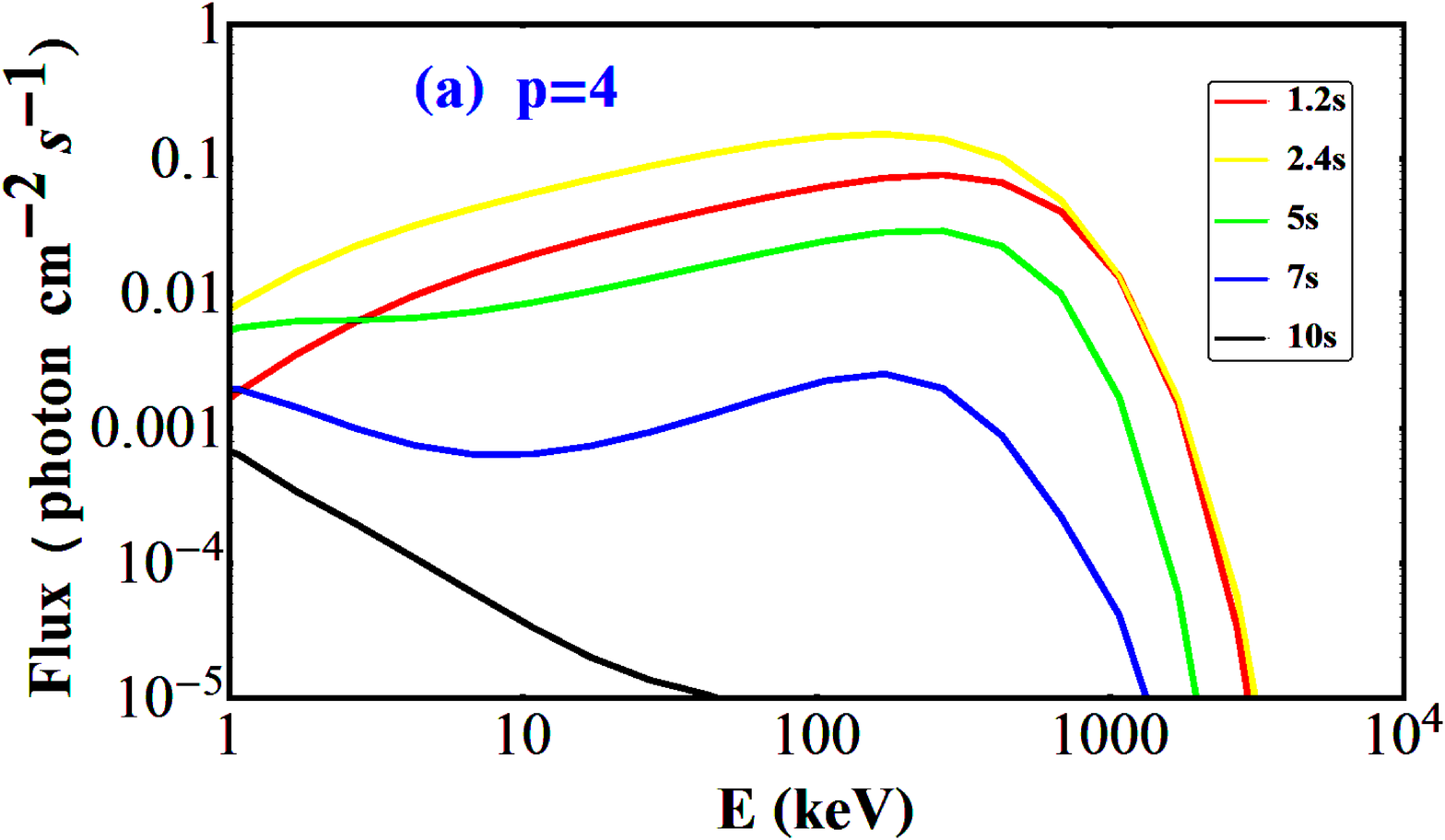}\ \ %
\includegraphics[angle=0,height=1.8in]{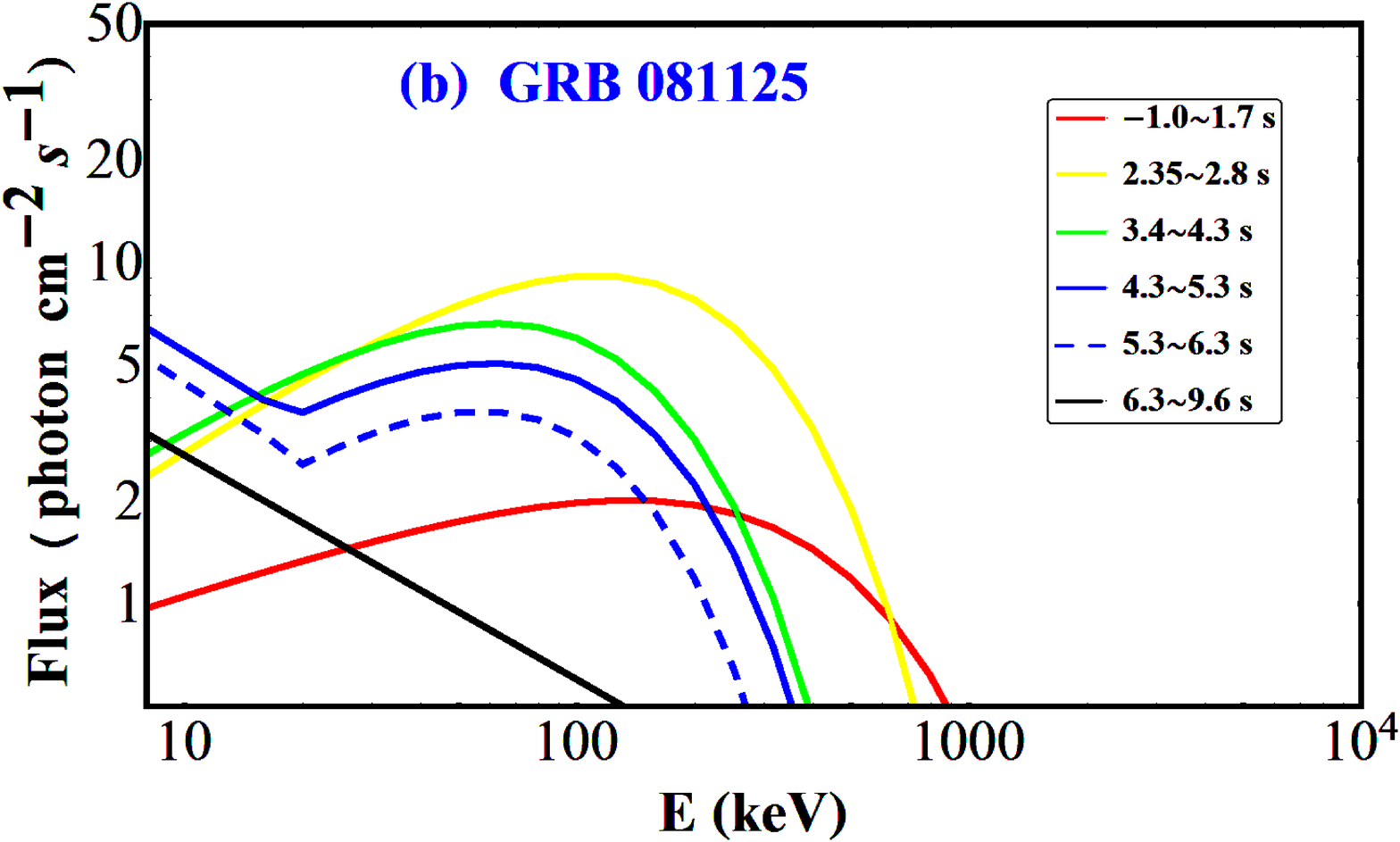}\ \ \ \ \ \ \ \ \ \ %
\centering\includegraphics[angle=0,height=2.1in]{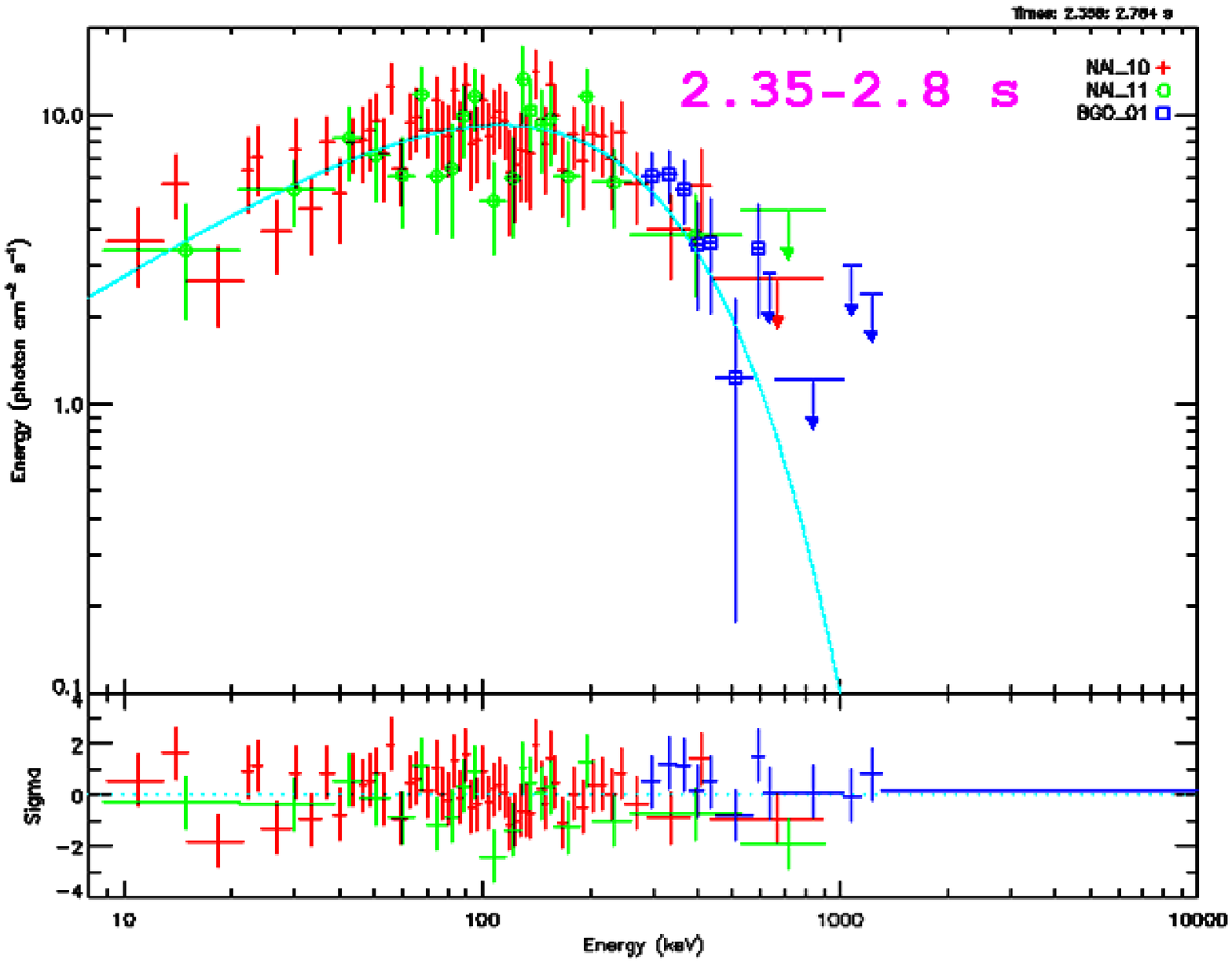} \ %
\includegraphics[angle=0,height=2.1in]{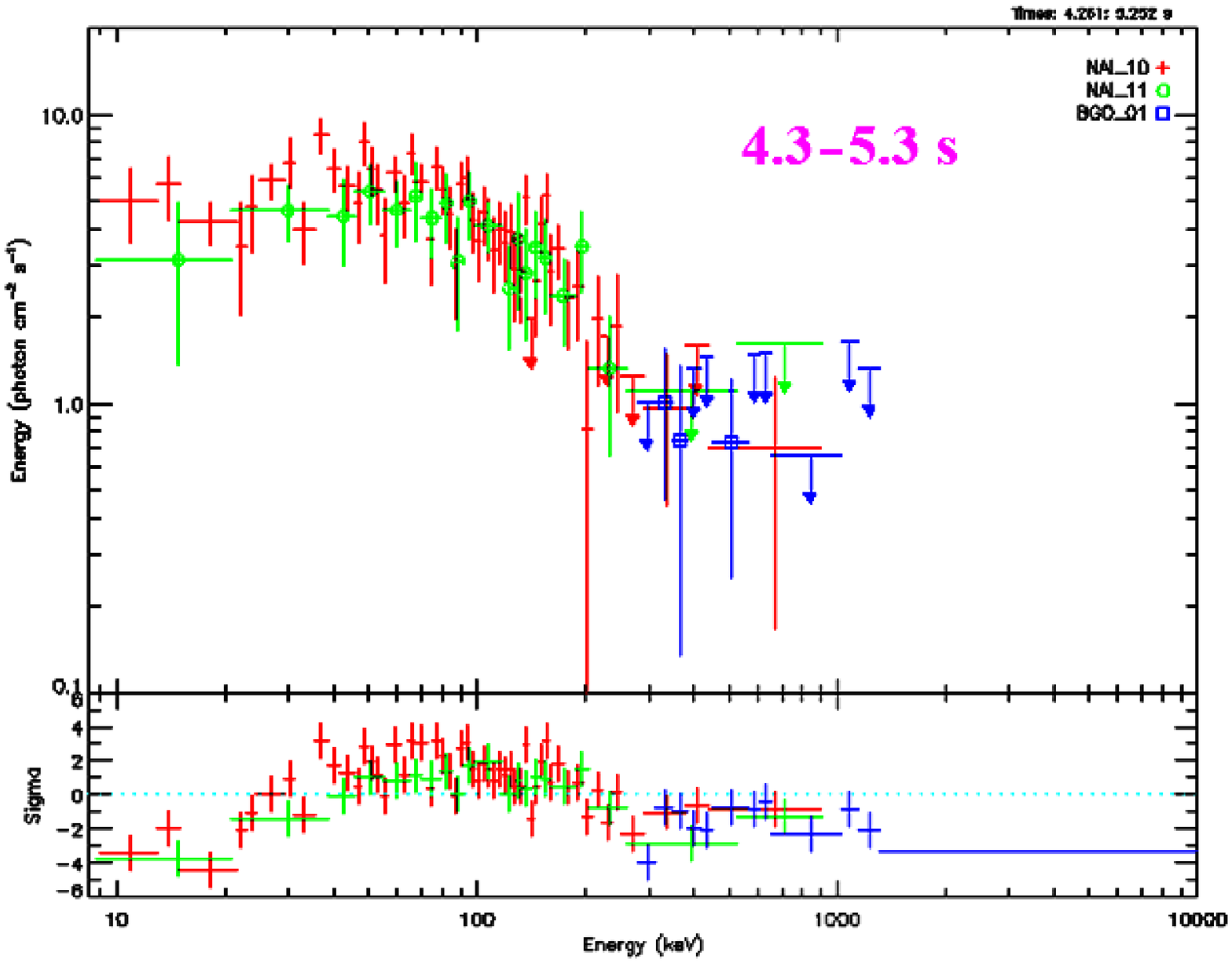} \ \ \ \ \ \ \ \ \ \ %
\includegraphics[angle=0,height=2.0in]{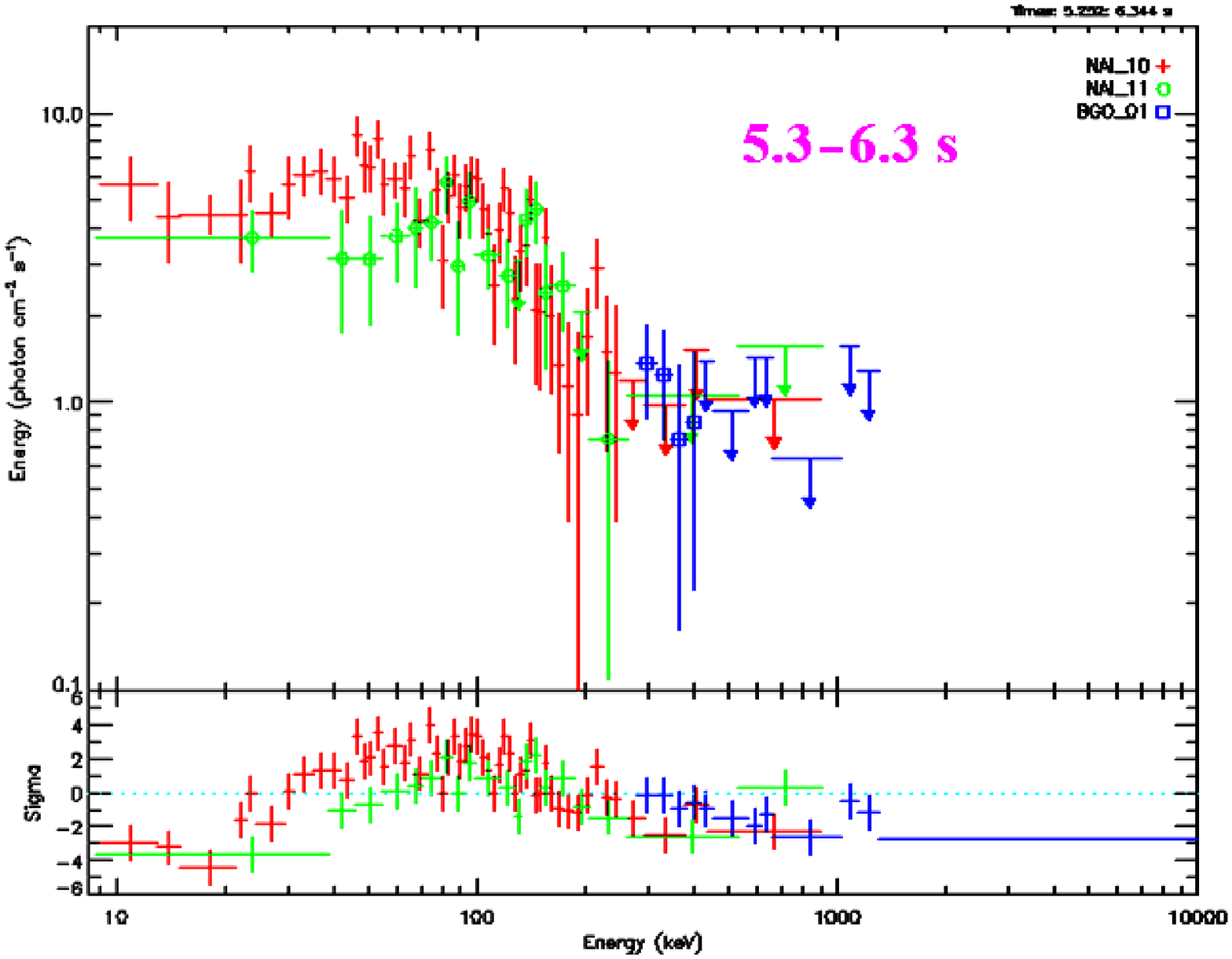} \ %
\includegraphics[angle=0,height=2.0in]{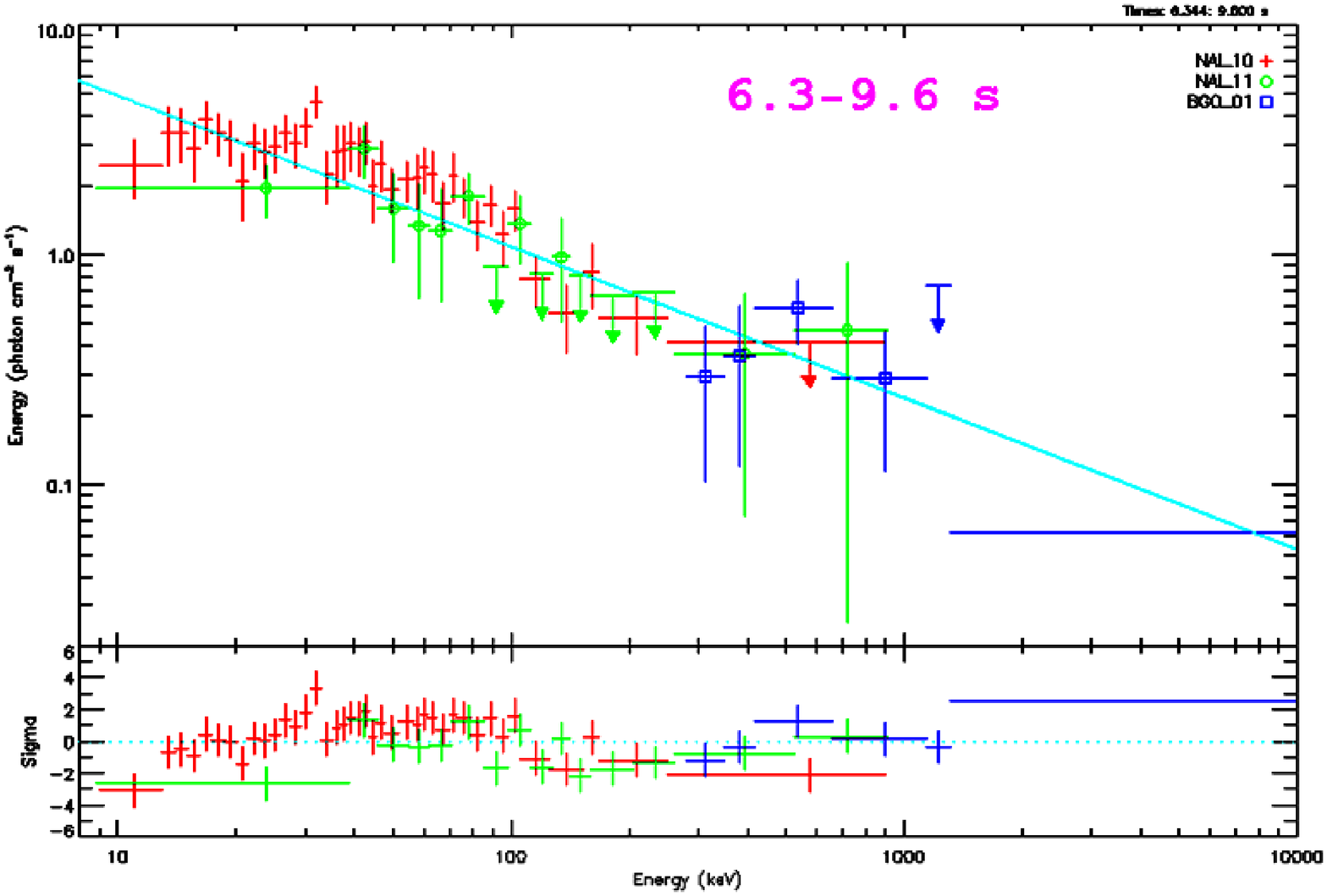}\ \ \ \ \ \ \ \ \ \ \ \ \ \
\ \ \ \ \ \ \ \ \ \ \ \ \ \ \ \ \ \ \ \ \ \ \ \ \ \ \ \ \ \ \ 
\caption{{}Comparison of the calculated time-resolved spectra for $\protect%
\theta _{c}\Gamma _{0}=1$, $p=4$ and the time-resolved spectral analysis
results of GRB 081125. Top left panel:\textit{\ }the calculated
time-resolved spectra for $\protect\theta _{c}\Gamma _{0}=1$ and $p=4$, same
as Figure 7a.\textit{\ }Top right panel: the best-fit model spectra of the
time-resolved spectra from GRB 081125\ for several different time
intervals.\ Middle and bottom panels: the spectral fits to the time-resolved
spectra for $2.35-2.8$ s (middle left) and $6.3-9.6$ s (bottom right), and
the observed spectra (fitted with the PL model) for $4.3-5.3$ s (middle
right) and $5.3-6.3$ s (bottom left).}
\end{figure*}

\begin{figure*}[ph]
\label{Fig_11} \centering\includegraphics[angle=0,height=2.0in]{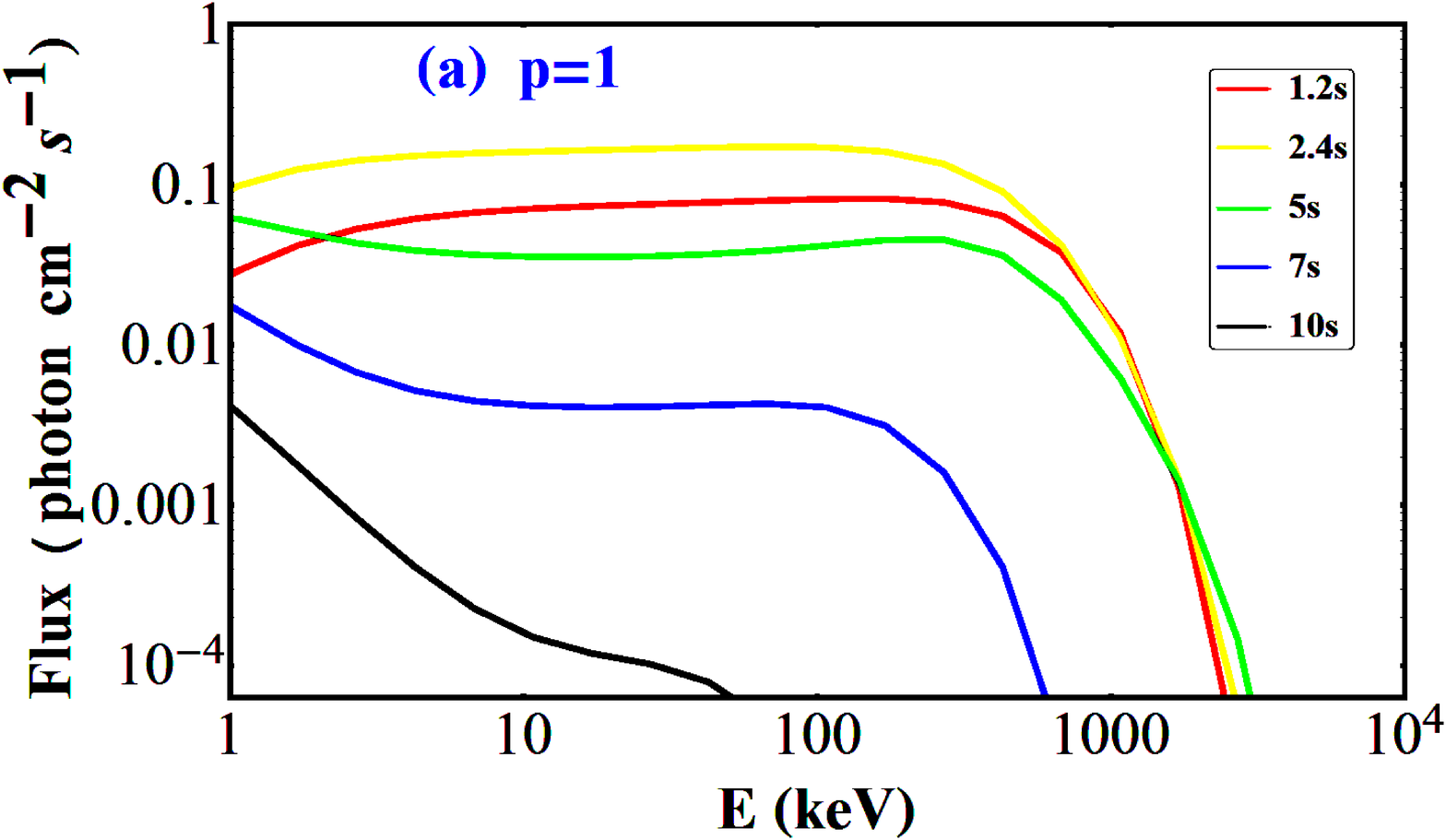} \
\ \ \centering\includegraphics[angle=0,height=2.0in]{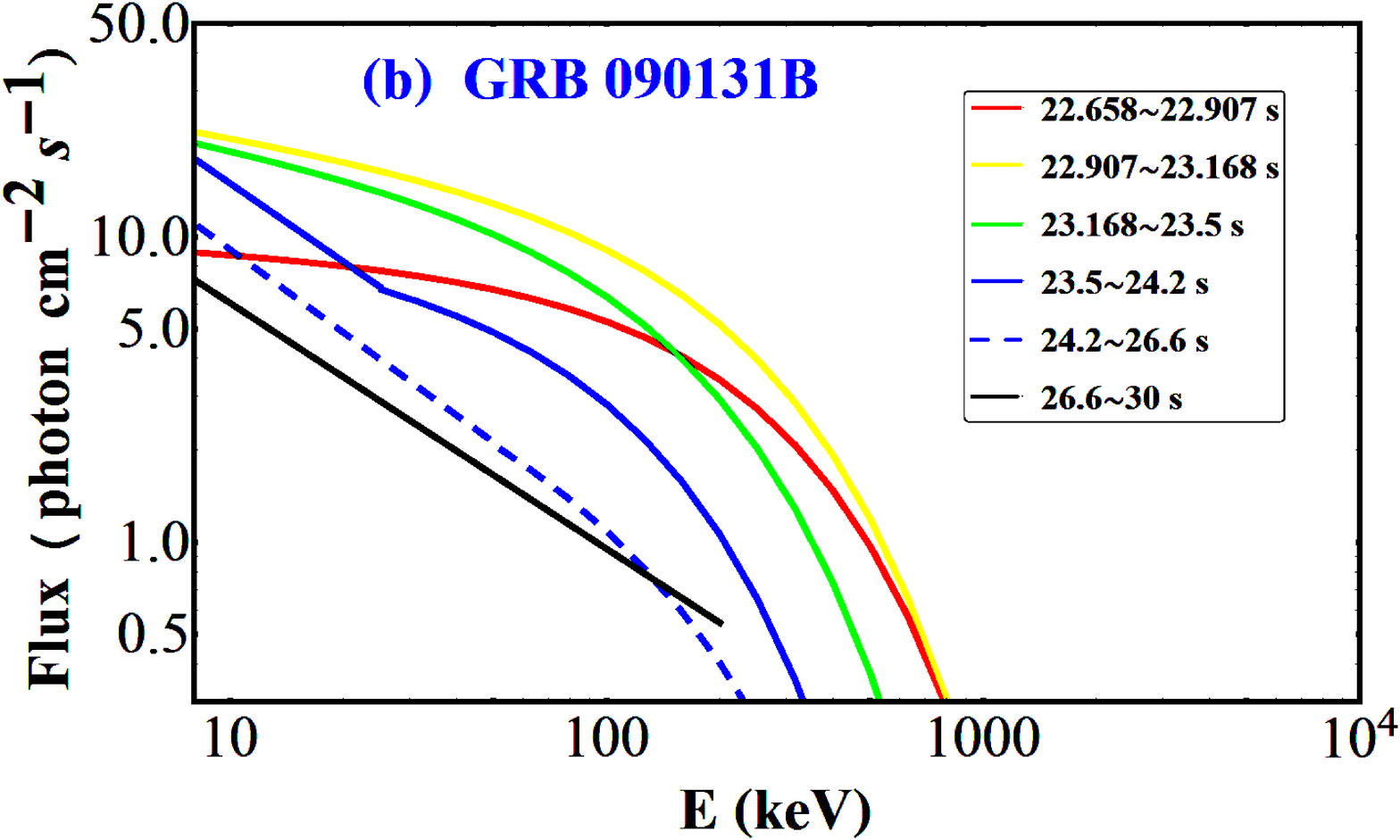} \ \ \ \ \ \ \
\ \ \ \ \includegraphics[angle=0,height=1.8in]{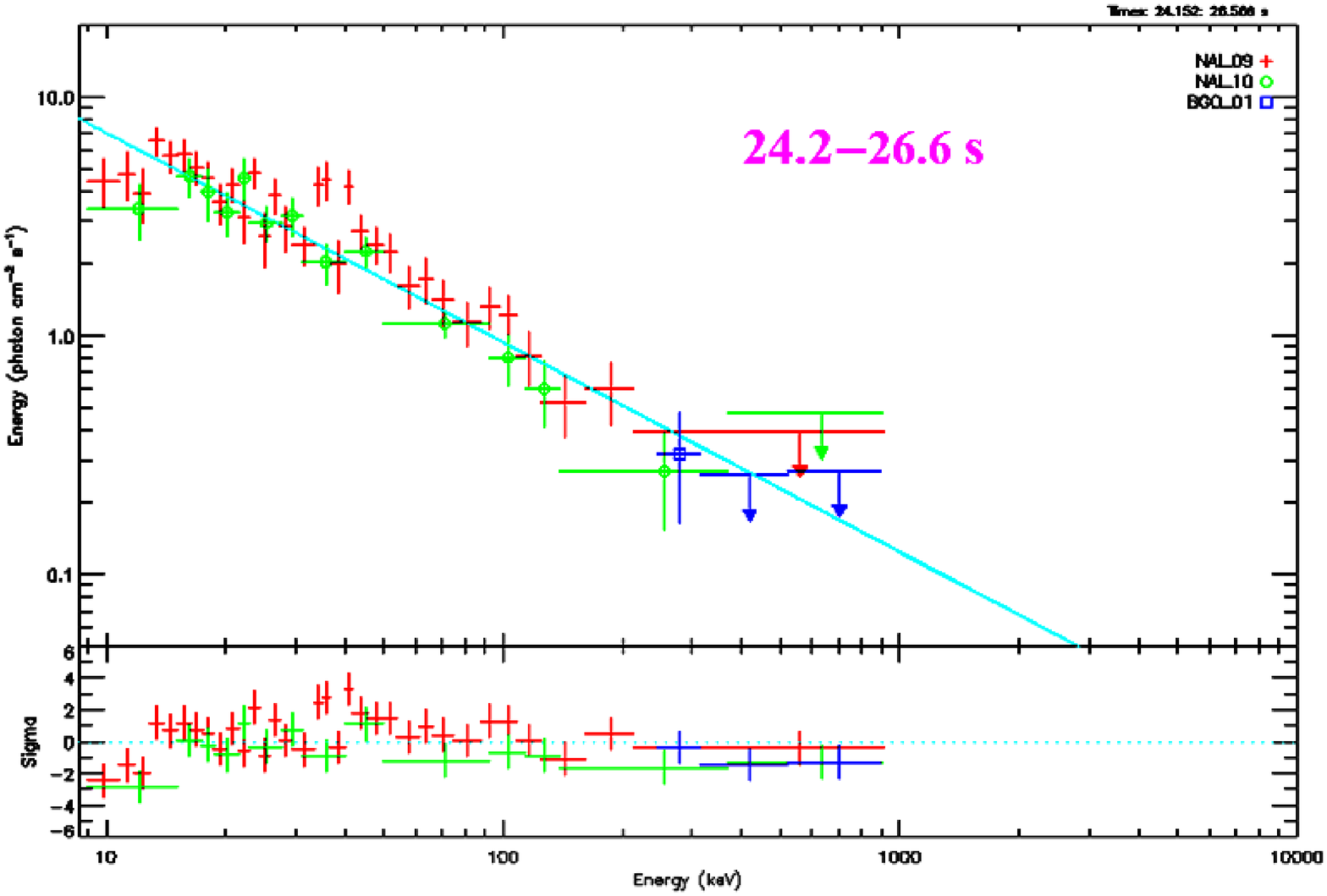} \ %
\includegraphics[angle=0,height=1.8in]{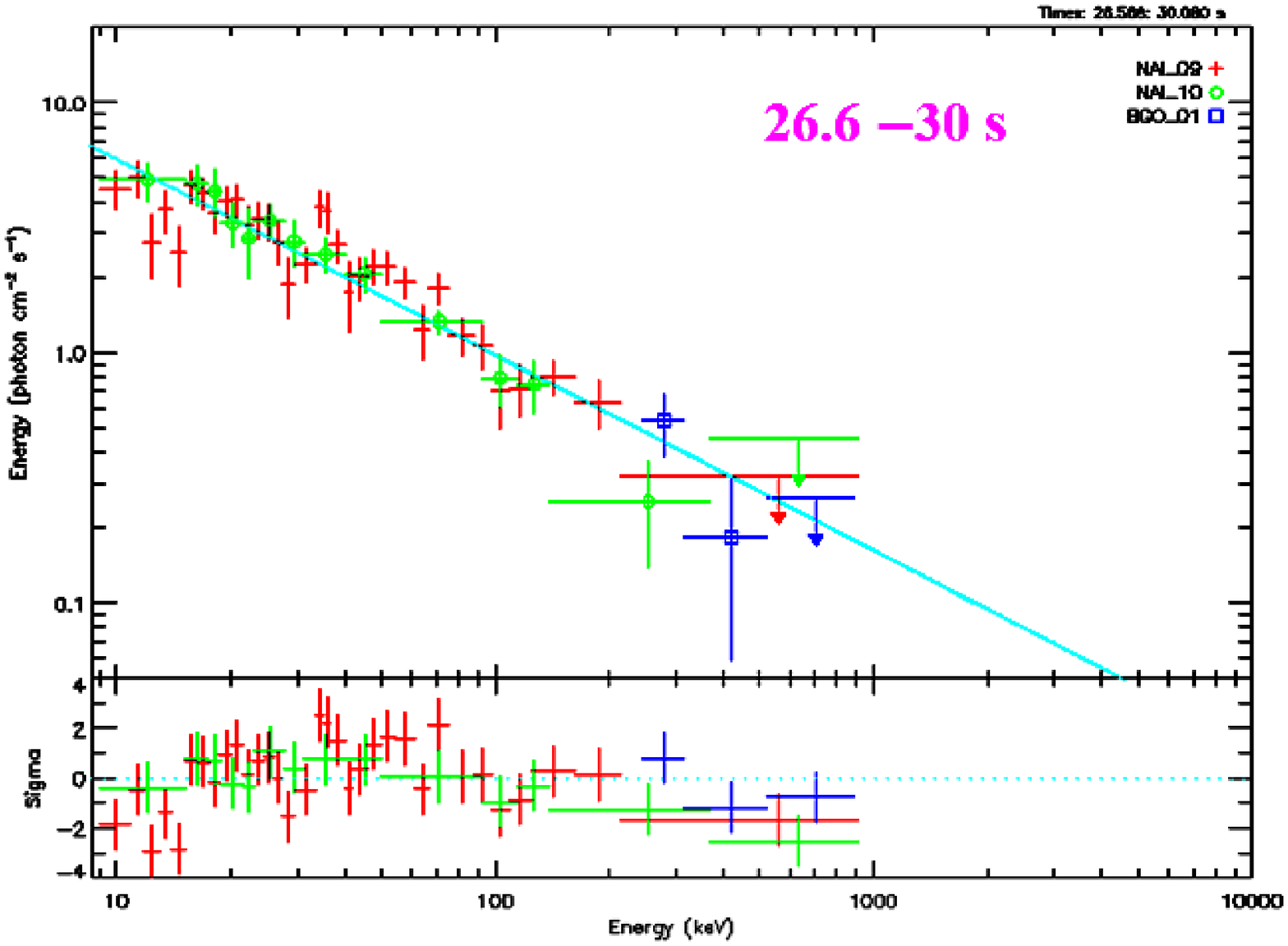} \ \ \ \ \ \ \ 
\caption{{}Comparison of the calculated time-resolved spectra for $\protect%
\theta _{c}\Gamma _{0}=1$, $p=1$ and the time-resolved spectral analysis
results of GRB 090131B. Top left panel:\textit{\ }the calculated
time-resolved spectra for $\protect\theta _{c}\Gamma _{0}=1$ and $p=1$, same
as Figure 6c.\textit{\ }Top right panel: \ the best-fit model spectra of the
time-resolved spectra from GRB 090131B\ for several different time
intervals.\ Bottom panels: the spectral fits to the time-resolved spectra
for $24.2-26.6$ s and $26.6-30$ s (from left to right), respectively.}
\end{figure*}

\begin{figure*}[ph]
\label{Fig_12} \centering\includegraphics[angle=0,height=2.0in]{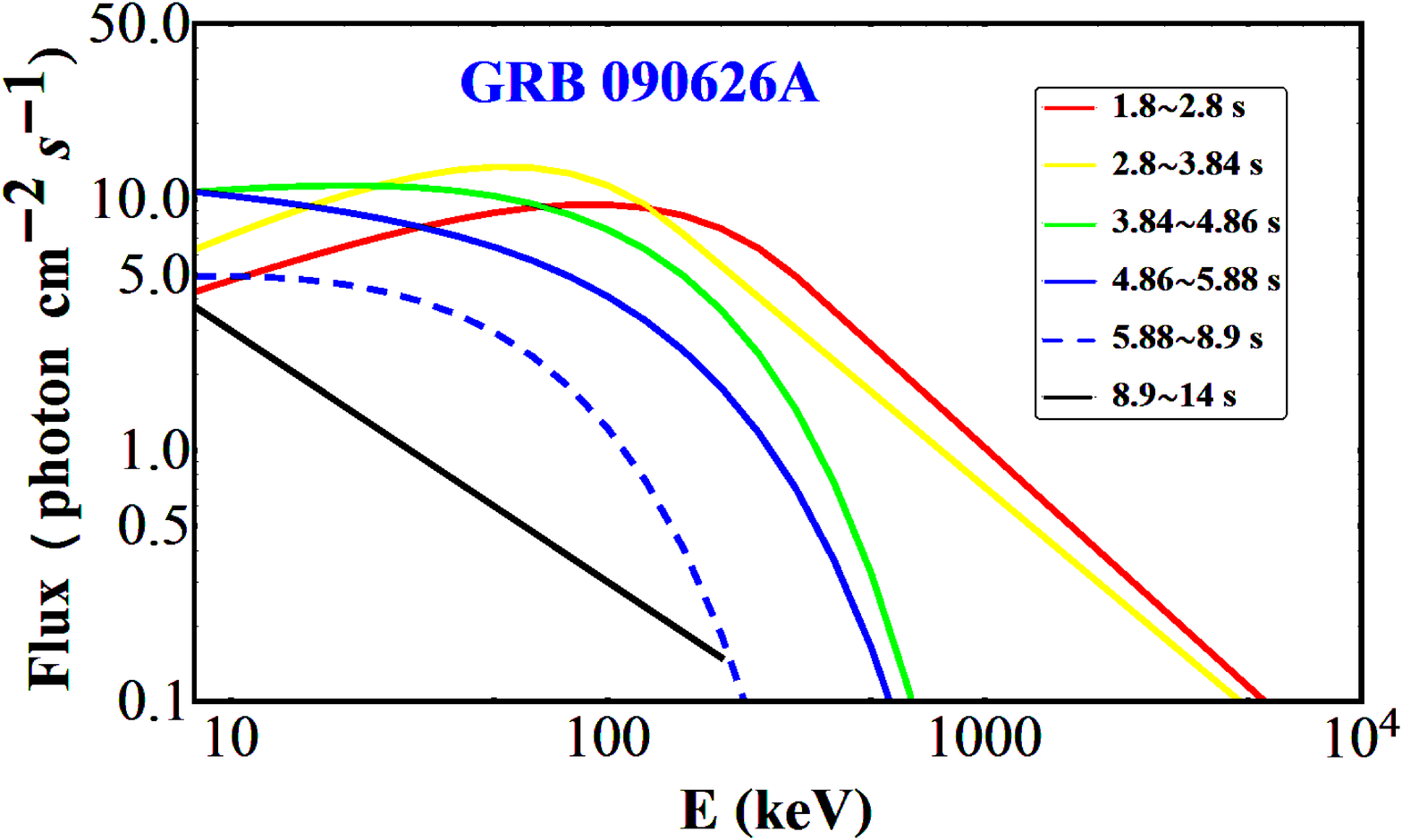} \
\ \centering\includegraphics[angle=0,height=2.0in]{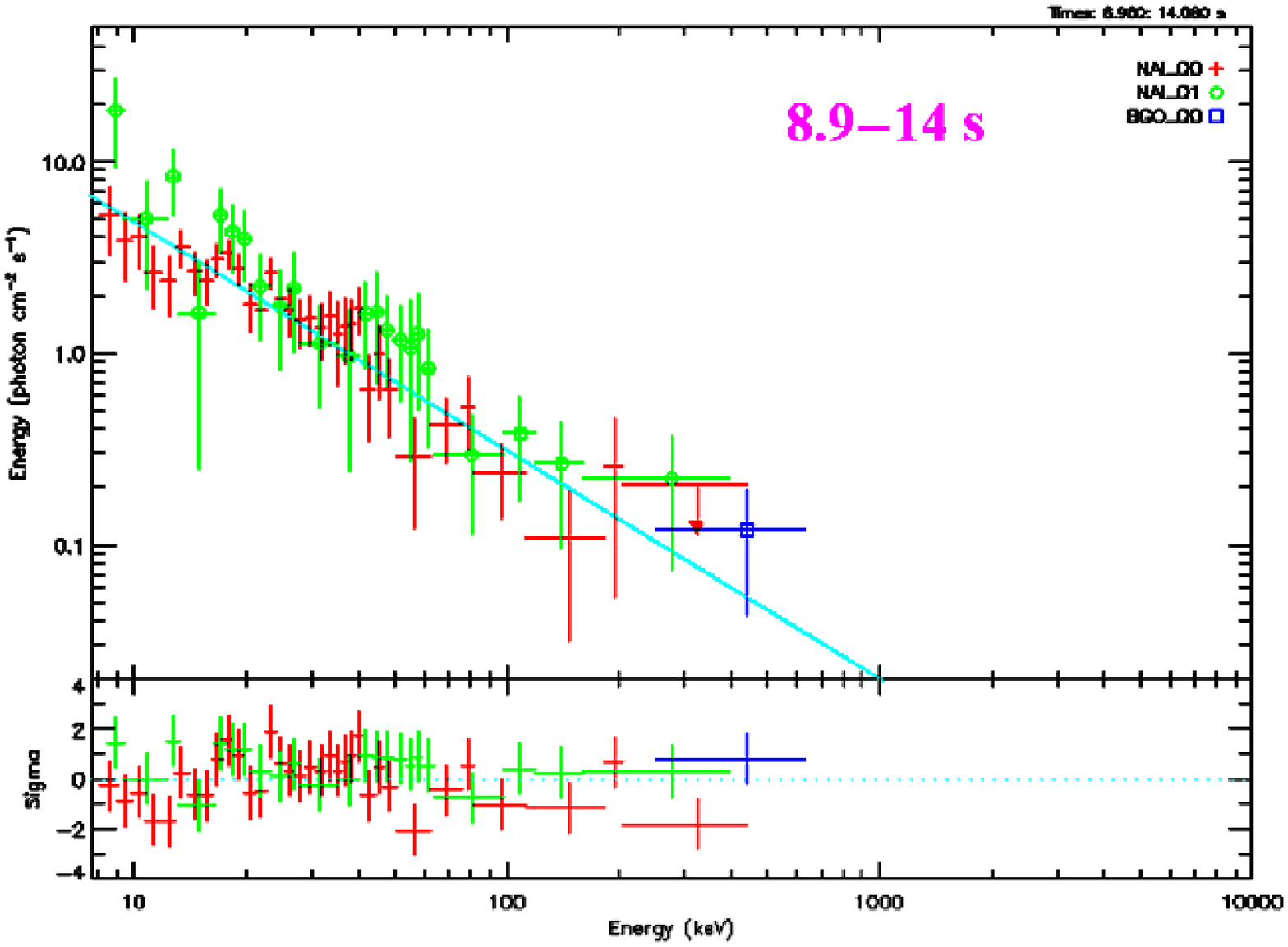} \ \ \ \ \ \ \ 
\caption{{}The time-resolved spectral analysis results of GRB 090626A. Left
panel:\textit{\ }the best-fit model spectra of the time-resolved spectra
from GRB 090626A\ for several different time intervals.\textit{\ }Right
panel: the spectral fits to the time-resolved spectrum for $8.9-14$ s. }
\end{figure*}

\begin{figure*}[ph]
\label{Fig_13} \centering\includegraphics[angle=0,height=2.0in]{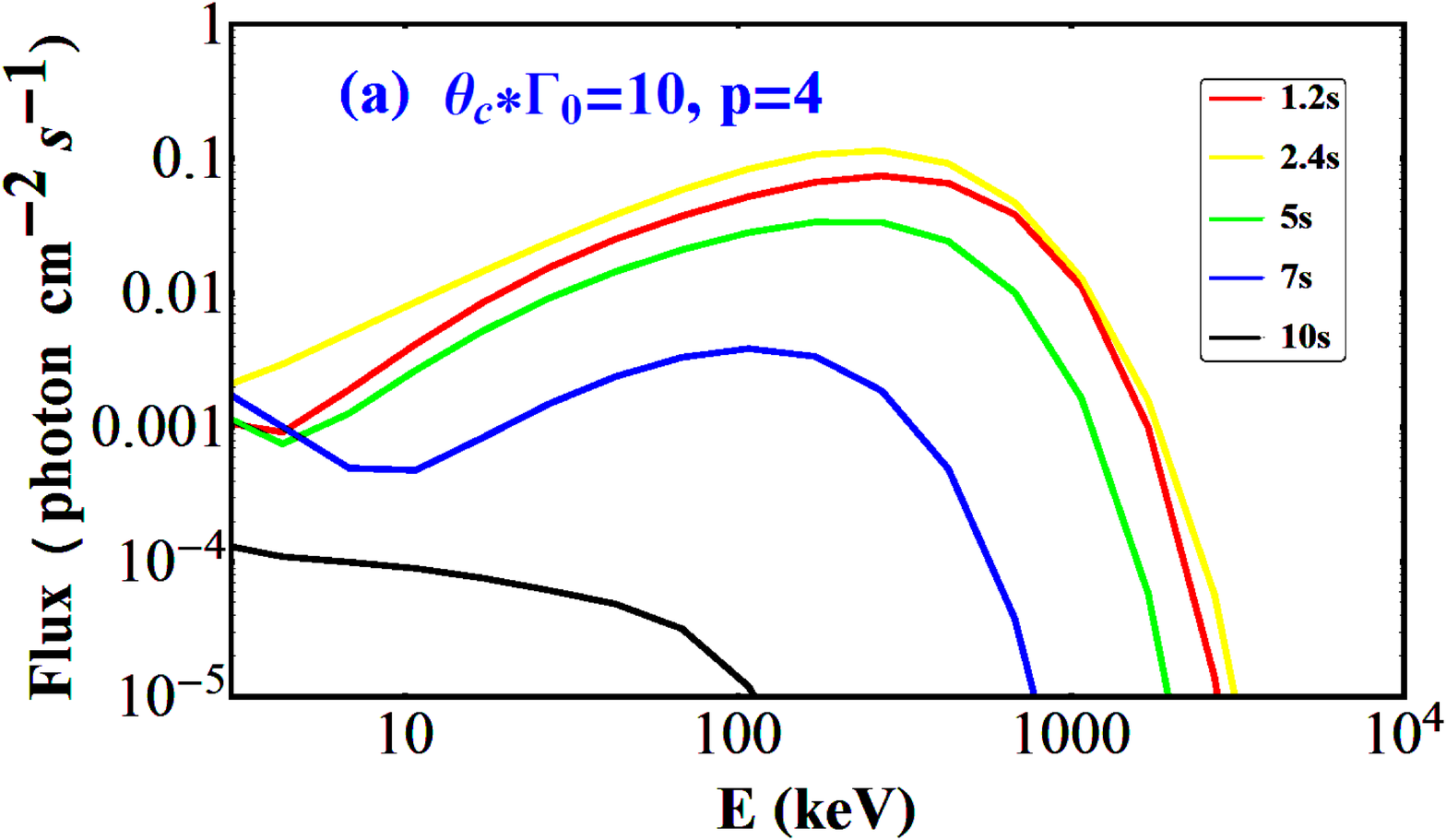} \
\ \centering\includegraphics[angle=0,height=2.0in]{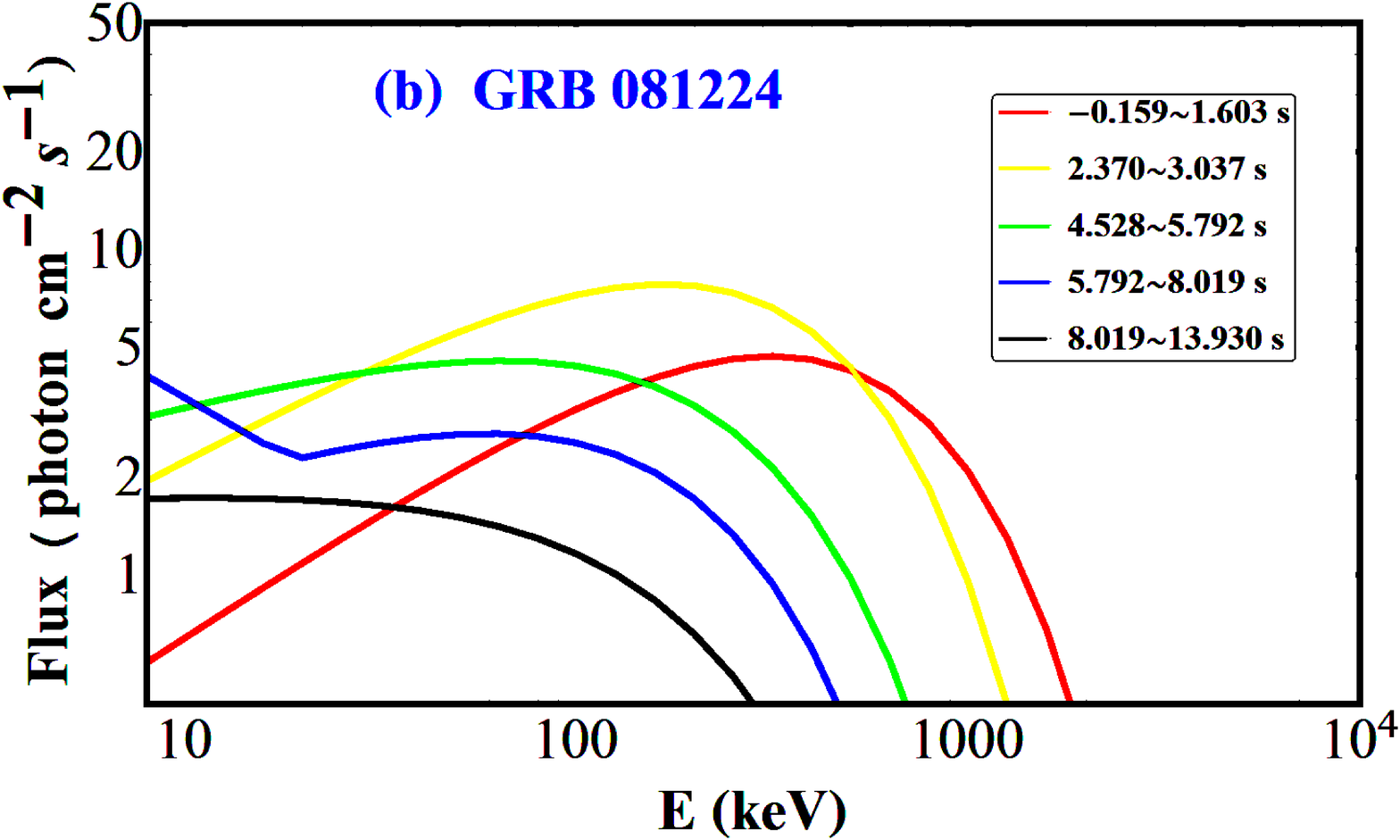} \centering%
\includegraphics[angle=0,height=1.8in]{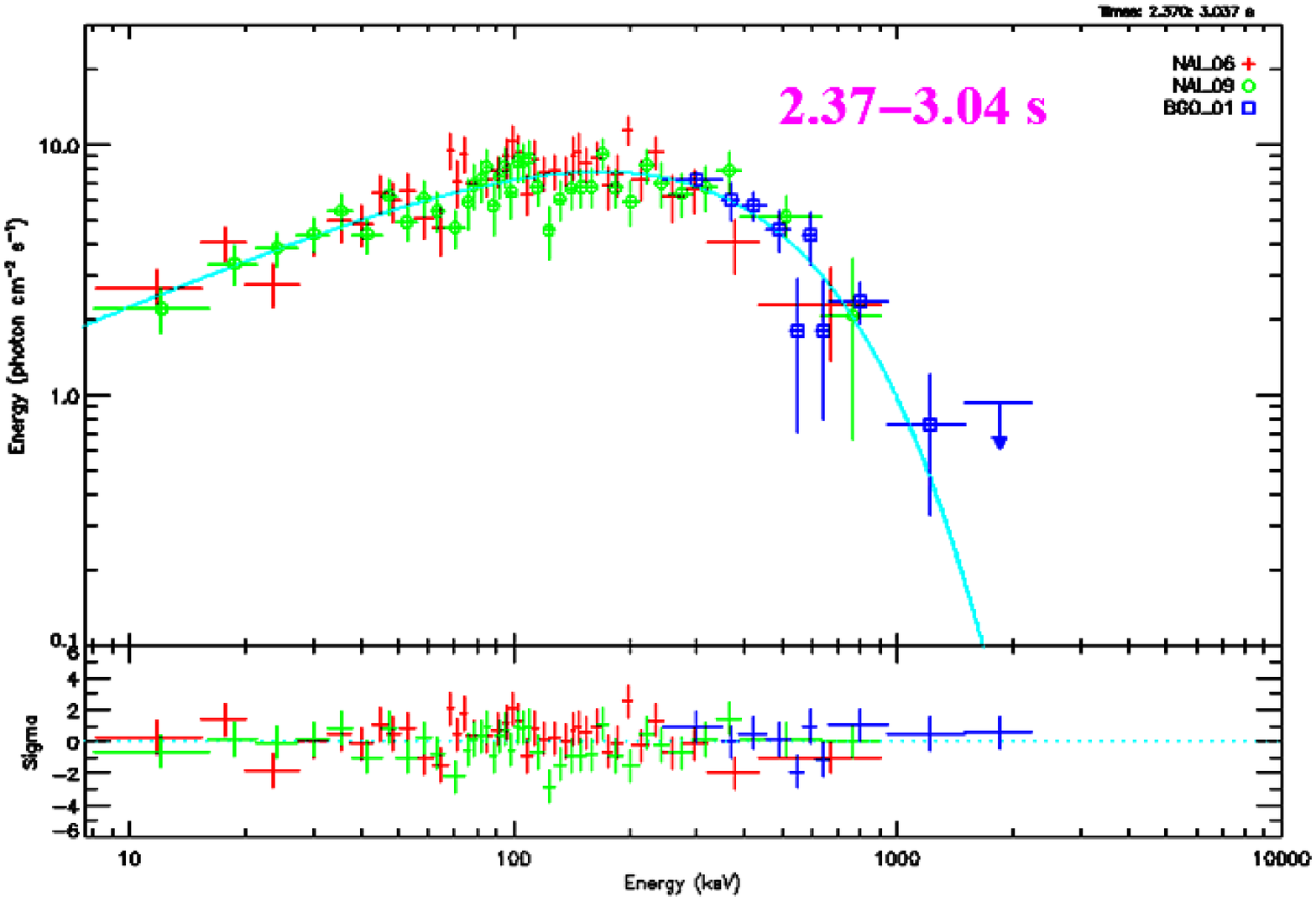} \ \centering%
\includegraphics[angle=0,height=1.8in]{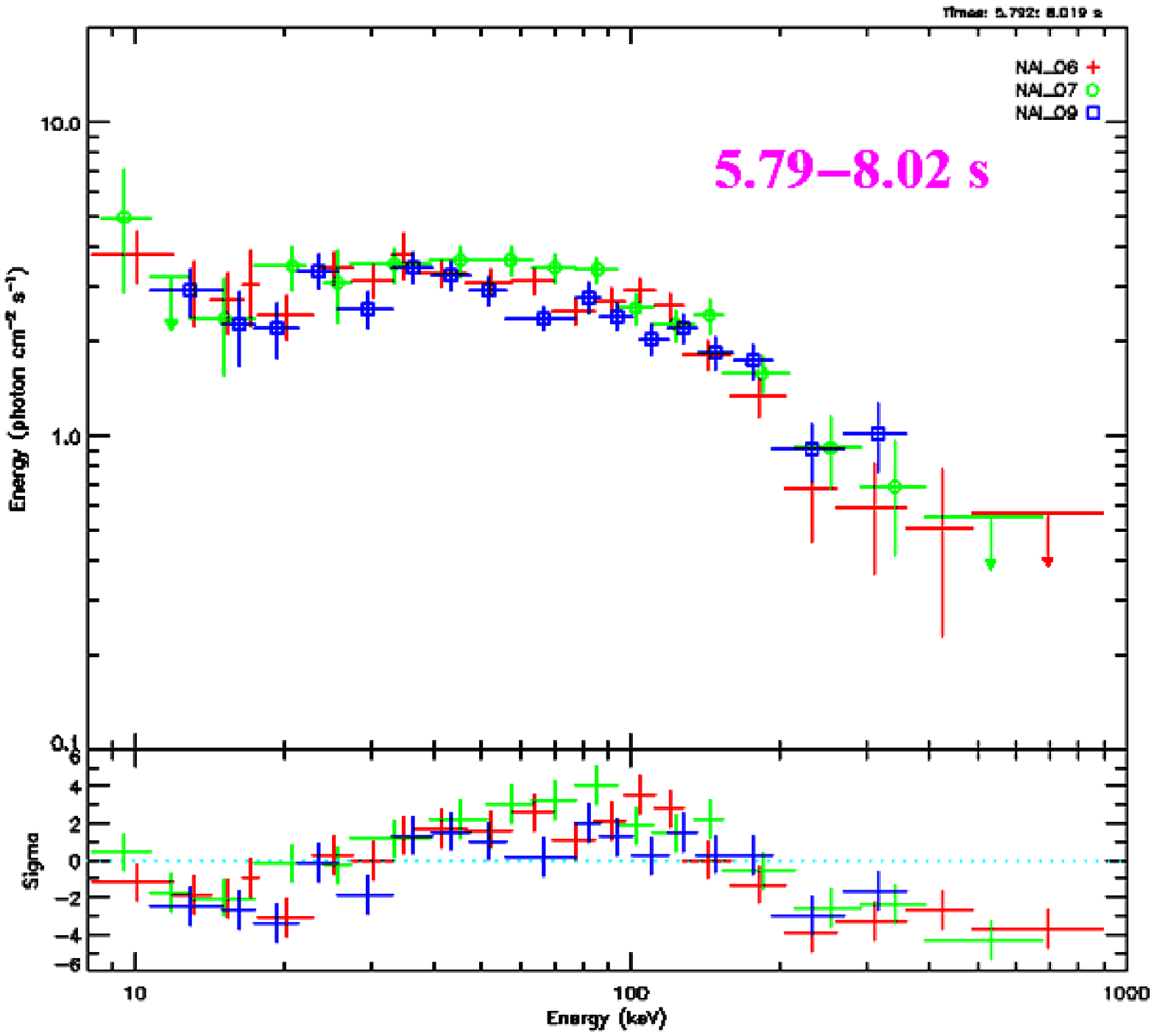} \ \centering%
\includegraphics[angle=0,height=1.8in]{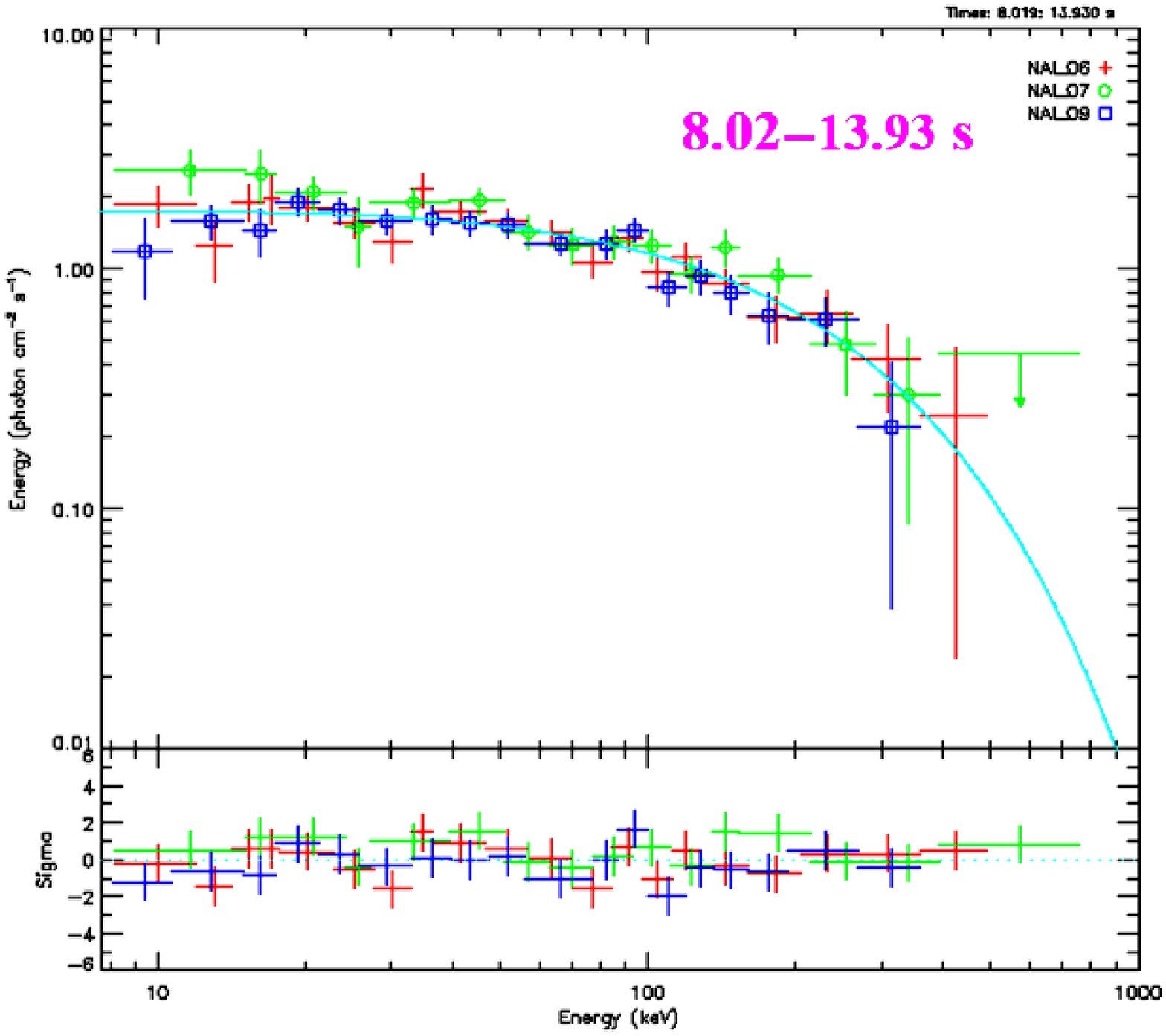}\ \ \ \ \ \ 
\caption{{}Comparison of the calculated time-resolved spectra for $\protect%
\theta _{c}\Gamma _{0}=10$, $p=4$ and the time-resolved spectral analysis
results of GRB 081224. Top left panel:\textit{\ }the calculated
time-resolved spectra for $\protect\theta _{c}\Gamma _{0}=10$ and $p=4$,
same as Figure 7b.\textit{\ }Top right panel: the best-fit model spectra of
the time-resolved spectra from GRB 081224\ for several different time
intervals.\ Bottom panels: the spectral fits to the time-resolved spectra
for $2.370-3.037$ s (left) and $8.019-13.930$ s (right), and the observed
spectrum (fitted with the PL model) for $5.792-8.019$ s (middle).}
\end{figure*}

\begin{figure*}[ph]
\label{Fig_14} \centering\includegraphics[angle=0,height=2.0in]{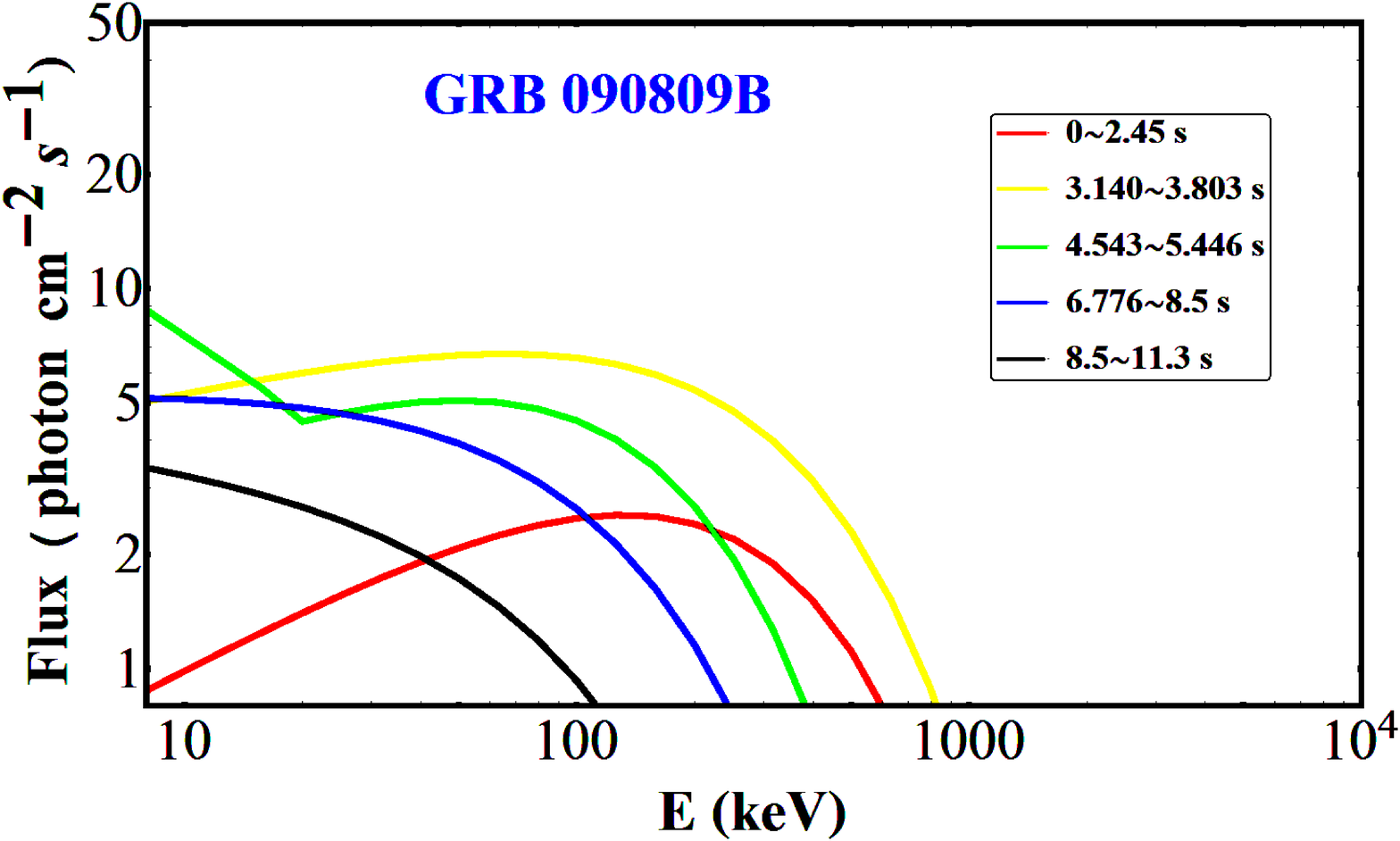} \
\ \ \ \ \ \ \ \ \ \ \ \ \ \ \ \ \ \ \ \ \ \ \ \ \ \ \ \ \ \ \ \ \ \ \ \ \ \
\ \ \ \ \ \ \ \ \ \ \ \ \ \ \ \ \ \ \ \ \ \ \ \ \ \ \ \ \ \ \ \ \ \ \ \ \ \ %
\includegraphics[angle=0,height=1.7in]{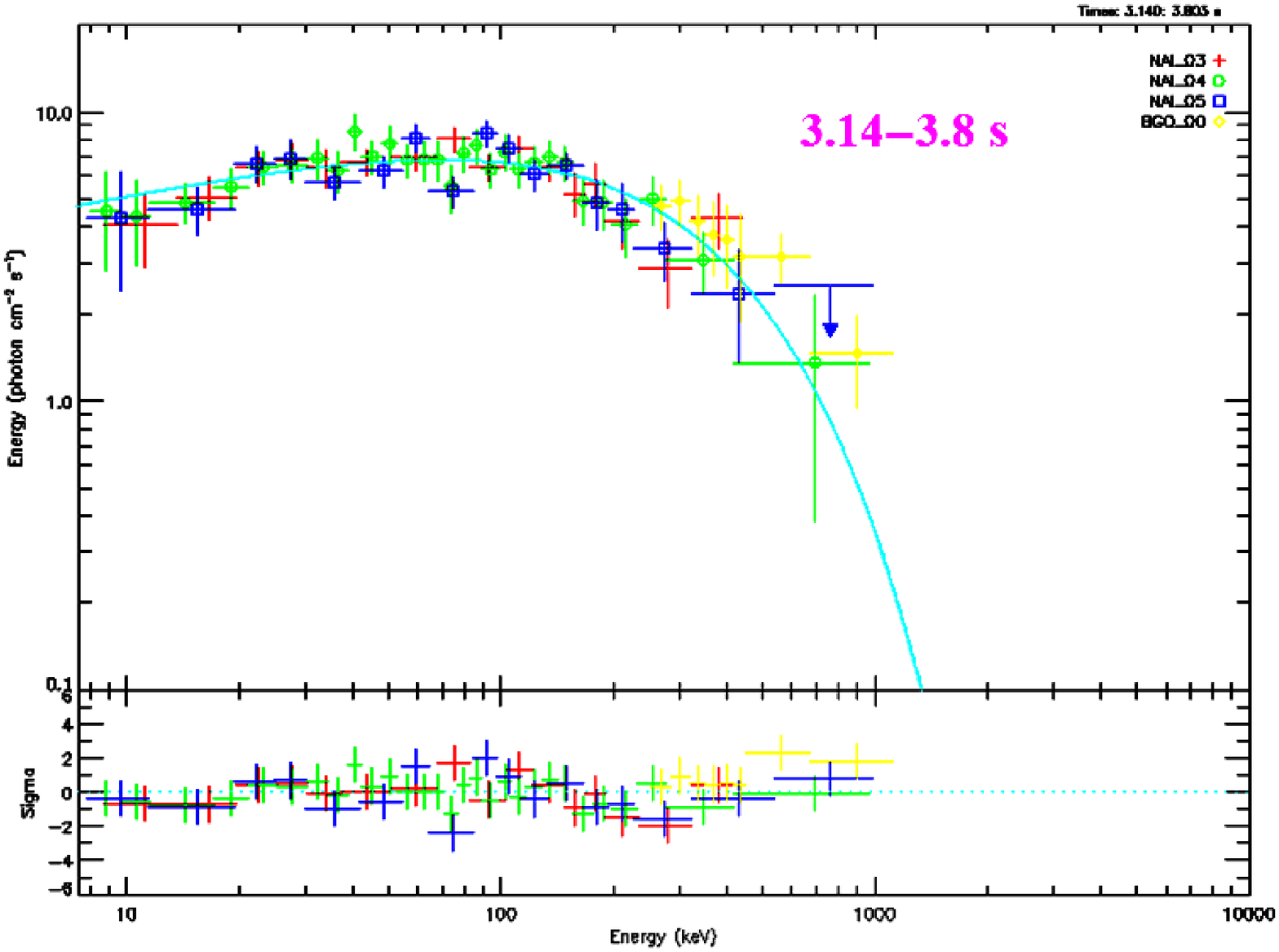} \ \centering%
\includegraphics[angle=0,height=1.7in]{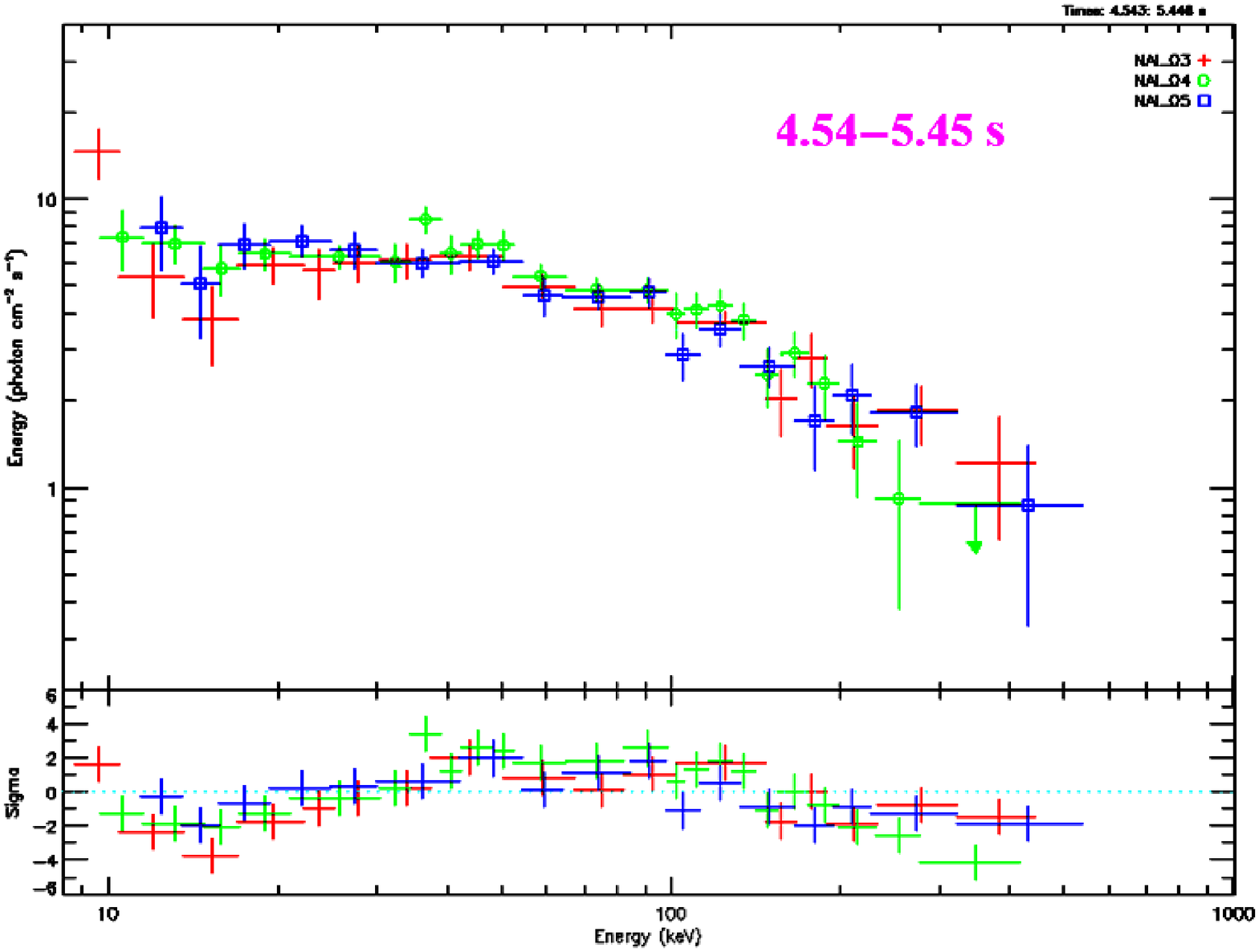} \ \centering%
\includegraphics[angle=0,height=1.7in]{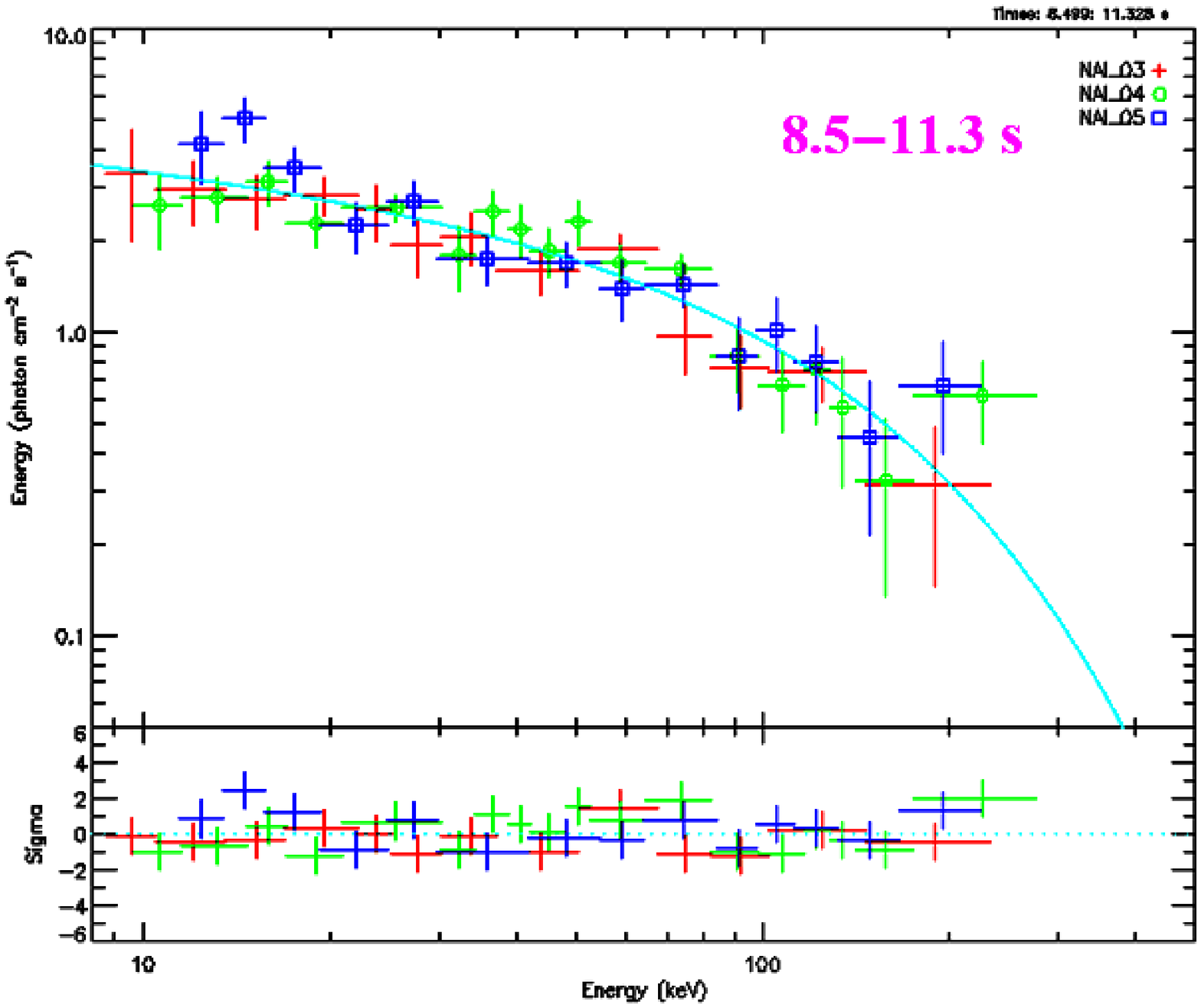}\ \ \ \ \ \ 
\caption{{}The time-resolved spectral analysis results of GRB 090809B.%
\textit{\ }Top panel: \ the best-fit model spectra of the time-resolved
spectra from GRB 090809B\ for several different time intervals.\ Bottom
panels: the spectral fits to the time-resolved spectra for $3.140-3.803$ s
(left) and $8.5-11.3$ s (right), and the observed spectrum (fitted with the
PL model) for $4.543-5.446$ s (middle).}
\end{figure*}

Previously, we calculated the time-resolved spectra of winds with variable
luminosity for a big set of the parameter space region. Here, we perform
time-resolved spectral analysis of several GRBs observed by \textit{Fermi}
GBM and possessing a pulse that has a rather good profile. By comparison, we
may know whether the photosphere model with an angular Lorentz factor
profile can explain the observed spectral evolution well. We select the GRBs
based on \citet{Lu2012}, and divide the GRBs into two categories:
multi-pulse GRBs (GRB 081125, GRB 090131B and GRB 090626A) which have
multiple pulses, and single-pulse GRBs (GRB 081224, GRB 090809B and GRB
110817A) which have single pulse\footnote{%
Note that several recent works study the spectral characteristics of the
single-pulse dominated GRBs \citep{Yu2018} and some special multi-pulse GRBs %
\citep[e.g.,][]{Lv2017,Wei2017,ZhaBB2018,Li2019b}.}.

\subsubsection{multi-pulse GRBs}

Figure 10 shows the time-resolved spectral analysis results of GRB 081125.
Three different empirical models are fitted to each observed time-resolved
spectrum, namely, the Band function (BAND), the cutoff power law (COMP)
model and a simple power law (PL). In the top right panel, we present the
model spectra of the best-fit models with corresponding parameters\ for $%
2.35-2.8$ s and $6.3-9.6$ s. The spectral fits to the time-resolved spectra
for $2.35-2.8$ s and $6.3-9.6$ s are illustrated in the middle left panel
and the bottom right panel, respectively. But we find that neither of the
three different empirical models can fit the time-resolved spectra very well
for $4.3-5.3$ s and $5.3-6.3$ s, since the time-resolved spectra seem to be
fitted better with the power law model on the low-energy end and the COMP
model on the high-energy end. The combined model spectra of the PL plus COMP
model with corresponding rough\footnote{%
Since we do not fit the cutoff energy for these two models but fix it at a
rough value.} best-fit parameters are shown in the top right panel also. The
observed spectra (fitted with the PL model) for $4.3-5.3$ s and $5.3-6.3$ s
are illustrated in the middle right panel and the bottom left panel,
respectively. We find that the observed spectral evolution is quite similar
to that for the case of $\theta _{c}\Gamma _{0}=1$ and $p=4$ (shown in the
top left panel, and same as Figure 7a). The spectra during the rising phase (%
$t=$ $1.2$ s, $2.4$ s for model, $2.35-2.8$ s for observation) are cutoff
power law with a little harder than the flattened shape below the peak,
during the decay phase ($t=5$ s, $7$ s, $10$ s for model, $4.3-5.3$ s, $%
5.3-6.3$ s and $6.3-9.6$ s for observation), a power law with negative index
shows up gradually. At $t=7$ s for model and $4.3-5.3$ s, $5.3-6.3$ s for
observation, the spectra are all the mix of the power law with negative
index on the low-energy end and the modified blackbody with a shallower
low-energy spectral index (larger than $-1$) on the high-energy end. While
the spectra at $t=10$ s\ for model and $6.3-9.6$ s for observation both are
fully the power law with negative index. Our best-fit results of the
time-resolved spectra from GRB 081125 are almost consistent with those of %
\citet{Yu2016}, where the best-fit models before $3.615$ s are all the COMP
model and the best-fit models after $3.615$ s are the PL model.

Similarly, Figure 11 shows the time-resolved spectral analysis results of
GRB 090131B. The best-fit results of the time-resolved spectra\ for several
different time intervals are presented in the top right panel. In the bottom
panels, the spectral fits to the time-resolved spectra for $24.2-26.6$ s
(left) and $26.6-30$ s (right) are illustrated. The observed spectral
evolution is found to be quite similar to that for the case of $\theta
_{c}\Gamma _{0}=1$ and $p=1$ (shown in the top left panel, and same as
Figure 6c). The spectra during the rising phase ($t=$ $1.2$ s, $2.4$ s for
model, $22.658-22.907$ s, $22.907-23.168$ s for observation) are cutoff
power law with the flattened shape below the peak, during the decay phase ($%
t=5$ s, $7$ s, $10$ s for model, $23.5-24.2$ s, $24.2-26.6$ s and $26.6-30$
s for observation), a power law with negative index shows up gradually. At $%
t=7$ s for model and $23.5-24.2$ s for observation, the spectra both are the
mix of the power law with negative index on the low-energy end and the
modified blackbody with the low-energy flattened shape on the high-energy
end. While the spectra at $t=10$ s\ for model and $24.2-26.6$ s, $26.6-30$ s
for observation (see the two bottom panels) are all fully the power law with
negative index. Our best-fit results of the time-resolved spectra from GRB
090131B are also almost consistent with those of \citet{Yu2016}, where the
best-fit models before $23.422$ s are all the COMP model and the best-fit
models after $23.898$ s are the PL model.

Also, Figure 12 shows the time-resolved spectral analysis results of GRB
090626A. The best-fit results of the time-resolved spectra\ for several
different time intervals are presented in the left panel, and the spectral
fits to the time-resolved spectrum for $8.9-14$ s in the right panel. The
observed spectral evolution is similar to that for the case of $\theta
_{c}\Gamma _{0}=1$ and $p=1$ (see Figure 6c), too. And the best-fit results
of the time-resolved spectra are also almost consistent with those of %
\citet{Yu2016}, where the best-fit models after $5.888$ s are the PL model.

\subsubsection{single-pulse GRBs}

\begin{figure*}[t]
\label{Fig_15} \centering\includegraphics[angle=0,height=2.0in]{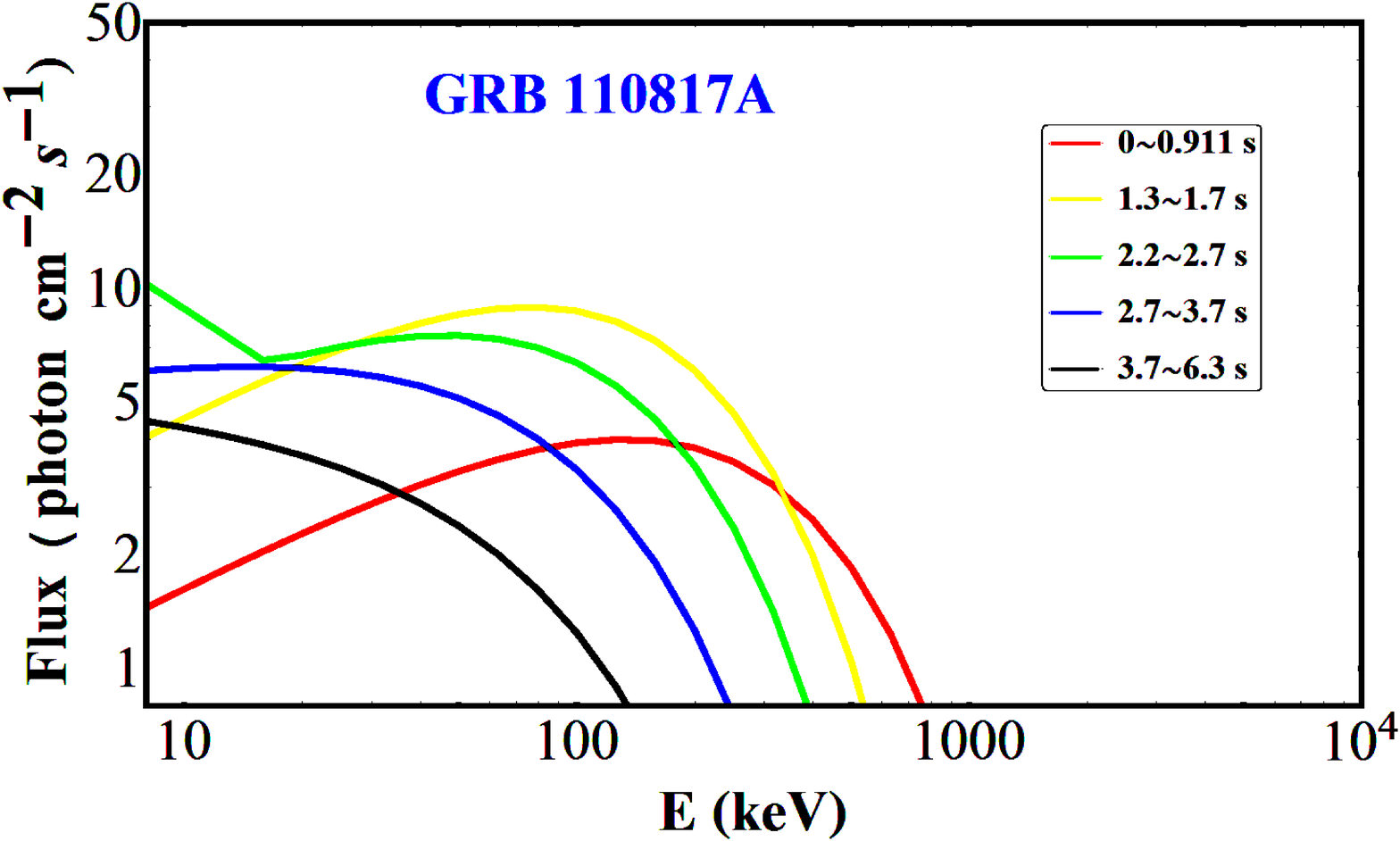} \
\ \ \ \ \ \ \ \ \ \ \ \ \ \ \ \ \ \ \ \ \ \ \ \ \ \ \ \ \ \ \ \ \ \ \ \ \ \
\ \ \ \ \ \ \ \ \ \ \ \ \ \ \ \ \ \ \ \ \ \ \ \ \ \ \ \ \ \ \ \ \ \ \ \ \ \ %
\includegraphics[angle=0,height=1.8in]{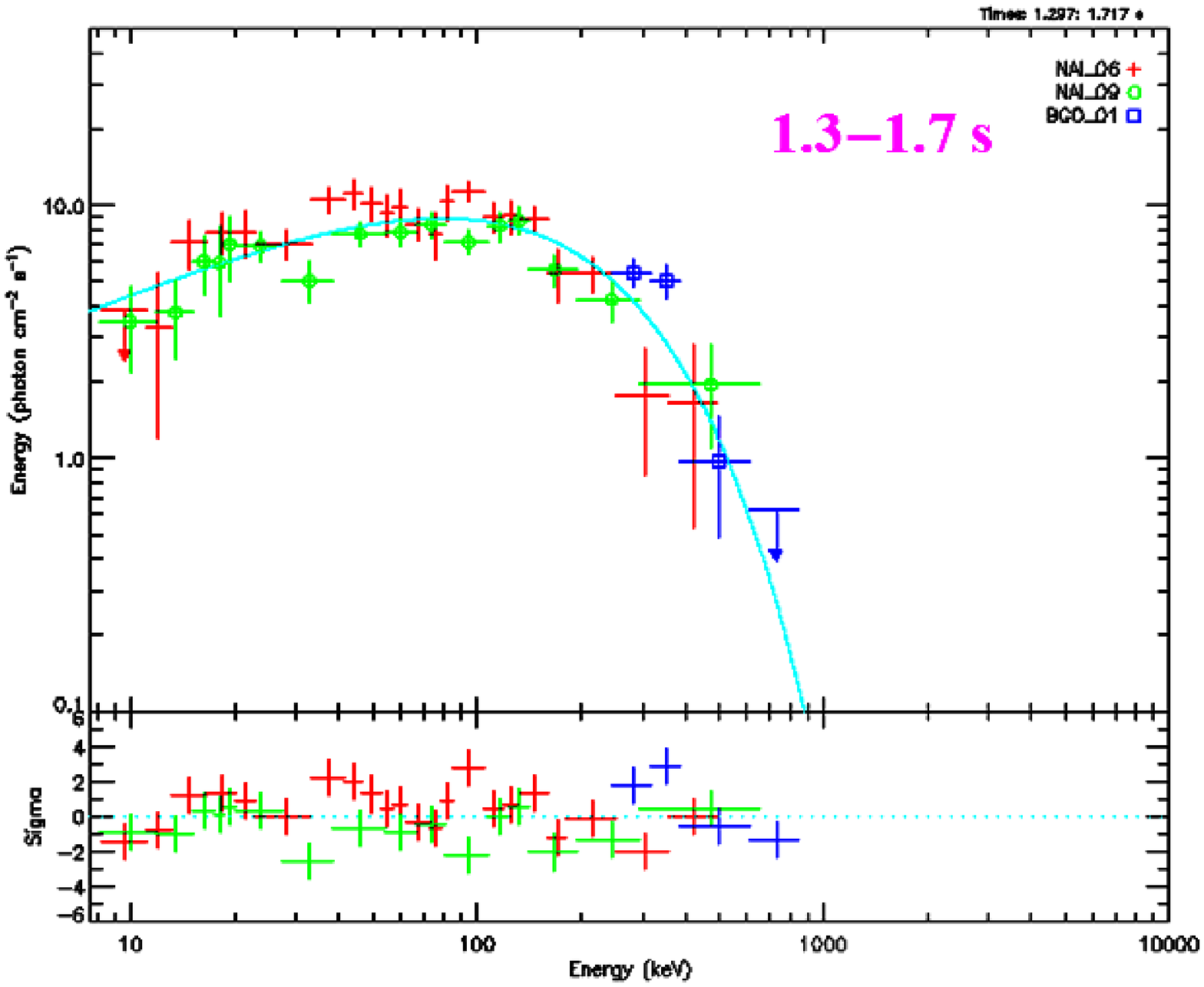} \ \centering%
\includegraphics[angle=0,height=1.8in]{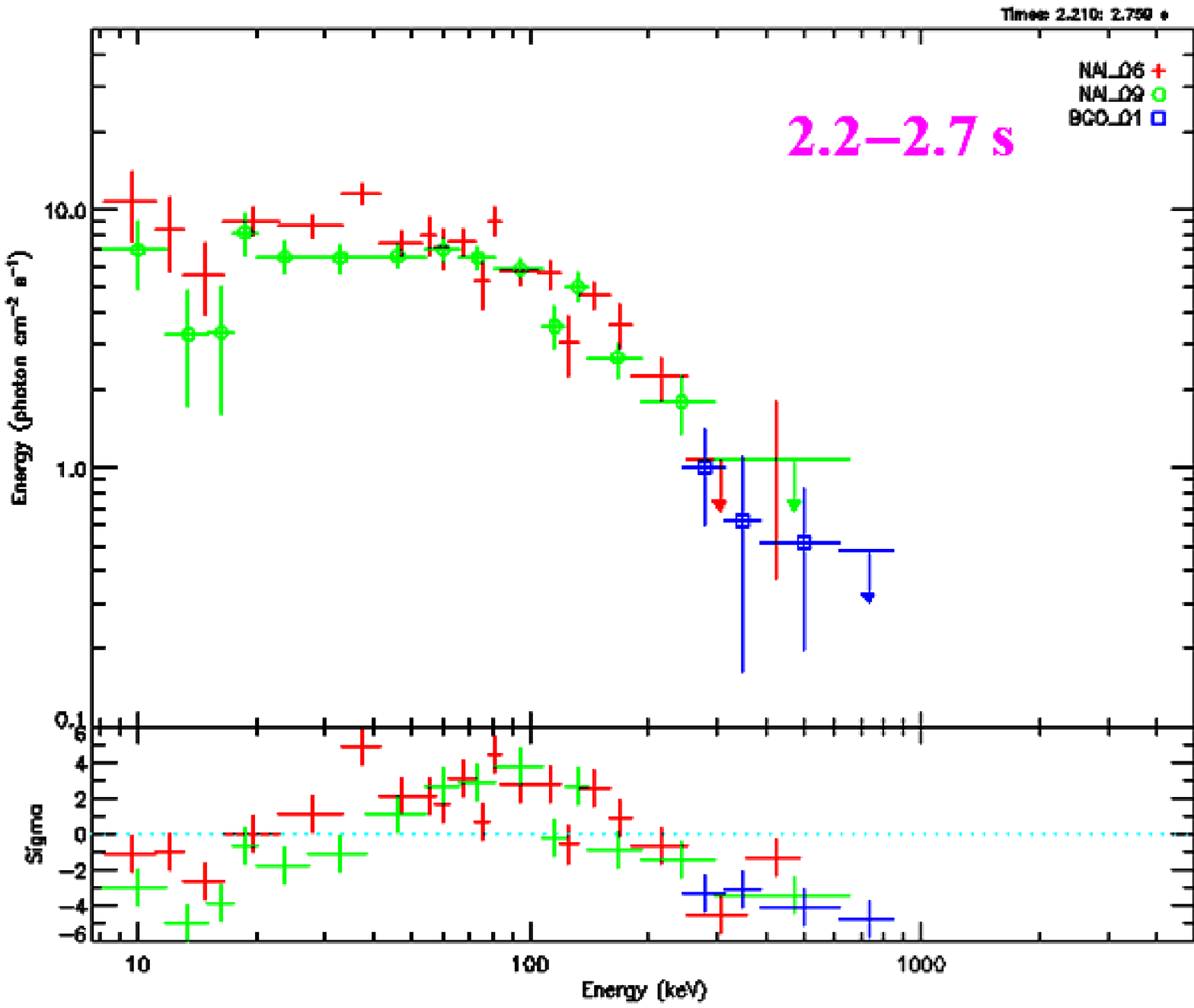} \ \centering%
\includegraphics[angle=0,height=1.8in]{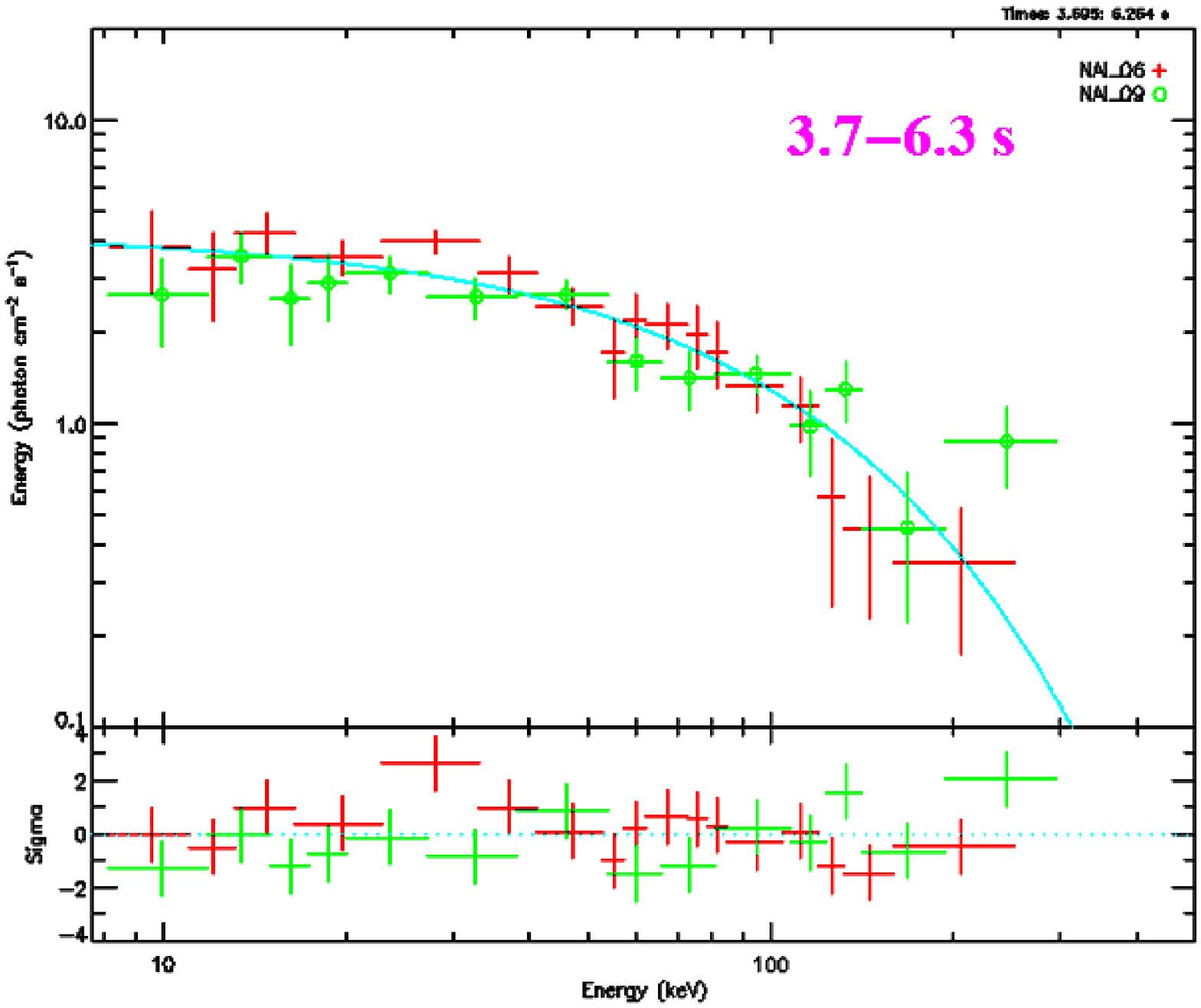}\ \ \ \ \ \ 
\caption{{}The time-resolved spectral analysis results of GRB 110817A.%
\textit{\ }Top panel: \ the best-fit model spectra of the time-resolved
spectra from GRB 110817A\ for several different time intervals.\ Bottom
panels: the spectral fits to the time-resolved spectra for $1.3-1.7$ s
(left) and $3.7-6.3$ s (right), and the observed spectrum (fitted with the
PL model) for $2.2-2.7$ s (middle).}
\end{figure*}

Figure 13 shows the time-resolved spectral analysis results of GRB 081224.
The best-fit results of the time-resolved spectra\ for several different
time intervals are presented in the top right panel. In the bottom panels,
the spectral fits to the time-resolved spectra for $2.370-3.037$ s (left)
and $8.019-13.930$ s (right), and the observed spectra (fitted with the PL
model) for $5.792-8.019$ s (middle) are illustrated. The observed spectral
evolution is found to be quite similar to that for the case of $\theta
_{c}\Gamma _{0}=10$ and $p=4$ (shown in the top left panel, and same as
Figure 7b). The spectra during the rising phase ($t=$ $1.2$ s, $2.4$ s for
model, $2.370-3.037$ s for observation) are still the cutoff power law with
a little harder than the flattened shape below the peak. At $t=7$ s for
model and $5.792-8.019$ s for observation, the spectra both are the mix of
the power law with negative index on the low-energy end and the modified
blackbody with a shallower low-energy spectral index (larger than $-1$) on
the high-energy end. However, the spectra at $t=10$ s\ for model and $%
8.019-13.930$ s for observation are not the power law with negative index,
but the modified blackbody with the flattened shape ($F_{\nu }\sim \nu ^{0}$%
) below the peak. The best-fit results of the time-resolved spectra are
almost consistent with those of \citet{Yu2016}, where the best-fit models
before $13.975$ s are all the COMP model and the best-fit low-energy
spectral index changes from $-0.2$ to $-1$.

Figures 14 and 15 present the time-resolved spectral analysis results of GRB
090809B and GRB 110817A, respectively. The observed spectral evolution for
each is similar to that for GRB 081224, thus the case of $\theta _{c}\Gamma
_{0}=10$ and $p=4$. The best-fit results of the time-resolved spectra are
also almost consistent with those of \citet{Yu2016}.

According to the above analyses, we can see that the photosphere model with
an angular Lorentz factor profile may explain the observed spectral
evolution well. In addition, the observed spectral evolutions for
multi-pulse GRBs and single-pulse GRBs seem to be different. The $\theta
_{c} $ for multi-pulse GRBs seems to be more narrow, while much wider for
the single-pulse GRBs.

\subsection{Luminosity profiles and $E_{p}$ evolution}

\begin{figure*}[ph]
\label{Fig_16} \centering\includegraphics[angle=0,height=5.2in]{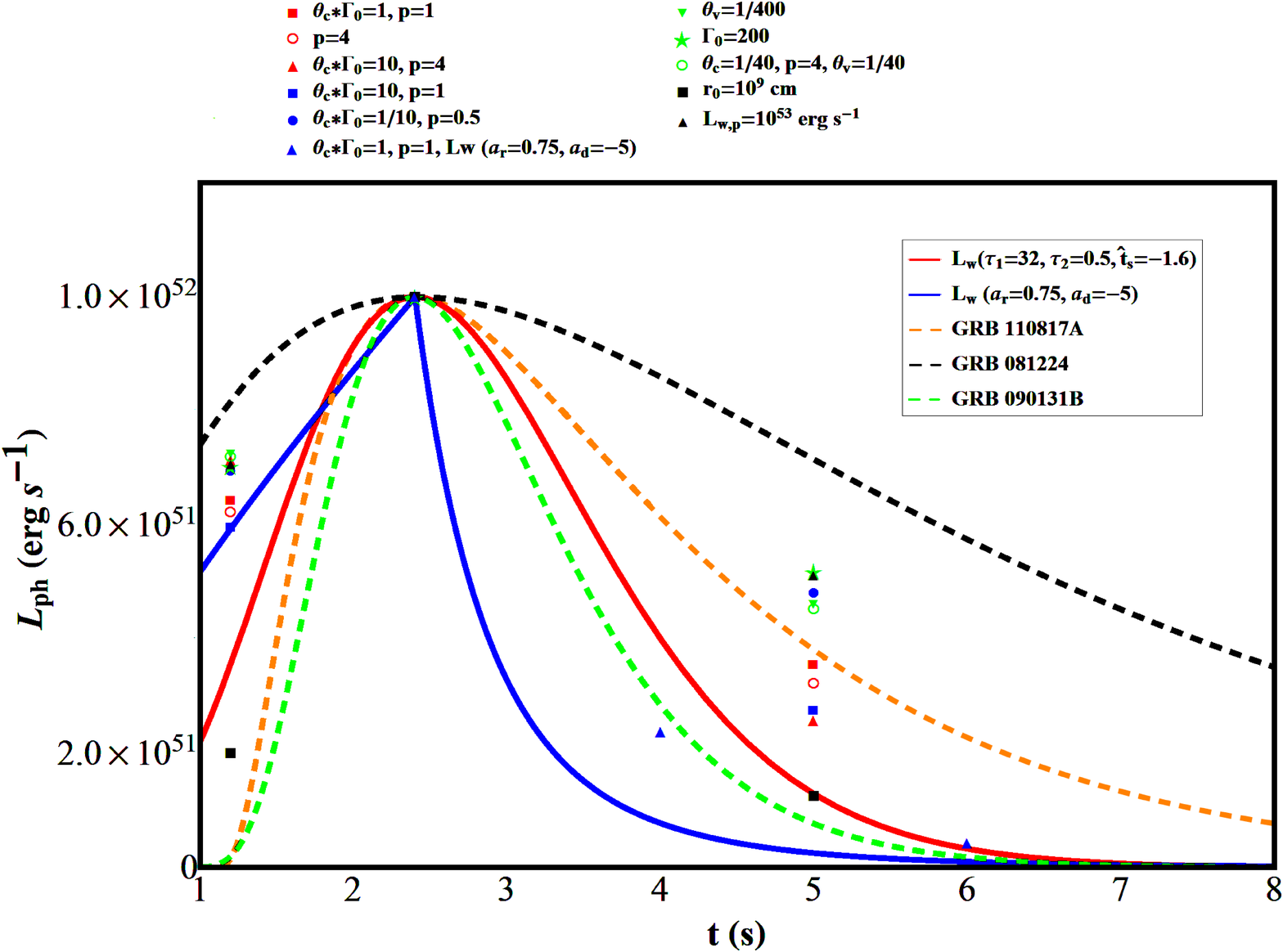}
\caption{The initial wind luminosity $L_{w}$ and photosphere luminosity $L_{%
\text{ph}}$ profiles. For comparison, the photosphere luminosity $L_{\text{ph%
}}$ profiles have all been normalized to $L_{w,\text{ }p}$ $=10^{52}$ erg s$%
^{-1}$ at $2.4$ s. The initial wind luminosity $L_{w}$ profiles for the
exponential model ($\protect\tau _{1}=32$, $\protect\tau _{2}$ $=0.5$, $\hat{%
t}_{s}=$ $-1.6$) and the broken power law model ($a_{r}=0.75$, $a_{d}=$ $-5$%
) are plotted by the red solid line and the blue solid line, respectively.
The photosphere luminosity $L_{\text{ph}}$ profiles for the exponential
model with various parameters are shown by the points of different shapes
and colors at $1.2$ s, $2.4$ s and $5$ s. While the photosphere luminosity $%
L_{\text{ph}}$ profile for the broken power law model with the typical
parameters ($\protect\theta _{c}\Gamma _{0}=1$, $p=1$, $\protect\theta _{v}$ 
$=0$ ) is shown by the blue triangle points at $2.4$ s, $4$ s and $6$ s. The
three dashed lines are the light curves for GRB 110817A (orange), GRB 081224
(black) and GRB 090131B (green), respectively.}
\end{figure*}

\begin{figure*}[t]
\label{Fig_17} \ \ \ \ \ \ \ \ \ \ \centering%
\includegraphics[angle=0,height=1.75in]{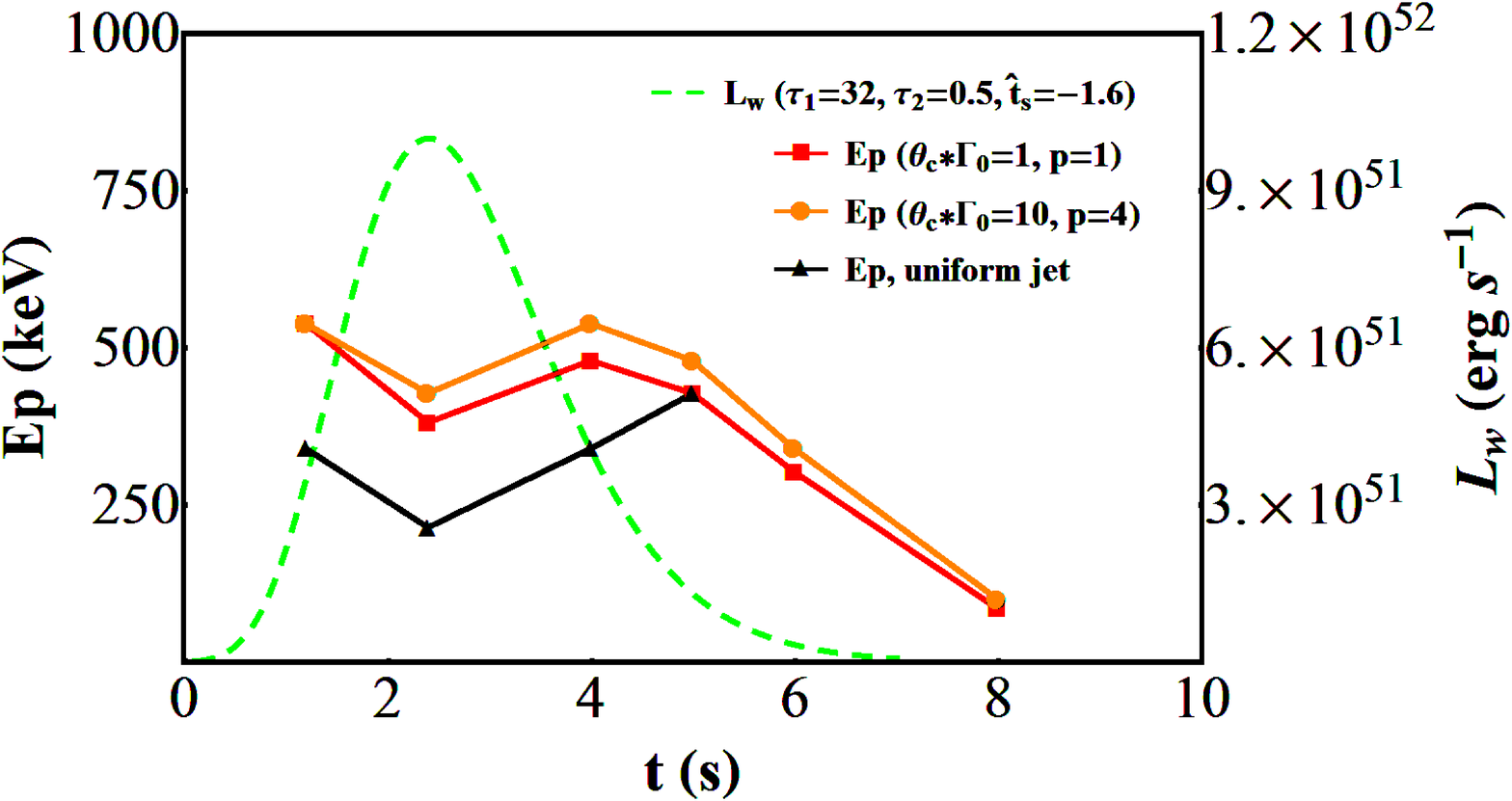}%
\includegraphics[angle=0,height=1.75in]{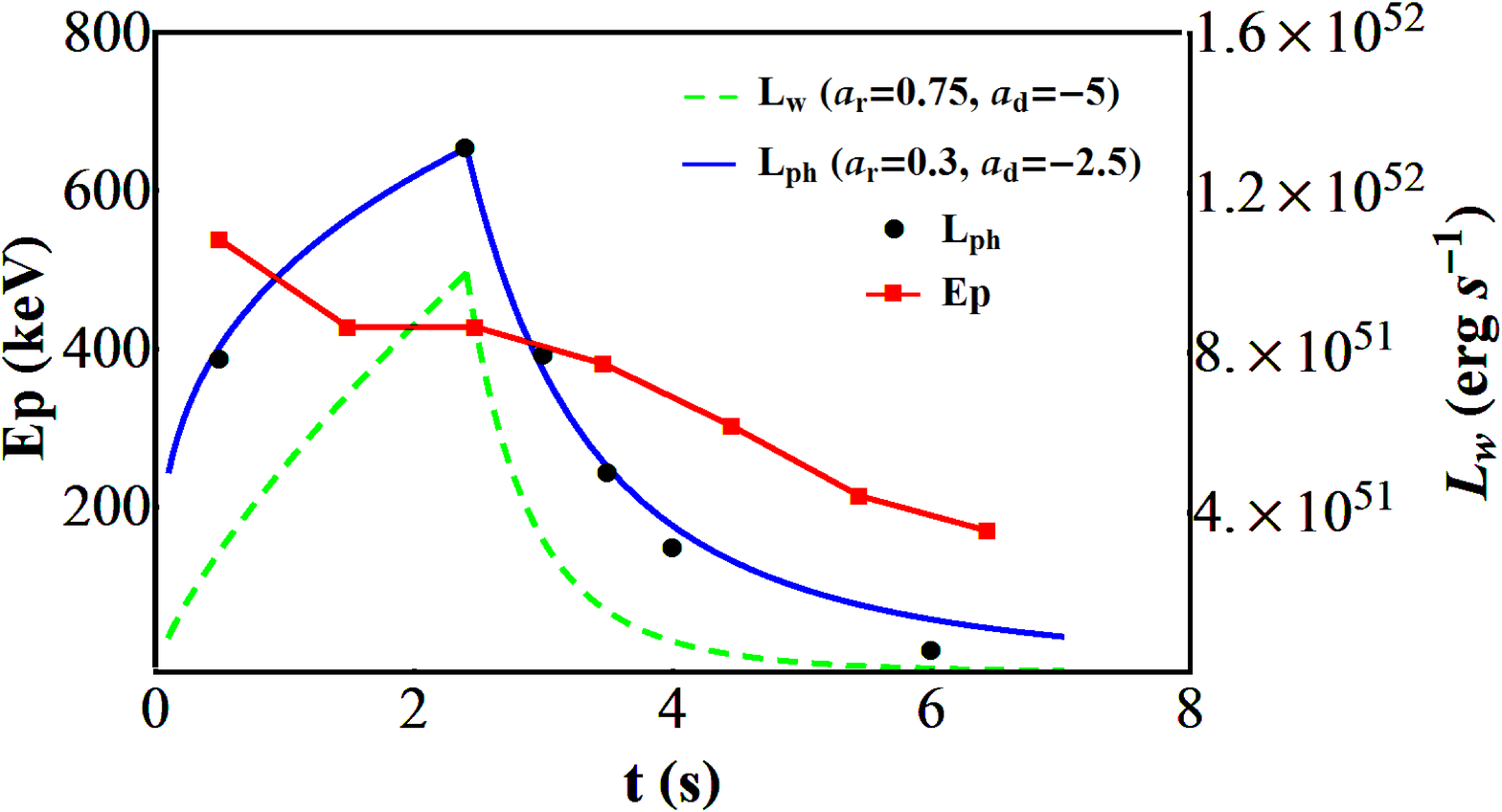} \ \ \ \ \ \ \ \ \ \ \ \ \
\ \ \ \ \ \ \ \ \ \ \ \ \ \ \ \ \ \ \ \ \ \ \ \ \ \ \ \ \ \ \ \ \ \ \ \ \ \
\ \ \ \ \ \ \ \ \ \ \ \ \ \ \ \ \ \ \ \ \ \ \ \ \ \ \ \ \ \ \ \ \ \ \ \ \ \
\ \ \ \ \ \ \ \ \ \ \ \ \ \ \ \ \ \ \ \ \ \ \ \ \ \ \ \ \ \ \ \ \ \ \ \ \ \
\ \ \ \ \ \ \ \ \ \ \ \ \ \ \ \ \ \ \ \ \ \ \ \ \ \ \ \ \ \ \ \ \ \ \ \ \ \
\ \ \ \ \ \ \ \ \ \ \ \ \ \ \ \ \ \ \ \ \ \ \ \ \ \ \ \ \ \ \ \ \ \ \ \ \ \
\ \ \ \ \ \ \ \ \ \ \ \ \ \ \ \ \ \ \ \ \ \ \ \ \ \ \ \ \ \ \ \ \ \ \ \ \ \
\ \centering\includegraphics[angle=0,height=1.75in]{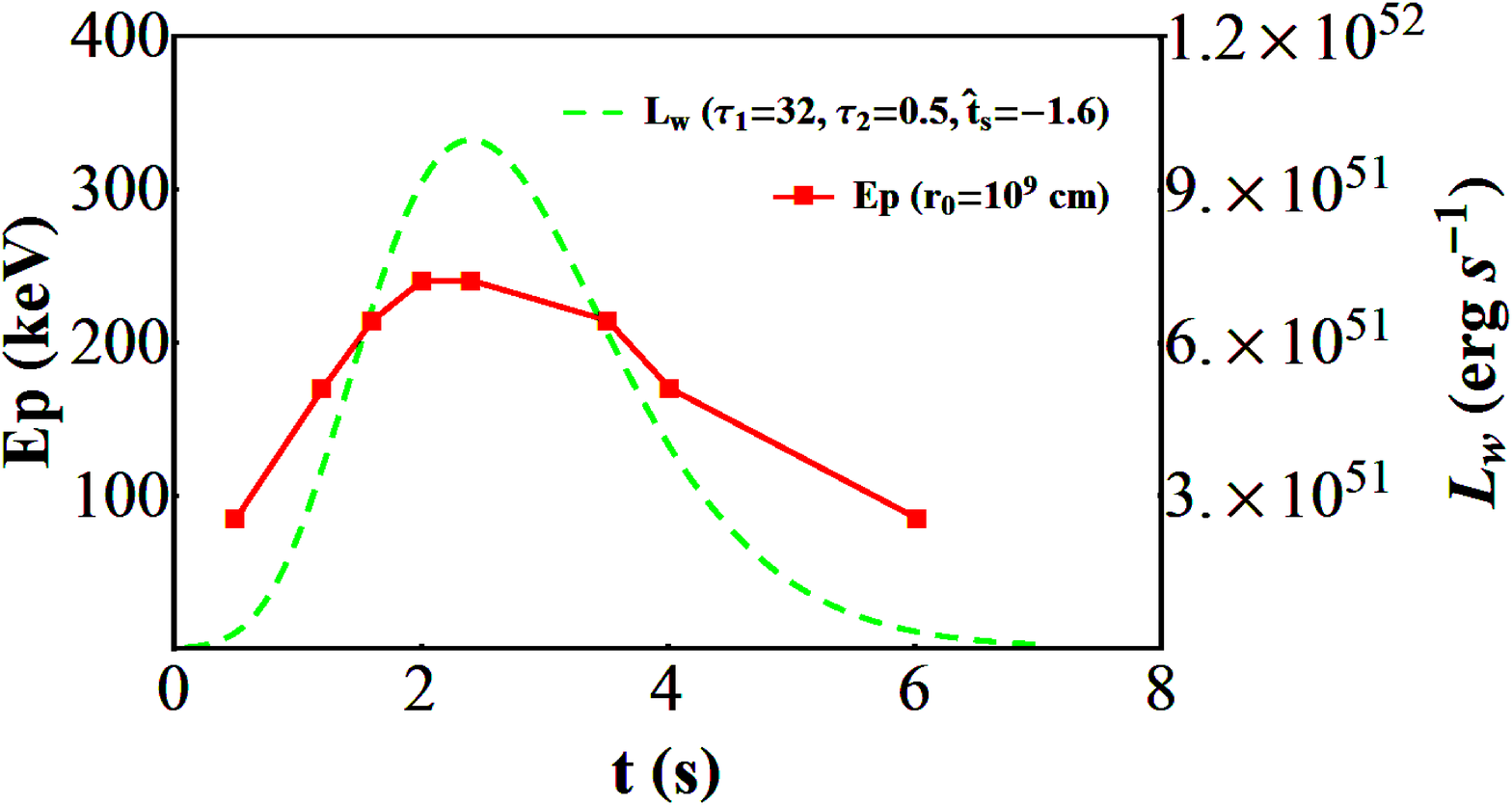}\ 
\caption{{}Evolution of $E_{p}$ for different parameters. Top left panel:
initial wind luminosity $L_{w}$ (green dashed line) and evolution of $E_{p}$
with \textquotedblleft $\protect\theta _{c}\Gamma _{0}=1$, $p=1$%
\textquotedblright (red solid line), \textquotedblleft $\protect\theta %
_{c}\Gamma _{0}=10$, $p=4$\textquotedblright (orange solid line) or a
uniform jet (black solid line) for the case \textquotedblleft $\protect\tau %
_{1}=32$, $\protect\tau _{2}=0.5$, $\hat{t}_{s}=-1.6$\textquotedblright .
Top right panel:\textit{\ }$E_{p}$ evolution (red solid line), the initial
wind luminosity $L_{w}$ (green dashed line) and the photosphere luminosity $%
L_{\text{ph}}$ (black circle points and blue solid line) for the case
\textquotedblleft $a_{r}=0.75$, $a_{d}=$ $-5$\textquotedblright . The $E_{p}$
is for the time-integrated spectra here, which have a time bin of $1$ $s$
(The $E_{p}$ at $1.5$ s is for the time interval of $1-2$ s, and so on). \
Bottom panel:\textit{\ }initial wind luminosity $L_{w}$ (green dashed line)
and evolution of $E_{p}$ with $r_{0}=10^{9}$ cm (red solid line) for the
case \textquotedblleft $\protect\tau _{1}=32$, $\protect\tau _{2}=0.5$, $%
\hat{t}_{s}=-1.6$\textquotedblright .}
\end{figure*}

The light curve is an important observational characteristic\ for GRBs.
Here, in Figure 16, we use the light curve profiles of a few GRBs to test
the reasonability of our initial wind luminosity $L_{w}$ profiles\footnote{%
Note that we mainly focus on the decay phase since the time scale for the
rise phase is quite short.}. Also, we explore the parameter dependencies of
the photosphere luminosity $L_{\text{ph}}$ profiles. For the initial wind
luminosity $L_{w}$ profile of $a_{r}=0.75$ and $a_{d}=$ $-5$, the
photosphere luminosity $L_{\text{ph}}$ profile (blue triangle points) is
close to the light curve profile of GRB 090131B. While for the initial wind
luminosity $L_{w}$ profile of $\tau _{1}=32$, $\tau _{2}$ $=0.5$ and $\hat{t}%
_{s}=$ $-1.6$, the photosphere luminosity $L_{\text{ph}}$ profiles (the
points of different shapes and colors for various parameters) are close to
the light curve profile of GRB 110817A. As for the parameter dependencies,
the photosphere luminosity $L_{\text{ph}}$ falls down more rapidly for a
wider jet core ($\theta _{c}\Gamma _{0}=10$, $p=1$ or $\theta _{c}\Gamma
_{0}=10$, $p=4$) or larger $r_{0}$ ($r_{0}=10^{9}$ cm), and more slowly for
a narrower jet core ($\theta _{c}\Gamma _{0}=1/10$, $p=0.5$), smaller $%
\Gamma _{0}$ ($\Gamma _{0}=200$), non-zero viewing angle ($\theta _{\text{v}%
} $ $=1/400$ or $\theta _{c}=1/40$, $p=4$, $\theta _{\text{v}}$ $=1/40$) or
larger $L_{w,p}$ ($L_{w,p}$ $=10^{53}$ erg s$^{-1}$).

Furthermore, the evolution of $E_{p}$ is crucial to judge whether a GRB
prompt emission model is better. Observationally, the hard-to-soft evolution
and intensity-tracking patterns have been identified %
\citep{Ford1995,Lia1996,Lu2010,Lu2012}. For the photosphere emission model
of a uniform jet, \citet{Deng2014} showed that the observed hard-to-soft\
evolution pattern cannot be reproduced, and the $E_{p}-L$ tracking pattern
can be reproduced when the dimensionless entropy $\eta $ depends positively
on $L_{w}$. Here, by considering the structured jet and including the $R_{%
\text{ph}}<R_{s}$ regime, we can reproduce the hard-to-soft evolution and
intensity-tracking patterns better with the photosphere model.

Based on the numerical results of the time-resolved spectra presented in
Figure 6c ($\theta _{c}\Gamma _{0}=1$, $p=1$) and Figure 7b ($\theta
_{c}\Gamma _{0}=10$, $p=4$), we plot the evolution of $E_{p}$ with respect
to wind luminosity ($\tau _{1}=32$, $\tau _{2}=0.5$, $\hat{t}_{s}=-1.6$) in
the top left panel of Figure 17. An approximate hard-to-soft evolution is
shown, except for a slight increase after the peak of $L_{w}$. The evolution
is similar for other Lorentz factor profiles and viewing angles considered
above. Now, we perform some analytical discussions. For the regime $R_{\text{%
ph}}>R_{s}$ (relatively large $L_{w}$ near the peak), the observed
temperature can be expressed as 
\begin{eqnarray}
T_{\text{ph}}(\theta ) &\propto &L_{w}^{1/4}r_{0}^{-1/2}[R_{\text{ph}%
}(\theta )/R_{s}(\theta )]^{-2/3}  \notag \\
&\propto &L_{w}^{1/4}r_{0}^{-1/2}\{[L_{w}/\Gamma ^{3}(\theta )]/[\Gamma
(\theta )r_{0}]\}^{-2/3}  \notag \\
&\propto &L_{w}^{-5/12}r_{0}^{1/6}\Gamma ^{8/3}(\theta ).
\end{eqnarray}

Here, the observed temperature is angle-dependent. Compared with the case of
a uniform jet (black solid line), the $L_{w}-E_{p}$ anti-correlation is much
weaker. And if we consider the time-integrated spectra, the anti-correlation
may almost disappear. Then as the $L_{w}$ falls down, one enters the regime $%
R_{\text{ph}}<R_{s}$, and 
\begin{equation}
T_{\text{ph}}(\theta )\propto L_{w}^{1/4}r_{0}^{-1/2}.
\end{equation}

Now, we have $E_{p}\propto L_{w}^{1/4}$, which means $E_{p}$ decreases.
Thus, a approximate hard-to-soft evolution shows up. For\ the
time-integrated spectra of the wind luminosity $a_{r}=0.75$, $a_{d}=$ $-5$
which has a steeper decay phase, the hard-to-soft evolution is rather well
as showed in the top right panel of Figure 17. For this wind luminosity, the
photosphere luminosity $L_{\text{ph}}$ in the decay phase has an index of $%
a_{d}=$ $-2.5$ which is consistent with the average decay phase index $%
d=2.44\pm 0.12$ of a large sample of GRBs in \citet{Koce2003}. In addition,
with a larger $r_{0}$ ($r_{0}=10^{9}$ cm) the evolution of $E_{p}$ acts as
the intensity-tracking pattern well, as showed in the bottom panel of Figure
17. This is because the $R_{\text{ph}}<R_{s}$ regime works for the whole
wind profile (For the same reason, the intensity-tracking pattern can be
obtained with smaller peak luminosity $L_{w,p}=10^{51}$ erg s$^{-1}$). And
the much larger $r_{0}$ is consistent with the rather high mean value $%
\left\langle r_{0}\right\rangle \sim 10^{8.5}$ cm deduced in \citet{Pe2015}.

\section{CONCLUSIONS AND DISCUSSION}

\label{sec:con}

In this paper, we investigate the time-resolved spectra and $E_{p}$
evolutions of photospheric emission from a structured jet. To be more
realistic, a continuous wind with a time-dependent wind luminosity has been
considered. The following conclusions are drawn. (1) The photosphere
spectrum near the peak luminosity is similar to the spectrum of the cutoff
power-law model, which is the best-fit model for a large percentage of the
time-resolved spectra in GRBs \citep[e.g.,][]{Kan2006,Yu2016}. (2) The
photosphere spectrum near the peak luminosity can have a flattened shape ($%
F_{\nu }$ $\sim $ $\nu ^{0}$) below the peak, consistent with the average
low-energy spectral index ($\alpha $ $\sim -1$) for the time-resolved
spectra observed in GRBs \citep[e.g.,][]{Kan2006,Yu2016}. Also, the
distribution of the low-energy spectral index for our photosphere model is
similar to that observed ($-2\lesssim $ $\alpha \lesssim 0$). For $\theta
_{c}\Gamma _{0}=1/10$ and $p=0.5$, $\alpha $ $\sim -1.8$; for $\theta
_{c}\Gamma _{0}=10$ and $p=4$, $\alpha $ $\sim 0$. (3) Judged by the width
of the jet core, the spectral evolutions during the decay phase for our
photosphere model can be mainly divided into two types. A power law with
negative index gradually emerges for narrower core ($\theta _{c}\Gamma
_{0}=1 $), while a modified blackbody with a flattened shape ($F_{\nu }\sim
\nu ^{0} $) below the peak shows up for wider core ($\theta _{c}\Gamma
_{0}=10$, $p=4$). Based on the time-resolved spectral analysis of several
GRBs observed by \textit{Fermi} GBM and possessing a pulse that has a rather
good profile, we find that the above-mentioned two kinds of spectral
evolutions during the decay phase do seem to exist. The spectral evolution
for the multi-pulse GRBs is similar to that for narrower jet core, while the
single-pulse GRBs similar to wider jet core. (4) For this photosphere model,
we can reproduce the two types of observed $E_{p}$ evolution patterns rather
well. For the typical parameters, we get the hard-to-soft evolution; and for
a larger $r_{0}$ ($r_{0}=10^{9}$ cm) or smaller $L_{w,p}$ ($L_{w,p}=10^{51}$
erg s$^{-1}$), we have the intensity-tracking pattern. From the above, by
considering the geometrical broadening for structured jet, we reproduce the
observed time-resolved spectra, the spectral evolutions and $E_{p}$
evolutions well for the GRBs best fitted by the cutoff power-law model for
the peak-flux spectrum or the time-integrated spectrum.

Photospheric emission for spherically symmetric outflows has been
investigated by several authors \citep{Pe2008,Belo2011,Pe2011,Deng2014}. But
hydrodynamic simulations for a jet propagating through the envelope of the
progenitor star \citep{ZhaWoo2003,Mizu2006,Mor2007,Lazz2009,Naga2011} show
that the jet should have lateral structure and rapid time variability. Thus,
in this paper we consider an angular Lorentz factor profile and a continuous
wind with a time-dependent wind luminosity. \citet{Lund2013} showed that,
with an inner-constant and outer-decreasing angular Lorentz factor profile
and\textbf{\ }steady-state jet, the photospheric spectrum can reproduce the
observed average low-energy photon index $\alpha \approx -1$. But whether
the time-resolved spectra, the spectral evolutions and $E_{p}$ evolutions
for more reasonable energy injection of continuous wind can match the
observations needs to be further considered, which have been carefully
treated in this paper. And we find that they match well for the GRBs best
fitted by the cutoff power-law model for the peak-flux spectrum or the
time-integrated spectrum. We only give a rough comparison here since we
mainly focus on the model calculations, the complete fit to the data with
the model will be further explored in future works.

In this work, we assumed a local thermal radiation spectrum for each
independently-evolving angular fluid element and ignored the sideway
diffusion effect of photons at certain angular distance. The sideway
diffusion can cause a smearing out effect on temperature and lead to a
non-thermal spectrum due to inverse Compton radiation for jets with $\theta
_{c}\sim 1/\Gamma _{0}$ (Ito et al 2013, Lundman et al 2013). Such effect
unfortunately can not be calculated by the approach in this paper. We thus
caution the spectral calculations performed in this paper when $\theta _{c}$%
\ $\sim $\ $1/\Gamma _{0}$, and especially when $\theta _{c}$\ $\ll $\ $%
1/\Gamma _{0}$.

In addition, we consider non-dissipative fireball dynamics here. The radial
distributions of the Lorentz factor and the comoving temperature in
dissipative outflows are significantly different \citep{Gian2012,Belo2013}.
Energy dissipation in the area of moderate optical depth has been proposed
by many authors, with various dissipative mechanisms such as shocks %
\citep{Pe2005,Pe2006,Lazz2010}, magnetic reconnection %
\citep{Gian2006,Gian2007,Beni2017} and proton--neutron nuclear collisions %
\citep{Belo2010,Vur2011}. Then, relativistic electrons are generated that
upscatter the thermal photons to shape the non-thermal spectrum above the
peak energy. Namely, we may get the Band function spectrum with the observed
low-energy photon index if the subphotospheric dissipation and the geometric
broadening (for structured jet) coexist. So, decided by whether the
dissipation exists, we may obtain the two kinds of spectra (COMP or Band)
for the peak-flux spectrum or the time-integrated spectrum within the
framework of the photosphere model.

\acknowledgments We thank the referee for helpful suggestions. We thank Wei
Deng for the helpful discussions. We acknowledge the use of the public data
from the \textit{Fermi} data archives. This work is partially supported by
the National Natural Science Foundation of China (grant Nos. 11603076,
11673068, 11725314, and~U1831122), the Youth Innovation Promotion
Association (2017366), the Key Research Program of Frontier Sciences (grant
No. QYZDB-SSW-SYS005), the Strategic Priority Research Program
\textquotedblleft Multi-waveband gravitational wave
Universe\textquotedblright (grant No. XDB23000000) of Chinese Academy of
Sciences, and the \textquotedblleft 333 Project\textquotedblright\ and the
Natural Science Foundation (grant No. BK20161096) of Jiangsu Province.
B.B.Z. acknowledge the support from the National Thousand Young Talents
program of China.

\bigskip

\bigskip


\begin{thebibliography}{Zhang \& M{\'e}sz{\'a}ros(2002)}
\bibitem[Abdo et al.(2009)]{Abdo2009} Abdo, A.~A., Ackermann, M., Ajello,
M., et al.\ 2009, ApJL, 706, L138

\bibitem[Abramowicz et al.(1991)]{Abra1991} Abramowicz, M.~A., Novikov,
I.~D., \& Paczynski, B.\ 1991, \apj, 369, 175

\bibitem[Acuner \& Ryde(2018)]{Acun2018} Acuner, Z., \& Ryde, F.\ 2018, %
\mnras, 475, 1708

\bibitem[Ai et al.(2018)]{Ai2018} Ai, S., Gao, H., Dai, Z.-G., et al.\ 2018, %
\apj, 860, 57

\bibitem[Axelsson et al.(2012)]{Axel2012} Axelsson, M., Baldini, L.,
Barbiellini, G., et al.\ 2012, ApJL, 757, L31

\bibitem[Axelsson \& Borgonovo(2015)]{AxBo2015} Axelsson, M., \& Borgonovo,
L.\ 2015, \mnras, 447, 3150

\bibitem[Band et al.(1993)]{Band1993} Band, D., Matteson, J., Ford, L., et
al.\ 1993, \apj, 413, 281

\bibitem[B{\'e}gu{\'e} \& Pe'er(2015)]{Be2015} B{\'e}gu{\'e}, D., \& Pe'er,
A.\ 2015, \apj, 802, 134

\bibitem[Beloborodov(2010)]{Belo2010} Beloborodov, A.~M.\ 2010, \mnras, 407,
1033

\bibitem[Beloborodov(2011)]{Belo2011} Beloborodov, A.~M.\ 2011, \apj, 737, 68

\bibitem[Beloborodov(2013)]{Belo2013} Beloborodov, A.~M.\ 2013, \apj, 764,
157

\bibitem[Beloborodov(2017)]{Belo2016} Beloborodov, A.~M.\ 2017, \apj, 838,
125

\bibitem[Beniamini et al.(2015)]{Beni2015} Beniamini, P., Nava, L., Duran,
R.~B., \& Piran, T.\ 2015, \mnras, 454, 1073

\bibitem[Beniamini \& Giannios(2017)]{Beni2017} Beniamini, P., \& Giannios,
D.\ 2017, \mnras, 468, 3202

\bibitem[Beniamini \& Nakar(2019)]{BeNa2019} Beniamini, P., \& Nakar, E.\
2019, \mnras, 482, 5430

\bibitem[Beniamini et al.(2019)]{Beni2019} Beniamini, P., Petropoulou, M.,
Barniol Duran, R., \& Giannios, D.\ 2019, \mnras, 483, 840

\bibitem[Crider et al.(1997)]{Cri1997} Crider, A., Liang, E.~P., Smith,
I.~A., et al.\ 1997, ApJL, 479, L39

\bibitem[Dai \& Gou(2001)]{Dai2001} Dai, Z.~G., \& Gou, L.~J.\ 2001, \apj,
552, 72

\bibitem[Deng \& Zhang(2014)]{Deng2014} Deng, W., \& Zhang, B.\ 2014, \apj,
785, 112

\bibitem[Fan \& Piran(2006)]{Fan2006} Fan, Y., \& Piran, T.\ 2006, \mnras,
369, 197

\bibitem[Fan et al.(2012)]{Fan2012} Fan, Y.-Z., Wei, D.-M., Zhang, F.-W., \&
Zhang, B.-B.\ 2012, ApJL, 755, L6

\bibitem[Ford et al.(1995)]{Ford1995} Ford, L.~A., Band, D.~L., Matteson,
J.~L., et al.\ 1995, \apj, 439, 307

\bibitem[Gao \& Zhang(2015)]{Gao2015} Gao, H., \& Zhang, B.\ 2015, \apj,
801, 103

\bibitem[Geng et al.(2018a)]{Geng18} Geng, J.-J., Huang, Y.-F., Wu, X.-F.,
Zhang, B., \& Zong, H.-S.\ 2018a, \apjs, 234, 3

\bibitem[Geng et al.(2018b)]{Geng2018} Geng, J.-J., Dai, Z.-G., Huang,
Y.-F., et al.\ 2018b, \apjl, 856, L33

\bibitem[Geng et al.(2019)]{Geng2019} Geng, J.-J., Zhang, B., K{\"o}lligan,
A., Kuiper, R., \& Huang, Y.-F.\ 2019, \apjl, 877, L40

\bibitem[Ghirlanda et al.(2010)]{Ghir2010} Ghirlanda, G., Nava, L., \&
Ghisellini, G.\ 2010, \aap, 511, A43

\bibitem[Ghirlanda et al.(2013)]{Ghir2013} Ghirlanda, G., Pescalli, A., \&
Ghisellini, G.\ 2013, \mnras, 432, 3237

\bibitem[Ghirlanda et al.(2019)]{Ghir2019} Ghirlanda, G., Salafia, O.~S.,
Paragi, Z., et al.\ 2019, Science, 363, 968

\bibitem[Giannios(2006)]{Gian2006} Giannios, D.\ 2006, \aap, 457, 763

\bibitem[Giannios \& Spruit(2007)]{Gian2007} Giannios, D., \& Spruit, H.~C.\
2007, \aap, 469, 1

\bibitem[Giannios(2012)]{Gian2012} Giannios, D.\ 2012, \mnras, 422, 3092

\bibitem[Goldstein et al.(2012)]{Gold2012} Goldstein, A., Burgess, J.~M.,
Preece, R.~D., et al.\ 2012, \apjs, 199, 19

\bibitem[Goodman(1986)]{Good1986} Goodman, J.\ 1986, ApJL, 308, L47

\bibitem[Guetta et al.(2001)]{Gue2001} Guetta, D., Spada, M., \& Waxman, E.\
2001, \apj, 557, 399

\bibitem[Guiriec et al.(2013)]{Gui2013} Guiriec, S., Daigne, F., Hasco{\"e}%
t, R., et al.\ 2013, \apj, 770, 32

\bibitem[Guiriec et al.(2011)]{Gui2011} Guiriec, S., Connaughton, V.,
Briggs, M.~S., et al.\ 2011, ApJL, 727, L33

\bibitem[Hou et al.(2018)]{Hou2018} Hou, S.-J., Zhang, B.-B., Meng, Y.-Z.,
et al.\ 2018, \apj, 866, 13

\bibitem[Ito et al.(2013)]{Ito2013} Ito, H., Nagataki, S., Ono, M., et al.\
2013, \apj, 777, 62

\bibitem[Kaneko et al.(2006)]{Kan2006} Kaneko, Y., Preece, R.~D., Briggs,
M.~S., et al.\ 2006, \apjs, 166, 298

\bibitem[Kino et al.(2004)]{Kin2004} Kino, M., Mizuta, A., \& Yamada, S.\
2004, \apj, 611, 1021

\bibitem[Kobayashi et al.(1997)]{Koba1997} Kobayashi, S., Piran, T., \&
Sari, R.\ 1997, \apj, 490, 92

\bibitem[Kocevski et al.(2003)]{Koce2003} Kocevski, D., Ryde, F., \& Liang,
E.\ 2003, \apj, 596, 389

\bibitem[Lan et al.(2019)]{Lan2019} Lan, M.-X., Geng, J.-J., Wu, X.-F., \&
Dai, Z.-G.\ 2019, \apj, 870, 96

\bibitem[Larsson et al.(2015)]{Lar2015} Larsson, J., Racusin, J.~L., \&
Burgess, J.~M.\ 2015, ApJL, 800, L34

\bibitem[Lazzati et al.(1999)]{Lazz1999} Lazzati, D., Ghisellini, G., \&
Celotti, A.\ 1999, \mnras, 309, L13

\bibitem[Lazzati et al.(2009)]{Lazz2009} Lazzati, D., Morsony, B.~J., \&
Begelman, M.~C.\ 2009, ApJL, 700, L47

\bibitem[Lazzati \& Begelman(2010)]{Lazz2010} Lazzati, D., \& Begelman,
M.~C.\ 2010, \apj, 725, 1137

\bibitem[Lazzati et al.(2013)]{Lazz2013} Lazzati, D., Morsony, B.~J.,
Margutti, R., \& Begelman, M.~C.\ 2013, \apj, 765, 103

\bibitem[Lazzati et al.(2018)]{Lazzati2018} Lazzati, D., Perna, R., Morsony,
B.~J., et al.\ 2018, Physical Review Letters, 120, 241103

\bibitem[Li et al.(2018)]{LiB2018} Li, B., Li, L.-B., Huang, Y.-F., et al.\
2018, \apjl, 859, L3

\bibitem[Li(2019b)]{Li2019b} Li, L.\ 2019, \apjs, 242, 16

\bibitem[Li(2019a)]{Li2019} Li, L.\ 2019, arXiv:1905.02340

\bibitem[Li et al.(2019)]{LiL2019} Li, L.-B., Geng, J.-J., Huang, Y.-F., \&
Li, B.\ 2019, arXiv:1901.08266

\bibitem[Liang \& Kargatis(1996)]{Lia1996} Liang, E., \& Kargatis, V.\ 1996, %
\nat, 381, 49

\bibitem[Lin et al.(2018)]{Lin2018} Lin, D.-B., Liu, T., Lin, J., et al.\
2018, \apj, 856, 90

\bibitem[Lu et al.(2010)]{Lu2010} Lu, R.-J., Hou, S.-J., \& Liang, E.-W.\
2010, \apj, 720, 1146

\bibitem[Lu et al.(2012)]{Lu2012} Lu, R.-J., Wei, J.-J., Liang, E.-W., et
al.\ 2012, \apj, 756, 112

\bibitem[Lundman et al.(2013)]{Lund2013} Lundman, C., Pe'er, A., \& Ryde,
F.\ 2013, \mnras, 428, 2430

\bibitem[L{\"u} et al.(2017)]{Lv2017} L{\"u}, H.-J., L{\"u}, J., Zhong,
S.-Q., et al.\ 2017, \apj, 849, 71

\bibitem[Lyman et al.(2018)]{Lyman2018} Lyman, J.~D., Lamb, G.~P., Levan,
A.~J., et al.\ 2018, Nature Astronomy, 2, 751

\bibitem[MacFadyen \& Woosley(1999)]{Mac1999} MacFadyen, A.~I., \& Woosley,
S.~E.\ 1999, \apj, 524, 262

\bibitem[Meng et al.(2018)]{Meng2018} Meng, Y.-Z., Geng, J.-J., Zhang,
B.-B., et al.\ 2018, \apj, 860, 72

\bibitem[M{\'e}sz{\'a}ros \& Rees(2000)]{Me2000} M{\'e}sz{\'a}ros, P., \&
Rees, M.~J.\ 2000, \apj, 530, 292

\bibitem[Mizuta et al.(2006)]{Mizu2006} Mizuta, A., Yamasaki, T., Nagataki,
S., \& Mineshige, S.\ 2006, \apj, 651, 960

\bibitem[Mizuta, Nagataki \& Aoi(2011)]{Mizu2011} Mizuta, A., Nagataki, S.,
\& Aoi, J.\ 2011, \apj, 732, 26

\bibitem[Mooley et al.(2018)]{Mooley2018} Mooley, K.~P., Deller, A.~T.,
Gottlieb, O., et al.\ 2018, \nat, 561, 355

\bibitem[Morsony, Lazzati \& Begelman(2007)]{Mor2007} Morsony, B.~J.,
Lazzati, D., \& Begelman, M.~C.\ 2007, \apj, 665, 569

\bibitem[Nagakura et al.(2011)]{Naga2011} Nagakura, H., Ito, H., Kiuchi, K.,
\& Yamada, S.\ 2011, \apj, 731, 80

\bibitem[Nava et al.(2011)]{Nava2011} Nava, L., Ghirlanda, G., Ghisellini,
G., \& Celotti, A.\ 2011, \aap, 530, A21

\bibitem[Norris et al.(2005)]{Norr2005} Norris, J.~P., Bonnell, J.~T.,
Kazanas, D., et al.\ 2005, \apj, 627, 324

\bibitem[Paczynski(1986)]{Pac1986} Paczynski, B.\ 1986, ApJL, 308, L43

\bibitem[Pe'er et al.(2005)]{Pe2005} Pe'er, A., M{\'e}sz{\'a}ros, P., \&
Rees, M.~J.\ 2005, \apj, 635, 476

\bibitem[Pe'er et al.(2006)]{Pe2006} Pe'er, A., M{\'e}sz{\'a}ros, P., \&
Rees, M.~J.\ 2006, \apj, 642, 995

\bibitem[Pe'er(2008)]{Pe2008} Pe'er, A.\ 2008, \apj, 682, 463

\bibitem[Pe'er \& Ryde(2011)]{Pe2011} Pe'er, A., \& Ryde, F.\ 2011, \apj,
732, 49

\bibitem[Pe'er et al.(2015)]{Pe2015} Pe'er, A., Barlow, H., O'Mahony, S., et
al.\ 2015, \apj, 813, 127

\bibitem[Pescalli et al.(2016)]{Pes2016} Pescalli, A., Ghirlanda, G.,
Salvaterra, R., et al.\ 2016, \aap, 587, A40

\bibitem[Piran(1999)]{Pi1999} Piran, T.\ 1999, \physrep, 314, 575

\bibitem[Preece et al.(1998)]{Pree1998} Preece, R.~D., Briggs, M.~S.,
Mallozzi, R.~S., et al.\ 1998, ApJL, 506, L23

\bibitem[Preece et al.(2000)]{Pree2000} Preece, R.~D., Briggs, M.~S.,
Mallozzi, R.~S., et al.\ 2000, \apjs, 126, 19

\bibitem[Rees \& Meszaros(1994)]{Ree1994} Rees, M.~J., \& Meszaros, P.\
1994, ApJL, 430, L93

\bibitem[Rees \& M{\'e}sz{\'a}ros(2005)]{Ree2005} Rees, M.~J., \& M{\'e}sz{%
\'a}ros, P.\ 2005, \apj, 628, 847

\bibitem[Rossi et al.(2002)]{Ro2002} Rossi, E., Lazzati, D., \& Rees, M.~J.\
2002, \mnras, 332, 945

\bibitem[Ruffini et al.(2013)]{Ru2013} Ruffini, R., Siutsou, I.~A., \&
Vereshchagin, G.~V.\ 2013, \apj, 772, 11

\bibitem[Ryde(2004)]{Ry2004} Ryde, F.\ 2004, \apj, 614, 827

\bibitem[Ryde(2005)]{Ry2005} Ryde, F.\ 2005, ApJL, 625, L95

\bibitem[Ryde \& Pe'er(2009)]{Ry2009} Ryde, F., \& Pe'er, A.\ 2009, \apj,
702, 1211

\bibitem[Ryde et al.(2010)]{Ry2010} Ryde, F., Axelsson, M., Zhang, B.~B., et
al.\ 2010, ApJL, 709, L172

\bibitem[Ryde et al.(2017)]{Ry2017} Ryde, F., Lundman, C., \& Acuner, Z.\
2017, \mnras, 472, 1897

\bibitem[Thompson(1994)]{Thom1994} Thompson, C.\ 1994, \mnras, 270, 480

\bibitem[Toma et al.(2011)]{Toma2011} Toma, K., Wu, X.-F., \& M{\'e}sz{\'a}%
ros, P.\ 2011, \mnras, 415, 1663

\bibitem[Vurm et al.(2011)]{Vur2011} Vurm, I., Beloborodov, A.~M., \&
Poutanen, J.\ 2011, \apj, 738, 77

\bibitem[Vurm \& Beloborodov(2016)]{Vur2016} Vurm, I., \& Beloborodov,
A.~M.\ 2016, \apj, 831, 175

\bibitem[Wang et al.(2019)]{Wang2019} Wang, Y.-Z., Shao, D.-S., Jiang,
J.-L., et al.\ 2019, \apj, 877, 2

\bibitem[Wei et al.(2017)]{Wei2017} Wei, J.-J., Zhang, B.-B., Shao, L., Wu,
X.-F., \& M{\'e}sz{\'a}ros, P.\ 2017, \apjl, 834, L13

\bibitem[Yu et al.(2015)]{Yu2015} Yu, H.-F., van Eerten, H.~J., Greiner, J.,
et al.\ 2015, \aap, 583, A129

\bibitem[Yu et al.(2016)]{Yu2016} Yu, H.-F., Preece, R.~D., Greiner, J., et
al.\ 2016, \aap, 588, A135

\bibitem[Yu et al.(2018)]{Yu2018} Yu, H.-F., Dereli-B{\'e}gu{\'e}, H., \&
Ryde, F.\ 2018, arXiv:1810.07313

\bibitem[Zhang et al.(2011)]{ZhaBB2011} Zhang, B.-B., Zhang, B., Liang,
E.-W., et al.\ 2011, \apj, 730, 141

\bibitem[Zhang et al.(2018a)]{ZhaBB2018} Zhang, B.-B., Zhang, B.,
Castro-Tirado, A.~J., et al.\ 2018a, Nature Astronomy, 2, 69

\bibitem[Zhang et al.(2018b)]{ZhangBB18b} Zhang, B.-B., Zhang, B., Sun, H.,
et al.\ 2018b, Nature Communications, 9, 447

\bibitem[Zhang \& M{\'e}sz{\'a}ros(2002)]{ZhaB2002} Zhang, B., \& M{\'e}sz{%
\'a}ros, P.\ 2002, \apj, 581, 1236

\bibitem[Zhang et al.(2004)]{ZhaB2004} Zhang, B., Dai, X., Lloyd-Ronning,
N.~M., \& M{\'{e}}sz{\'{a}}ros, P.\ 2004, ApJL, 601, L119

\bibitem[Zhang et al.(2007)]{ZhaB2007} Zhang, B., Liang, E., Page, K.~L., et
al.\ 2007, \apj, 655, 989

\bibitem[Zhang(2011)]{ZhaB2011} Zhang, B.\ 2011, Comptes Rendus Physique,
12, 206

\bibitem[Zhang \& Yan(2011)]{Zhang2011} Zhang, B., \& Yan, H. 2011, \apj,
726, 90

\bibitem[Zhang,Woosley \& MacFadyen(2003)]{ZhaWoo2003} Zhang, W., Woosley,
S.~E., \& MacFadyen, A.~I.\ 2003, \apj, 586, 356
\end{thebibliography}
\end{document}